\documentclass[12pt]{article}
\pdfoutput=1

\usepackage{hyphenat}
\usepackage{putex}
\usepackage{graphicx}
\usepackage{caption}
\usepackage{amsmath, amsbsy, mathtools}
\usepackage{array}
\usepackage{subcaption}
\usepackage{epstopdf}
\usepackage{enumerate}
\usepackage{cite}
\usepackage{youngtab}
\usepackage{tensor}
\usepackage{slashed}
\usepackage[aligntableaux=center]{ytableau}
\usepackage[utf8]{inputenc}
\usepackage[
      colorlinks=true,
      linkcolor=blue,
      urlcolor=blue,
      filecolor=blacsk,
      citecolor=red,
      ]{hyperref}
\usepackage{dsfont}
\usepackage{slashed}
\usepackage{bm}
\usepackage{bbm}
\usepackage{booktabs}
\usepackage{colortbl}
\usepackage{comment}
\usepackage{xcolor}

\definecolor{99_1}{rgb}{0.65,0.,0.}
\definecolor{99_2}{rgb}{0.0504678,0.526626,0.627561}
\definecolor{99_3}{rgb}{0.752461,0.362306,0.125339}
\definecolor{99_4}{rgb}{0.435888,0.259065,0.71028}

\newcommand{\grp}[1]{\mathrm{#1}}

\newcommand{\grSU}{\grp{SU}}

\newcommand {\be} {\begin {equation}}
\newcommand {\ee} {\end {equation}}

\newcommand {\bes} {\begin {equation*}}
\newcommand {\ees} {\end {equation*}}

\newcommand{\Z}{\mathbb{Z}}

\newcommand{\beq}{\begin{equation}}
\newcommand{\eeq}{\end{equation}}

\def\ie{\begin{equation}\begin{aligned}}
\def\fe{\end{aligned}\end{equation}}

\numberwithin{equation}{section}

\def\<{\langle}
\def\>{\rangle}

\newcommand{\bra}[1]{\langle #1|}
\newcommand{\ket}[1]{|#1\rangle}
\newcommand{\braket}[1]{\langle#1\rangle}

\newcommand{\adj}[1]{\hat{#1}^{\dagger}}

\newcommand{\rk}{\rho}

\let\tr\relax
\DeclareMathOperator{\tr}{tr}
\DeclareMathOperator{\SU}{SU}
\DeclareMathOperator{\SO}{SO}
\DeclareMathOperator{\USp}{USp}
\DeclareMathOperator{\Spin}{Spin}
\DeclareMathOperator{\su}{\mathfrak{su}}
\DeclareMathOperator{\so}{\mathfrak{so}}
\let\sp\relax
\DeclareMathOperator{\sp}{\mathfrak{sp}}
\let\rk\relax
\DeclareMathOperator{\rk}{rk}

\def\H{\mathcal{H}}

\def\adj{\textbf{adj}}

\usepackage{tikz}
\usetikzlibrary{calc}
\usetikzlibrary{decorations.markings,decorations.pathreplacing}
\tikzstyle{tensor}=[circle,draw=black!50,thick, minimum size = 40pt]

\tikzstyle{square}=[fill=white, draw=black, shape=rectangle, rounded corners, minimum width=1.cm, minimum height=1.cm, inner sep=0pt, outer sep=0pt,thick]
\tikzstyle{bcircle}=[fill=white, draw=black, shape=circle, minimum width=1.cm, minimum height=1.cm, inner sep=0pt, outer sep=0pt, thick]
\tikzstyle{shade}=[dotted,fill=gray!30,rounded corners=5mm]
\tikzset{
	diagram/.style={
		baseline={([yshift=-.5ex]current bounding box.center)},
		every node/.style={scale=0.7},
		thick
	}
}

\let\bar\relax
\newcommand{\bar}[1]{\overline{#1}}

\graphicspath{{figures/}}

\begin{document}

\preprint{PUPT-2660\\MIT-CTP/5983}

\institution{MIT}{Center for Theoretical Physics -- a Leinweber Institute, Massachusetts Institute of Technology, Cambridge, MA 02139, USA}
\institution{PU}{Joseph Henry Laboratories, Princeton University, Princeton, NJ 08544, USA}
\institution{PCTS}{Princeton Center for Theoretical Science, Princeton University, Princeton, NJ 08544, USA}

\title{Vacuum structure of gapped QCD$_2$ theories\\from the infinite Hamiltonian lattice}

\authors{Ross Dempsey,\worksat{\MIT} Anna-Maria E.~Gl\"uck,\worksat{\PU} Silviu S.~Pufu,\worksat{\PU,\PCTS}\\[10pt] and Benjamin T.~S\o{}gaard\worksat{\PU}}

\abstract{Gapped two-dimensional gauge theories with massless fermions generically have rich vacuum structures consisting of many degenerate vacua related by the action of topological line operators. The algebra of such operators has been used to calculate ratios of vacuum expectation values of local operators and to predict nontrivial particle-soliton degeneracies. In this paper, we use recently-developed tensor network methods to study several examples of such theories via their Hamiltonian lattice descriptions. Our lattice results agree with all previously-made predictions. Furthermore, we identify the lattice strong-coupling states that can be adiabatically continued to the degenerate vacua in the continuum limit. We conjecture a procedure, referred to as a lattice decay rule, for how this identification works in general. This rule allows us to compute the continuum vacuum degeneracy by studying the lattice Hamiltonian in the strong-coupling limit.
}

\date{}

\maketitle

\tableofcontents
\section{Introduction}
In recent years, there has been substantial progress on using non-invertible topological line operators \cite{Bhardwaj:2017xup,Chang:2018iay} to determine the infrared (IR) phases of specific quantum field theories \cite{Komargodski:2020mxz,Nguyen:2021naa,Huang:2021zvu,Cordova:2024nux,Cordova:2024goh,Damia:2024kyt}. In particular, in two-dimensional QCD-like theories with {\em massless} fermionic matter, the IR phase is described by a coset model \cite{Bardakci:1987ee,Kutasov:1993gq,Kutasov:1994xq}. When this coset model has vanishing Virasoro central charge, the theory is gapped, and so the IR theory is a topological quantum field theory (TQFT)\@. This TQFT contains several topological line operators, which are generically non-invertible, and these line operators can be used to relate degenerate states of the theory. In particular, using the detailed structure of the topological lines, it has been possible to calculate ratios of expectation values of local operators in the degenerate vacua of the coset model \cite{Komargodski:2020mxz,Cordova:2024nux}, and also to predict nontrivial degeneracy patterns among massive excitations \cite{Cordova:2024nux}.

In this paper, we will study gapped two-dimensional QCD-like theories on the lattice and test some of these predictions. In doing so, we face a conceptual challenge: in known lattice regularizations of these theories \cite{Dempsey:2023fvm,Bergner:2024ttq,Dempsey:2024alw,Dempsey:2025wia}, the non-invertible symmetries that underlie these predictions are generically broken and vacuum degeneracies are lifted, re-emerging only in the continuum limit. One would thus expect that the vacua and excitations described in e.g. \cite{Komargodski:2020mxz,Cordova:2024nux}, while crisply distinguished and exactly degenerate in the continuum limit, are mixed with one another and thus much harder to identify on the lattice. However, we will find that by using uniform matrix product states (uMPSs) \cite{Vidal:2006ofj,Zauner-Stauber:2017eqw} and the link-enhanced matrix product operators (LEMPOs) introduced in \cite{Dempsey:2025wia}, it is possible to identify the lattice states that become degenerate vacua in the continuum limit, and furthermore to study the spectra of excitations above each of these vacua.

The vacuum structure of QCD-like theories relates to the physics of (de)confinement. For instance, the case of gauge group $G = \SU(N)$ and matter in the representation $\lambda = \adj$,\footnote{When $\lambda = \adj$, the infrared coset model has vanishing central charge for any gauge group, and so the IR is a TQFT and the theory is gapped.} i.e.~adjoint QCD$_2$ \cite{Dalley:1992yy}, has been studied in \cite{Komargodski:2020mxz}.\footnote{Part of the spectrum of this theory has been studied for a long time using lightcone quantization in \cite{Dalley:1992yy,Kutasov:1993gq,Bhanot:1993xp,Gross:1997mx,Katz:2013qua,Trittmann:2015oka,Dempsey:2021xpf,Dempsey:2022uie,Trittmann:2024jkf}.} For this model, it has long been known that the massless adjoint fermion leads to deconfinement of probe fundamental quarks \cite{Gross:1995bp,Gross:1997mx,Kutasov:1994xq,Antonuccio:1998uz}; indeed, a TQFT description of the IR precludes an area law for Wilson loops. As described in \cite{Komargodski:2020mxz}, this fact can be understood in terms of vacuum degeneracy. The theory has a $\mathbb{Z}_N$ one-form center symmetry that distinguishes $N$ distinct flux tube sectors \cite{Witten:1978ka}, sometimes called universes, and it was demonstrated in \cite{Komargodski:2020mxz} (see also \cite{Dempsey:2024ofo}) that there are $2^{N-1}$ degenerate vacua distributed among all of these sectors. This means the vacuum energy density in the presence of (say) a fundamental flux tube is the same as the vacuum energy density in the absence of that flux tube, so that the fundamental string is tensionless (and likewise with all $k$-strings). Importantly, as explained in \cite{Cherman:2019hbq}, the invertible symmetries and their mixed anomalies would only enforce the vanishing of the $\frac{N}{2}$-string tension for even $N$; in particular, the vanishing of the fundamental string tension is implied by invertible symmetries only for the $\SU(2)$ theory. A symmetry-based argument for the vanishing of the other string tensions must rely upon the non-invertible lines (and their anomalies). 

The conceptual distinction between invertible and non-invertible symmetries becomes a practical distinction on the lattice: currently known Hamiltonian lattice models for QCD-like theories can preserve the former but not the latter. As a consequence, the Hamiltonian lattice model introduced in \cite{Dempsey:2023fvm} for $\SU(2)$ adjoint QCD$_2$ has exactly vanishing fundamental string tension, but in a similar model for $\SU(3)$ adjoint QCD$_2$ the fundamental string tension is nonzero on the lattice and vanishes only in the continuum limit \cite{Dempsey:2024alw}.\footnote{Similarly, in Euclidean lattice models the string tension is nonzero on the lattice and can only be seen to vanish after taking continuum and chiral limits \cite{Bergner:2024ttq}.} The lack of non-invertible symmetries on the lattice\footnote{There do exist many lattice models with non-invertible symmetries; see, for instance, \cite{Aasen:2016dop,Aasen:2020jwb,Bhardwaj:2024kvy,Chatterjee:2024ych,Seifnashri:2024dsd,Seiberg:2023cdc,Koide:2021zxj,Li:2024gwx,Fidkowski:2025rsq,Gorantla:2024ocs}. However, no such model for QCD-like theories is presently available.} becomes a severe problem if we wish to study degenerate vacua within the same flux tube sector whose degeneracy is not protected by any invertible symmetry (which, for adjoint QCD$_2$, occurs for the first time in the $\SU(4)$ theory \cite{Komargodski:2020mxz}). Most of the several continuum vacua will differ in energy by an amount that scales with the system size, and so isolating them seems hopeless. As we will demonstrate, by working on an infinite lattice, we can address this problem using cluster decomposition. Uniform matrix product states (uMPSs) are generically cluster-decomposing (see \cite{Vanderstraeten:2019voi}), and by searching for points in the space of uMPS ans\"atze that are local minima of energy density, we are able to identify the lattice states that are adiabatically connected to the degenerate vacua in the continuum limit, including the ones that are not global minima of the energy density at finite lattice spacing $a$. This setup allows us to test many detailed predictions that have been made using non-invertible symmetries, and it provides us with a novel technique for studying the microscopic details of gauge theory states that are related to one another by these symmetries.

Indeed, since our method can adiabatically follow the degenerate vacua from the $ga\to 0$ limit out to finite lattice spacing, it is natural to track them all the way into the $ga\to\infty$ lattice strong-coupling limit. In this regime, the lattice Hamiltonian becomes much simpler, and we can work out the set of uMPSs that are local minima of the energy density without performing any numerical calculations. We formulate this procedure in more physical terms as a lattice decay rule, which posits that strong-coupling states that cannot decay into other states (in a sense made precise in Section~\ref{sec:lattice_decay}) are adiabatically connected to degenerate vacua in the continuum. This rule can be expressed solely in terms of the representation theory of $G$. It is surprising that it appears possible to determine the number of degenerate vacua, whose existence has heretofore only been demonstrated by studying coset models with non-invertible symmetries, in a limit on the lattice for which these symmetries should be most strongly violated.

To demonstrate our findings, we focus on four examples, two of which have been studied previously and two of which have not: 
\begin{center}
    \begin{tabular}{ccc}
        \toprule
        $G$ & $\bm \lambda$ & Literature \\
        \midrule
     $\SU(2)$ & $\bm 5$ & Studied in \cite{Cordova:2024nux}; numerically in \cite{Narayanan:2023jmi}\\
        $\SU(2)\times \SU(2)$ & $(\bm{3},\bm{3})$ & Not studied previously\\
        $\SU(2)\times \SU(2)$ & $(\bm{2},\bm{4})$ & Not studied previously\\
        $\SU(4)$ & $\bm{15}$ & Studied in \cite{Komargodski:2020mxz,Cordova:2024nux}; numerically in \cite{Dempsey:2022uie}\\
         \bottomrule
    \end{tabular}
    \captionof{table}{Gauge theories studied in the work.}
    \label{tab:theories}
\end{center}

\noindent In each of these theories, we observe nontrivial vacuum degeneracy in the continuum limit of our lattice model. We also calculate the chiral condensate $\langle \bar\psi_a \psi_a\rangle$ of the fermions $\psi$ in each vacuum, and find that the ratios of these condensates match the predictions made in \cite{Komargodski:2020mxz,Cordova:2024nux}. Finally, for all except the $\SU(4)$ theory (for which it is too numerically taxing), we calculate low-lying spectra above each vacuum, and observe non-trivial particle degeneracies as predicted by \cite{Cordova:2024nux} (along with indirect evidence of the particle-soliton degeneracies they predicted).

The rest of this paper is organized as follows. In Section~\ref{sec:ggt}, we describe the continuum theories we are studying and some salient details of their Hamiltonian lattice formulation. In Section~\ref{sec:mps}, we describe the infinite MPS methods we are using, and explain what we find in the lattice strong-coupling limit. In Section~\ref{sec:results}, we present our numerical results. We discuss and conclude in Section~\ref{sec:discussion}. Several technical details are relegated to the appendices.

\section{Gapped gauge theories}\label{sec:ggt}

In this section, we introduce the models studied in this work. To that end, we first discuss the continuum formulation of (1+1)D gauge theories coupled to Majorana fermions and the symmetries of these theories in Section~\ref{subsec:continuum_action}. In particular, we describe the different flux tube sectors distinguished by center one-form symmetry, and we give the numbers of degenerate vacua predicted from the coset model in the infrared. Then, in Section~\ref{subsec:lattice_hamiltonian}, we discuss the Hamiltonian lattice formulation of these theories. 

\subsection{Continuum model}
\label{subsec:continuum_action}
In the present work, we are studying $(1+1)$D QCD-like theories consisting of Yang-Mills theory with gauge group $G$ coupled to a Majorana fermion of mass $m$ transforming in a real representation $\bm \lambda$. The continuum action for these theories is given by 
\begin{equation}\label{eq:contiuum_action}
    S = \int d^2x\left(-\frac{1}{4g^2}F^{\mu\nu}_AF^A_{\mu\nu} +\frac{i}{2}\bar\psi_a\gamma^\mu (D_\mu\psi)_a - \frac{m}{2} \bar\psi_a \psi_a\right)\,,
\end{equation}
where $A=1,2,\ldots,\dim G$ are adjoint indices, and $a=1,2,\ldots,\dim\bm{\lambda}$ are indices of the representation $\bm \lambda$. The field strength and covariant derivative are defined as $F^A_{\mu\nu}=\partial_\mu A_\nu^A-\partial_\nu A_m^A+f^{ABC}A^B_\mu A^C_\nu$ and $(D_\mu \psi)_a= \partial_\mu\psi_a-iA_\mu^A T^A_{ab}\psi_b$, respectively, where $f^{ABC}$ are the structure constants of the Lie algebra $\mathfrak g$ of $G$ and $T^A_{ab}$ are the $\bm \lambda$-representation matrices. These are related by
\begin{equation}\label{eq:lambdarep}
    [T^A,T^B]= if^{ABC}T^C\,.
\end{equation}
We will work in conventions $\gamma^\mu = (\sigma_2,-i\sigma_3)$, so the metric is $\eta_{\mu\nu} =\operatorname{diag}(1,-1)$ and the chirality matrix is $\gamma^5=\gamma^0\gamma^1 = \sigma_1$. In these conventions, the Majorana condition amounts to the reality of the spinor components $\psi^*_a = \psi_a$, which allows us to choose the generators of the representation $\bm \lambda$ to be purely imaginary:
\begin{equation}
    \left(T^A_{ab}\right)^* = T^A_{ba} = -T^A_{ab}\,.
\end{equation}

We will mainly be interested in the theory \eqref{eq:contiuum_action} with massless Marjorana fermions ($m=0$). In this case, the theory has a $\Z_2$ chiral symmetry generated by $\psi^a \mapsto \gamma^5\psi^a$. There is also a $\Z_2$ fermion parity symmetry generated by $\psi^a \mapsto -\psi^a$, provided this transformation is not a gauge transformation, and an outer automorphism symmetry, provided $G$ has a nontrivial outer automorphism group (for instance, when $G=\SU(N)$ with $N>2$, the outer automorphism group is the $\Z_2$ charge conjugation symmetry). These zero-form symmetries may have mixed anomalies among each other, as has been demonstrated for adjoint QCD with gauge group $G=\SU(N)$ \cite{Cherman:2019hbq}. See Appendix~\ref{app:symmetries} for more details on the internal invertible symmetries of the theories in Table~\ref{tab:theories}.

In addition to the zero-form invertible symmetries of gauge theories coupled to matter, a generic theory \eqref{eq:contiuum_action} will also have a one-form symmetry \cite{Gaiotto:2014kfa} under which Wilson lines are charged. For most of the models in Table \ref{tab:theories}, the matter representation $\bm \lambda$ is invariant under the action of the center $Z(G)$ of the gauge group. In this case, QCD with matter in $\bm \lambda$ has a one-form symmetry equal to $Z(G)$.\footnote{In pure 2D Yang-Mills theory, the invertible one-form symmetry is $Z(G)$, but there is a richer non-invertible one-form symmetry as well \cite{Nguyen:2021naa}.} If instead part of $Z(G)$ acts non-trivially on $\bm \lambda$, then the one-form symmetry is explicitly broken to a subgroup $Z_{\bm \lambda}(G) \subset Z(G)$.
By diagonalizing the action of $Z_{\bm \lambda}(G)$, one can split the Hilbert space of the theory into so-called flux tube sectors, each labeled by a representation of $Z_{\bm \lambda}(G)$. These superselection sectors can be interpreted as sectors of the theory in which there is a background flux sourced by appropriate quarks and antiquarks at $-\infty$ and $+\infty$, respectively.

For the models in Table \ref{tab:theories}, the one-form symmetry and flux tube sectors are given in Table~\ref{tab:vacua}. The values of $p$ label the representations of $Z_{\bm\lambda}(G)$. Note that for $\SU(2) \times \SU(2) + \psi_{(\bm{2}, \bm{4})}$, the matter representation breaks the center symmetry $Z(\SU(2)\times \SU(2)) = \mathbb{Z}_2\times\mathbb{Z}_2$ to the diagonal $\mathbb{Z}_2$, and so the flux tube sectors are labeled by a single value indicating whether the background flux representations $(\bm \lambda_1, \bm \lambda_2)$ have the same or opposite $N$-ality. In the other cases, the matter representation acts trivially on the center of the group and so $Z_{\bm{\lambda}}(G) = Z(G)$.

At the massless point of the theory \eqref{eq:contiuum_action}, one might expect that the massless Majorana fermions give rise to gapless excitations. However, as mentioned in the introduction, for certain choices of $G$ and $\bm \lambda$ the IR dynamics can be gapped \cite{Delmastro:2021otj,Schellekens:1986mb}. In order to understand this phenomenon, we note that in two spacetime dimensions, the gauge coupling $g$ is relevant and the deep IR dynamics naively can be accessed by taking the limit $g\rightarrow \infty$. Under this assumption, the IR limit amounts to dropping the gauge kinetic term in the action \eqref{eq:contiuum_action}, which leaves only the action of the gauged massless Majorana fermion. Consequently, the resulting IR theory is described by the fermionic\footnote{Note that many papers, e.g. \cite{Komargodski:2020mxz,Cordova:2024nux}, study the bosonic version of this coset model, $\frac{\Spin(\dim\bm{\lambda})_1}{G_{I(\bm{\lambda})}}$.} coset\footnote{The coset \eqref{eq:coset} is for a simple gauge group $G$. If instead the gauge group is $G_1 \times G_2$ and the matter is in representation $(\bm{\lambda}_1, \bm{\lambda}_2)$, then the coset is 
\begin{equation}
\frac{\SO(\dim \bm{\lambda}_1 \times \dim \bm{\lambda}_2)_1}{(G_1)_{I(\bm{\lambda}_1) \dim \bm{\lambda}_2} \times (G_2)_{I(\bm{\lambda}_2) \dim \bm{\lambda}_1}}\,.
\end{equation}} CFT \cite{Witten:1983ar,Bardakci:1987ee}
\begin{equation}\label{eq:coset}
    \frac{\SO(\dim \bm \lambda)_1}{G_{I(\bm \lambda)}}\,,
\end{equation}
where $I(\bm{\lambda})$ is the Dynkin index\footnote{\label{fn:index}The Dynkin index $I(\bm \lambda)$ of a irreducible representation $\bm \lambda$ can be defined by the through the trace $\tr_{\bm \lambda} T^AT^B = I(\bm \lambda)\delta^{AB}$. Contracting the adjoint indices of the definition gives the relation $I(\bm \lambda)=\frac{C_2(\bm \lambda)\dim\bm \lambda}{\dim G}$, where $C_2(\bm \lambda)$ is the quadratic Casimir eigenvalue of $\bm \lambda$.} of the matter representation $\bm \lambda$. Depending on the gauge group $G$ and matter representation $\bm \lambda$, the coset \eqref{eq:coset} can be either a CFT describing gapless excitations or a TQFT consisting of only a finite number of degenerate ground states. In the latter case, the full theory is gapped, with massive excitations above each of the degenerate ground states.

The diagnostic for determining whether the coset \eqref{eq:coset} is a non-trivial CFT or a TQFT is its Virasoro central charge $c_\text{IR}$, which (for a simple gauge group) can be calculated via
\begin{equation}\label{eq:cir}
    c_\text{IR} = \frac{\dim \bm \lambda}{2} - \frac{I(\bm \lambda) \dim G}{I(\bm \lambda) + h^\vee}\,,
\end{equation}
where $h^\vee$ is the dual Coxeter number\footnote{The dual Coxeter number is the Dynkin index of the adjoint representation. For $\SU(N)$, we have $h^\vee = N$.} of the Lie algebra $\mathfrak g$. If the IR central charge vanishes, the coset is a TQFT, whereas when $c_\text{IR} > 0$, the coset is a non-trivial CFT. For each of the four theories in Table~\ref{tab:vacua}, one can check that the central charge of the IR coset model is zero.

\begin{table}[]
    \centering
    \begingroup
    \renewcommand{\arraystretch}{1.5}
    \scalebox{.92}{
    \begin{tabular}{ccccc}
        \toprule
        Theory & IR coset & \# of vacua & one-form sym. & Flux tube sectors \\
        \midrule
         $\SU(2) + \psi_{\bm{5}}$ & $\frac{\SO(5)_1}{\SU(2)_{10}}$ & $4=2+2$ & $\mathbb{Z}_2$ & $p=0,1$ \\
         $\SU(2) \times \SU(2) + \psi_{(\bm{3}, \bm{3})}$ & $\frac{\SO(9)_1}{\SU(2)_6\times \SU(2)_6}$ & $8=2+2+2+2$ & $\mathbb{Z}_2 \times \mathbb{Z}_2 $ & $p=(p_1, p_2), \, p_i=0,1$ \\
         $\SU(2) \times \SU(2) + \psi_{(\bm{2}, \bm{4})}$ & $\frac{\SO(8)_1}{\SU(2)_2\times \SU(2)_{10}}$ & $6=3+3$ & $\mathbb{Z}_2$ & $p=0,1$ \\
         $\SU(4) + \psi_{\bm{15}}$ & $\frac{\SO(15)_1}{\SU(4)_{4}}$ & $8=2+2+2+2$ & $\mathbb{Z}_4$ & $p=0,1,2,3$ \\       
         \bottomrule
    \end{tabular}
    }
    \endgroup
    \caption{One-form symmetry and IR data for each of the theories in Table \ref{tab:theories}. The partition of the total number of vacua indicate how the vacua are distributed across the flux tube sectors.}
    \label{tab:vacua}
\end{table}

For gapped gauge theories, the IR coset \eqref{eq:coset} can also be used to compute the number of degenerate vacua, as demonstrated in \cite{Delmastro:2021otj}. In particular, one can determine the partition function of an IR TQFT by studying the branching functions corresponding to the affine algebra embedding $G_{I(\bm \lambda)}\subset\SO(\dim \bm \lambda)_1$ (see Appendix \ref{app:count_checks} for more details). The predicted number of vacua for each of the gauge theories studied in this work is reported in Table~\ref{tab:vacua}. For all these theories, the vacuum degeneracy is too large to be explained by the invertible zero- and one-form symmetries, and so it can only be explained through the use of the non-invertible symmetries of \eqref{eq:coset}.

\subsection{Lattice model}
\label{subsec:lattice_hamiltonian}

To formulate a Hamiltonian lattice realization of the continuum QCD model \eqref{eq:contiuum_action}, we generalize the construction in \cite{Dempsey:2023fvm,Dempsey:2024alw} of the lattice Hamiltonian in adjoint QCD$_2$.
In this setup, the matter is represented by $\dim \bm \lambda$ real lattice fermions $\chi^a_n$ on every lattice site $n$, and these fermions obey the canonical anti-commutation relation $\{\chi_n^a,\chi^b_m\}=\delta_{nm}\delta^{ab}$. The gauge field lives on links and is represented by the parallel propagator $U^{ab}_n$ and its conjugate momenta, the electric field variables $L^A_n$ and $R^A_n$. Using staggered fermions, the lattice Hamiltonian has to be realized on an even number of lattice sites $N$, and it takes the form\footnote{Note that the sum runs from 1 to $N-1$ because we are summing over the links of an $N$-site open lattice.}
\begin{equation}\label{eq:lattice_ham}
    H = \sum_{n=1}^{N-1} \left[\frac{ag^2}{2}L^A_nL^A_n -\frac{i}{2} (a^{-1} + (-1)^nm) \chi_n^aU^{ab}_n\chi^b_{n+1}\right]\,,
\end{equation}
where $a$ is the lattice spacing. In addition to the Hamiltonian \eqref{eq:lattice_ham}, we also have the local Gauss law constraint
\begin{equation}\label{eq:gauss}
    G^A_n\equiv L_n^A-R_{n-1}^A-Q^A_n=0\,,\qquad\text{with}\qquad Q^A_n \equiv \frac{1}{2}\chi^a_n T^A_{ab}\chi_b\,,
\end{equation}
where the charge $Q^A_n$ is defined such that it satisfies the Lie algebra $[Q_n^A,Q^B_m]=\delta_{nm} i f^{ABC} Q^C_n$. This Hamiltonian and Gauss law appeared previously \cite{Dempsey:2023fvm, Dempsey:2024alw} for the case of adjoint matter, and further details can be found in those works. 

The fermionic matter Hilbert space has dimension $2^{N \dim \bm \lambda/2}$. This Hilbert space can be decomposed into representations of the group of gauge transformations, $G^{\otimes N}$; the decomposition procedure is explained in Appendix~\ref{app:matter_rep}, following \cite{Dempsey:2023fvm}. The result is
\begin{equation}\label{eq:hmatter}
    \mathcal H_\text{matter} = \bigoplus_{i=1}^{2^{Nn_0(\bm{\lambda})/2}} (\bm R(\bm \lambda),\bm R(\bm \lambda),\ldots,\bm R(\bm \lambda))\,,
\end{equation}
where $n_0(\bm{\lambda})$ is the number of zero weights in the representation $\bm \lambda$, and $\bm R(\bm \lambda)$ is a (possibly reducible) representation of $G$.
\begin{table}[]
    \centering
    \begin{tabular}{cp{.5cm}cp{.5cm}cp{.3cm}c}
        \toprule
        Theory & & $\bm R(\bm \lambda)$ & & $n_0$ & & $Z_{\bm \lambda}(G)$ irrep\\
        \midrule
         $\SU(2) + \psi_{\bm{5}}$ & & $\bm 4$ & & 1 & & $p=1$ \\
         $\SU(2) \times \SU(2) + \psi_{(\bm{3}, \bm{3})}$ & & $(\bm2,\bm4)\oplus(\bm4,\bm2)$ & & 1 & & $p = (1,1)$\\
         $\SU(2) \times \SU(2) + \psi_{(\bm{2}, \bm{4})}$ & & $(\bm2,\bm4)\oplus(\bm3,\bm1)\oplus (\bm 1,\bm5)$ & & 0 & & $p =0$ \\
         $\SU(4) + \psi_{\bm{15}}$ & & $\bm {64}$ & & 3 & & $p=2$\\    
         \bottomrule
    \end{tabular}
    \caption{One-site representations and their properties for the lattice Hamiltonian \eqref{eq:lattice_ham}.}
    \label{tab:lattice}
\end{table}
The on-site representations $\bm R(\bm \lambda)$ for the theories considered here are given in Table~\ref{tab:lattice}, and the definition of $\bm R(\bm \lambda)$ along with more examples are given in Appendix~\ref{app:matter_rep}.

The gauge degrees of freedom consist of a boson on $G$ for every link, so their Hilbert space is given by
\begin{equation}
    \mathcal H_\text{gauge}=L^2(G)^{\otimes N}\,.
\end{equation} 
By the Peter-Weyl theorem \cite{Peter:1927thq}, the Hilbert space $L^2(G)$ is spanned by the matrix elements of the irreducible representations of $G$. Thus, as explained in \cite{Dempsey:2023fvm,Dempsey:2024alw}, we can label the basis states of the gauge Hilbert space by a representation $\bm{r}$ of $G$ along with an index of $\bm{r}$ and an index of $\bm{\overline{r}}$ (i.e., the indices of the $\bm{r}$-representation matrix).

All physical states of the theory must obey the Gauss law \eqref{eq:gauss}, so we need to characterize the physical subspace $\H_\text{phys} \subset \mathcal H_\text{matter} \otimes \mathcal H_\text{gauge}$ that is annihilated by $G_n^A$. We follow the construction in \cite{Dempsey:2023fvm,Dempsey:2024alw} and build gauge-invariant combinations of the gauge field and matter states by using Clebsch-Gordan coefficients to contract the $\bm{R}(\bm{\lambda})$ indices on sites with the indices of the representations on neighboring links. This procedure gives rise to the physical Hilbert space  $\H_\text{phys}$, which is of the form 
\begin{equation}\label{eq:hphys}
    \H_\text{phys} = \bigoplus_{i=1}^{2^{Nn_0(\bm{\lambda})/2}}\H'\,, \qquad \H'\subset (\bm R(\bm \lambda),\bm R(\bm \lambda),\ldots,\bm R(\bm \lambda)) \otimes \H_\text{gauge}
\end{equation}
and is spanned by the basis
\begin{equation}\label{eq:phys_states}
    \ket{\ldots,(\bm r_n)_{e_n},(\bm r_{n+1})_{e_{n+1}},\ldots;i} \qquad \text{with}\qquad \bm r_{n+1}\subset \bm r_{n}\otimes \bm R(\bm{\lambda})\,,\qquad i =1,2,\ldots,2^{Nn_0(\bm{\lambda})/2}\,.
\end{equation}
Here, the multiplicity index $i$ stems from the multiplicity in \eqref{eq:hmatter}. The other labels of the ket are the representations $\bm{r}_n$ labeling basis states on the links, and the multiplicity indices $e_n$ of the Clebsch-Gordan symbols used to build gauge-invariant states; see \cite{Dempsey:2024alw} for details of the tensor contractions used to form the basis states. These Clebsch-Gordan symbols describe the embedding of $\bm{r}_n$ into $\bm{r}_{n-1}\otimes \bm{R}(\bm{\lambda})$,\footnote{Note that $\bm{R}(\bm{\lambda})$ can be reducible. When this is the case, we take the Clebsch-Gordan symbol to carry an additional label that specifies which irrep in $\bm{R}(\bm{\lambda})$ is being used.} and so the label $e_n$ is needed whenever $\bm{r}_n$ appears multiple times in this tensor product. Note that the choices of representations on links are not independent: in order to construct a gauge-invariant state, it must be the case that $\bm{r}_n$ appears at least once in the tensor product $\bm{r}_{n-1}\otimes \bm{R}(\bm{\lambda})$.

When formulated on a translation-invariant lattice, the lattice Hamiltonian \eqref{eq:lattice_ham} possesses analogues of all of the invertible zero- and one-form symmetries of the continuum model, as has been argued previously for adjoint matter \cite{Dempsey:2023fvm,Dempsey:2024alw}. We will focus here on the $\Z_2$ zero-form chiral symmetry and the $Z_{\bm{\lambda}}(G)$ one-form symmetry, which we explained in the continuum in Section~\ref{subsec:continuum_action}; some additional details can be found in Appendix~\ref{app:symmetries}. On the lattice, the $\Z_2$ chiral symmetry is generated by one-site translations.\footnote{Of course, on a lattice with more than two sites this is not a $\Z_2$ transformation; in the language of \cite{Seiberg:2023cdc} one says that the continuum chiral symmetry emanates from one-site lattice translations.} The lattice one-form symmetry has generators $U_{n,g}$, where $g\in Z_{\bm{\lambda}}(G)$; when $n$ is odd, it acts on a state of the form \eqref{eq:phys_states} by a phase equal to the action of $g$ on the representation $\bm{r}_n$.\footnote{Although $g$ acts on $\bm{r}_n$ as a matrix, $g$ is a center element, and so this matrix must commute with all other matrices in the representation; thus, by Schur's lemma, it must be a phase times the identity matrix, and this is the phase we are referring to.} This phase is the same for any odd site $n$, because $\bm{r}_{n+2m}$ appears in $\bm{r}_n \otimes \bm{R}(\bm{\lambda})^{\otimes 2m}$ and $Z_{\bm{\lambda}}(G)$ is sent to the identity by the representation $\bm{R}(\bm{\lambda})^{\otimes 2}$.\footnote{This follows from the construction in Appendix~\ref{app:matter_rep}.} In order to make $U_{n,g}$ a topological operator on the lattice, we must define it to act by the same phase for even $n$, but here we see the opportunity for an anomaly between the chiral and center symmetry: if the image of $g$ under the representation $\bm{R}(\bm{\lambda})$ is not the identity, then a one-site translation does not commute with $U_{n,g}$.

To illustrate the physical basis \eqref{eq:phys_states} and the lattice symmetries and anomalies, we will look at how this works explicitly in the $\SU(2)+\psi_{\bm 5}$ theory (similar details can be found for other theories in Appendix~\ref{app:lattice_detail}). In this case, the basis \eqref{eq:phys_states} specializes to
\begin{equation}\label{eq:phys_su25}
    \ket{\ldots,l_n,l_{n+1},\ldots;i} \qquad \text{with}\qquad |l_{n+1}-l_n| = \frac12\text{ or }\frac{3}{2}\,,\qquad i =1,2,\ldots,2^{N/2}\,,
\end{equation}
where $l_n$ is the spin label for the $\SU(2)$ representation on the $n$th link.
The center of $\SU(2)$ is $\Z_2$ and the $\bm{5}$ representation acts trivially on the $\Z_2$ elements, so we have a $\Z_2$ one-form symmetry. And indeed, we clearly see that the basis \eqref{eq:phys_su25} splits into two sectors: one with integer-spin representations on odd links, and the other with half-integer-spin representations on odd links. We label the flux tube sectors how the representations on odd links represent the $Z_{\bm{\lambda}}(G)$, so these are the $p = 0$ and $p = 1$ sectors respectively. We clearly see that these sectors are interchanged by a one-site lattice translation, which means that the chiral and center symmetries have a mixed anomaly in this theory.

By contrast with the continuum invertible symmetries, there is no known analogue of the non-invertible symmetries of the continuum theory on the lattice, and so the lattice model generically breaks these symmetries. This means in particular that the vacuum degeneracy in the continuum, most of which is enforced only by the non-invertible symmetries, will be lifted on the lattice. Nevertheless, in the next section we will explain how we utilize new infinite tensor network methods to identify lattice states that are adiabatically connected to the degenerate vacua of the continuum theory.

\section{Infinite MPS and vacuum degeneracy}\label{sec:mps}

A key feature of our setup is that we study the lattice Hamiltonian \eqref{eq:lattice_ham} directly on an infinite spatial lattice. This is what allows us to isolate the different degenerate vacua of a gapped gauge theory on the lattice. To work on an infinite lattice, we will employ the method developed in \cite{Dempsey:2025wia}.

In Section~\ref{sec:lempos}, we briefly describe this method. In Section~\ref{sec:adiabatic}, we explain how we use this method to identify lattice states that are adiabatically connected to degenerate vacua in the continuum. In Section~\ref{sec:lattice_decay}, we explain a simple pattern that describes these states in the lattice strong-coupling limit $ga\to\infty$. In Section~\ref{sec:strong_coupling}, we study the strong-coupling limit further and derive strong-coupling expansions that we compare with our numerics in Section~\ref{sec:results}.

\subsection{Uniform matrix product states and vacuum degeneracy}\label{sec:lempos}

In the context of quantum many-body problems, it is possible to study translation-invariant Hamiltonians directly on an infinite lattice using infinite tensor network methods. In particular, a uMPS \cite{Vidal:2006ofj} forms an ansatz for the ground state of such a Hamiltonian. A uMPS consists of repeating unit cells of $k$ lattice sites each, and it is parametrized by $k$ rank-three tensors $A^{(i)}$ (where $i=1,\ldots,k$ indexes the sites of the unit cell). Each tensor has a physical index (denoted $s_i^{(n)}$ for the tensor in the $n$th cell) for the physical Hilbert space on site $i$, as well as two virtual indices (which we don't write explicitly) for auxiliary Hilbert spaces. These virtual indices are contracted between neighboring tensors, and so the uMPS state on the infinite lattice is given by
\begin{equation}
\begin{split}
    \ket{A^{(1)}, \cdots, A^{(k)}} &= \sum_{\{s^{(n)}_i\}} \prod_{n=-\infty}^{\infty} \left[ A^{(1)}_{s^{(n)}_1} \cdots A^{(k)}_{s^{(n)}_k} \times \left(\ket{s^{(n)}_1}\otimes\cdots\otimes\ket{s^{(n)}_k}\right) \right]\\
    &=\begin{tikzpicture}[diagram]
			\node[style=square] at (0,0) (A1) {$A^{(k)}$};
			\node[style=square] at (1.5,0) (A2) {$A^{(1)}$};
			\node[style=square] at (3,0) (A3) {$A^{(2)}$};
            \node[style=square] at (5,0) (A4) {$A^{(k)}$};
			\node[style=square] at (6.5,0) (A5) {$A^{(1)}$};
			\node at (0,-1) {$\ket{s_k^{(n-1)}}$};
			\node at (1.5,-1) {$\ket{s_1^{(n)}}$};
			\node at (3,-1) {$\ket{s_2^{(n)}}$};
            \node at (5,-1) {$\ket{s_k^{(n)}}$};
			\node at (6.5,-1) {$\ket{s_1^{(n+1)}}$};
			\node at (-1.,0) (dots1) {$\cdots$};
			\node at (4.,0) (dots2){$\cdots$};
            \node at (7.5,0) (dots3){$\cdots$};
			\node at (-1.,-1.) {$\cdots$};
			\node at (4.,-1.) {$\cdots$};
            \node at (7.5,-1.) {$\cdots$};
			\draw (A1) -- (A2) -- (A3);
            \draw (A4) -- (A5);
			\draw (A1) -- ++(0,-.8);
			\draw (A2) -- ++(0,-.8);
			\draw (A3) -- ++(0,-.8);
            \draw (A4) -- ++(0,-.8);
			\draw (A5) -- ++(0,-.8);
			\draw (A1) -- ++(-0.75,0);
			\draw (A3) -- ++(0.75,0);
            \draw (A4) -- ++(-0.75,0);
			\draw (A5) -- ++(0.75,0);
		\end{tikzpicture}\,,
\end{split}
    \label{eq:uMPS}
\end{equation}
where we have suppressed the virtual indices for clarity. The entries of the tensors $A^{(i)}$ constitute a finite set of variational parameters, and we can increase the number of parameters by enlarging the virtual spaces. The ground states of gapped, translation-invariant Hamiltonians can be well-approximated by uMPSs, and the approximation can be made arbitrarily accurate by using sufficiently large virtual spaces \cite{Vidal:2006ofj}.

If the Hamiltonian has a global symmetry, then one can use symmetric uMPSs to prepare ans\"atze with definite charge \cite{Perez-Garcia:2008hhq, Sanz:2009lax,Singh_u1}. In \cite{Dempsey:2025wia}, it was shown that symmetric uMPSs are also a natural tool for Hamiltonian lattice gauge theories. In particular, using a generalization of a matrix product operator (MPO) called a link-enhanced MPO (LEMPO), which exists only for symmetric (u)MPSs, it is possible to encode the Hamiltonian of a lattice gauge theory in a simple and manifestly translation-invariant way. This allows us to apply infinite tensor network methods to gauge theories.

In addition to the obvious speedup of working directly with an infinite number of sites rather than studying finite lattices of increasing size and then extrapolating, infinite tensor network methods offer several distinct advantages. 
For our present discussion, the most important property we gain is cluster decomposition \cite{weinberg}. A translation-invariant vacuum state $\ket{v}$ satisfies cluster decomposition if, for any two local\footnote{For the purposes of this discussion, a local operator is one that acts non-trivially on only a finite span of MPS tensors.} operators $\mathcal{O}_1$ and $\mathcal{O}_2$, we have
\begin{equation}\label{eq:cluster_decomposition}
    \lim_{n\to\infty} \braket{v|\mathcal{O}_1(0)\mathcal{O}_2(n)|v} = \braket{v|\mathcal{O}_1(0)|v}\,\braket{v|\mathcal{O}_2(0)|v}\,.
\end{equation}
If $\ket{v_1}$ and $\ket{v_2}$ are both cluster-decomposing vacua, then a superposition $\alpha_1 \ket{v_1} +\alpha_2 \ket{v_2}$ generically will not be. Thus, unlike for a finite system, for which states of the Hilbert space can mix arbitrarily---and in particular, degenerate vacua in the continuum, which are no longer protected by the non-invertible symmetry once we are on a lattice, \emph{will} mix arbitrarily---cluster decomposition forbids such mixing on an infinite lattice and keeps different vacuum states sharply distinguished.

uMPSs are excellent ans\"atze for studying the special class of cluster-decomposing states because they are generically cluster-decomposing. We can give a brief argument (following \cite{Vanderstraeten:2019voi}) for this property of uMPSs in terms of the transfer matrix $T$, defined by contracting one unit cell's worth of the uMPS tensors for a bra and ket:
\begin{equation}
    \begin{tikzpicture}[diagram,yscale=.6,xscale=.6]
        \draw (-1,0) -- (1,0) (-1,-2) -- (1,-2);
        \draw[rounded corners,fill=white] (-.5,-2.5) rectangle node {$T$} (.5,.5);
    \end{tikzpicture}
    \;=\;
    \begin{tikzpicture}[diagram,yscale=.6]
        \node[style=square] at (1.5,0) (A2) {$A^{(1)}$};
        \node[style=square] at (3,0) (A3) {$A^{(2)}$};
        \node[style=square] at (5,0) (A4) {$A^{(k)}$};
        \node[style=square] at (1.5,-2) (A22) {$A^{(1)}$};
        \node[style=square] at (3,-2) (A32) {$A^{(2)}$};
        \node[style=square] at (5,-2) (A42) {$A^{(k)}$};
        \node at (4.,-1.) {$\cdots$};
        \node at (4.,0) (dots2){$\cdots$};
        \node at (4.,-2) (dots3){$\cdots$};
        \draw (A2) -- (A22) (A3) -- (A32) (A4) -- (A42) (A2) -- (A3) (A22) -- (A32) (A3) -- (dots2) (dots2) -- (A4) (A32) -- (dots3) (A42) -- (dots3);
        \draw (A2) -- ++(-.8,0) (A22) -- ++(-.8,0) (A4) -- ++(.8,0) (A42) -- ++(.8,0);
    \end{tikzpicture}\,.
\end{equation}
The norm of the uMPS involves an infinite power of $T$, so if the spectral norm of $T$ differs from 1, then the norm of uMPS will either vanish or diverge. Thus, we normalize the uMPS tensors such that the largest-magnitude eigenvalue of $T$ is 1. We denote the left- and right-eigenvectors of $T$ with eigenvalue 1 by $E_L$ and $E_R$ respectively. If all other eigenvalues of $T$ have magnitude strictly less than 1, then the uMPS is said to be injective \cite{Vanderstraeten:2019voi}, and cluster decomposition follows immediately:
\begin{equation}
\begin{split}
    \lim_{n\to\infty} \;&\begin{tikzpicture}[diagram,yscale=.6,baseline=-.7cm]
        \draw (-.7,0) -- (2.7,0) (3.3,0) -- (6.7,0);
        \draw (-.7,-2) -- (2.7,-2) (3.3,-2) -- (6.7,-2);
        \draw[rounded corners,fill=white] (-.4,-2.5) rectangle node {$T$} (.4,.5);
        \draw[rounded corners,fill=white] (.6,-2.5) rectangle node {$T_1$} (1.4,.5);
        \draw[rounded corners,fill=white] (1.6,-2.5) rectangle node {$T$} (2.4,.5);
        \node[scale=.8] at (-1,-.025) {$\cdots$};
        \node[scale=.8] at (-1,-2.025) {$\cdots$};
        \node[scale=.8] at (3,-.025) {$\cdots$};
        \node[scale=.8] at (3,-2.025) {$\cdots$};
        \node[scale=.8] at (7,-.025) {$\cdots$};
        \node[scale=.8] at (7,-2.025) {$\cdots$};
        \begin{scope}[xshift=1cm]
        \draw[rounded corners,fill=white] (2.6,-2.5) rectangle node {$T$} (3.4,.5);
        \draw[rounded corners,fill=white] (3.6,-2.5) rectangle node {$T_2$} (4.4,.5);
        \draw[rounded corners,fill=white] (4.6,-2.5) rectangle node {$T$} (5.4,.5);
        \end{scope}
        \draw[decorate,decoration={brace,amplitude=5pt,mirror,raise=1ex}] (1.6,-2.3) -- (4.4,-2.3);
        \node at (3,-3.3) {$n$};
    \end{tikzpicture} \\[.5em]
    ={}& \begin{tikzpicture}[diagram,yscale=.6]
        \draw (0,0) -- (2,0);
        \draw (0,-2) -- (2,-2);
        \draw[rounded corners,fill=white] (-.4,-2.5) rectangle node {$E_L$} (.4,.5);
        \draw[rounded corners,fill=white] (.6,-2.5) rectangle node {$T_1$} (1.4,.5);
        \draw[rounded corners,fill=white] (1.6,-2.5) rectangle node {$E_R$} (2.4,.5);
        \node at (2.75,-1) {$\times$};
        \begin{scope}[xshift=.5cm]
        \draw (3,0) -- (5,0);
        \draw (3,-2) -- (5,-2);
        \draw[rounded corners,fill=white] (2.6,-2.5) rectangle node {$E_L$} (3.4,.5);
        \draw[rounded corners,fill=white] (3.6,-2.5) rectangle node {$T_2$} (4.4,.5);
        \draw[rounded corners,fill=white] (4.6,-2.5) rectangle node {$E_R$} (5.4,.5);
        \end{scope}
    \end{tikzpicture}\,.
\end{split}
\end{equation}
Here $T_1$ and $T_2$ are defined like $T$ but with the local operators $\mathcal{O}_1$ and $\mathcal{O}_2$, respectively, inserted between the bra and ket. We see that a uMPS will generically obey cluster decomposition, because one would have to fine-tune the $A^{(i)}$ matrices in order to have a degenerate largest eigenvalue in the transfer matrix spectrum.\footnote{By a similar argument, the boundary conditions imposed on an injective uMPS do not affect any of the bulk properties, which is what allows us to be agnostic toward these boundary conditions in \eqref{eq:uMPS}.} This means that uMPSs are well-tailored to study the subset of quantum states that we are interested in.

Thus, to find cluster-decomposing vacuum states on the lattice, we can start from a uMPS ansatz and then vary its parameters so as to minimize its energy density. In this work, we use the VUMPS algorithm \cite{Zauner-Stauber:2017eqw}, as implemented in \texttt{MPSKit.jl}  \cite{MPSKit2025}, for this task. Depending on initial conditions, it is possible for VUMPS to converge to a local minimum of energy density rather than the global minimum. 

Generally, in the context of optimization, local minima are viewed as spurious outcomes to be avoided. However, in our case they are a feature. Indeed, we aim to study the states of gapped gauge theories that become degenerate vacua in the continuum, but we know that they will not all be degenerate on the lattice. Thus, the local minimizers found by VUMPS are not necessarily spurious: they can become the vacua we are interested in as we dial $ga\to 0$. In practice, if we find a lattice state that is not a global minimum of energy density on the lattice, but whose energy density approaches the global minimum as we decrease the lattice spacing $a$, then we will conjecture that this state is adiabatically connected to one of the degenerate vacua in the continuum. In the following, we will call such states lattice vacua of the theory. 

\begin{figure}
    \begin{subfigure}[t]{0.5\textwidth}
        \centering
        \includegraphics[width=\textwidth]{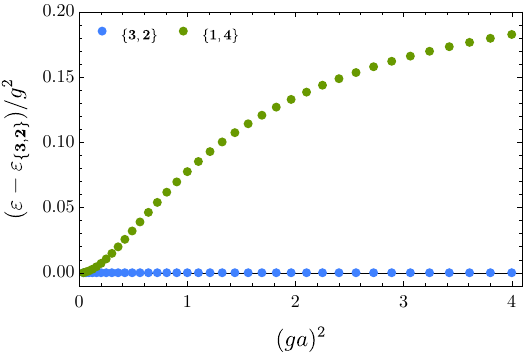}
        \caption{$\SU(2)+\psi_{\bm 5}$}
        \label{fig:energydifferences_su25}
    \end{subfigure}
    \hfill
    \begin{subfigure}[t]{0.5\textwidth}
        \centering
        \includegraphics[width=\textwidth]{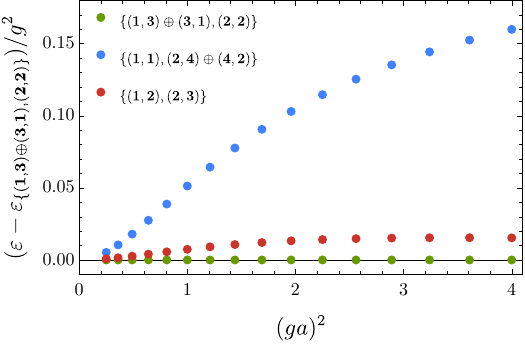}
        \caption{$\SU(2)\times\SU(2)+\psi_{(\bm 3,\bm3)}$}
    \end{subfigure}\\[3em]
    \begin{subfigure}[t]{0.5\textwidth}
        \centering
        \includegraphics[width=\textwidth]{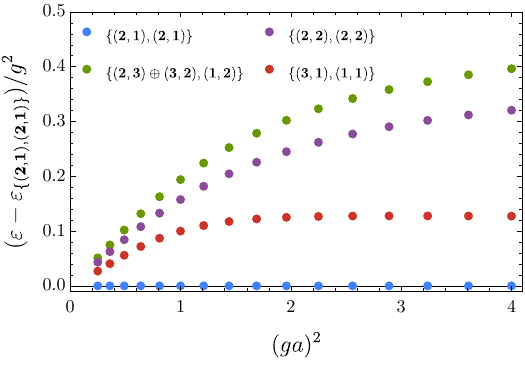}
        \caption{$\SU(2)\times\SU(2)+\psi_{(\bm 2,\bm4)}$}
        \label{fig:su2su224_energy}
    \end{subfigure}
    \hfill
    \begin{subfigure}[t]{0.5\textwidth}
        \centering
        \includegraphics[width=\textwidth]{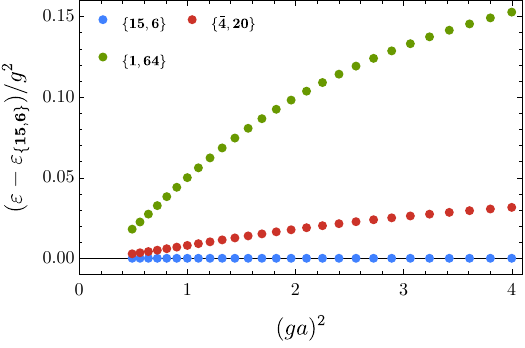}
        \caption{$\SU(4)+\psi_{\bm{15}}$}
    \end{subfigure}\\[2em]
    \caption{Difference in energy density between the lattice vacua as a function of the dimensionless lattice spacing, showing that all states become degenerate vacua in the continuum limit. Here $\varepsilon_{\{\bm{r}_1,\bm{r}_2\}}$ is the energy density of the lattice vacuum that is adiabatically connected to the strong-coupling state with link representations $\{\bm{r}_1,\bm{r}_2\}$. Lattice vacua that are related to those plotted by exact lattice symmetries (one-site translation and $G$ outer-automorphism) are omitted (see color coding in Table \ref{tab:lattice_vacua}). }
    \label{fig:energydifferences}
\end{figure}

To illustrate this point, we can look at the $\SU(2)+\psi_{\bm 5}$ example. In the continuum, there are four degenerate vacua (two in each flux tube sector), as listed in Table~\ref{tab:vacua}. For this theory, the two flux tube sectors are related by chiral symmetry due to a mixed anomaly, as we explained in Section~\ref{subsec:lattice_hamiltonian}. Thus, we can restrict ourselves to the $p=0$ flux tube sector, since the energy densities in the $p = 1$ flux tube sector will be the same.

Using VUMPS with a LEMPO representation of the Hamiltonian \eqref{eq:lattice_ham}, we are able to find two distinct and \emph{non-degenerate} lattice vacua\footnote{Note that many of the lattice vacua are not global minima of energy density, and hence could be considered false vacua. For this reason, we can only distinguish these states on infinite lattices.} at finite lattice spacing. The difference in the energy densities of these two states is plotted (using a  labeling scheme that will be explained in Section~\ref{sec:adiabatic}) in Figure~\ref{fig:energydifferences_su25}  as a function of the dimensionless lattice spacing $ga$.\footnote{A simple (but inefficient) method for obtaining these states is to repeatedly run VUMPS on randomly initialized uMPSs until all local minima are found. Empirically, we find that for $ga=1$ and normally distributed entries in the initial uMPS tensors, VUMPS converges to the state labeled $\{\bm 1,\bm 4\}$ in roughly  $\sim 60\%$ of cases and to the state labeled $\{\bm 3,\bm 2\}$ in the $\sim 40\%$ of cases. In Section~\ref{sec:adiabatic}, we will give a systematic method for finding the lattice vacua.} The non-degeneracy of the lattice vacua clearly shows that the non-invertible symmetries of the continuum model are broken at finite lattice spacing. However, the energy difference appears to vanish in the continuum limit $ga\rightarrow 0$, indicating that the non-invertible symmetries are restored and that the states we have found on the lattice are adiabatically connected to the degenerate vacua in the continuum. 

In the other panels of Figure~\ref{fig:energydifferences}, we show analogous plots of the distinct energy densities for the other theories in Table \ref{tab:theories}. We omit data for vacua that are degenerate due to exact lattice symmetries; the patterns of these lattice degeneracies (as opposed to those which emerge only in the continuum limit) are indicated by the color coding in Table~\ref{tab:lattice_vacua} and explained further in Appendix~\ref{app:symmetries}. 
For each theory, we find evidence that all the lattice vacua are becoming degenerate as $ga\to 0$.

\subsection{Adiabatic continuity to the strong-coupling limit}\label{sec:adiabatic}
In Figure~\ref{fig:energydifferences}, we see that we can follow the adiabatic evolution of the vacua from very close to the continuum limit, where they are nearly degenerate, out to rather large values of the lattice spacing. In this section, we will explore the limits of these vacua in the lattice strong-coupling limit $ga\rightarrow \infty$. At leading order in this limit, the matter part of the Hamiltonian~\eqref{eq:lattice_ham} vanishes and only the gauge-kinetic part remains:
\begin{equation}\label{eq:strong_ham}
    H \overset{ga\rightarrow \infty}{\xrightarrow{\hspace*{1cm}}}  \frac{ag^2}{2}\sum_{n=1}^NL^A_nL^A_n\,.
\end{equation}
Thus, in this limit, the basis states \eqref{eq:phys_states} are all energy eigenstates. If we restrict ourselves to (two-site) translation-invariant basis states, then the representations on links must repeat in a two-fold pattern, with $\bm{r}_{2n+1} = \bm{r}_1$ and $\bm{r}_{2n} = \bm{r}_2$.\footnote{If $G$ has two irreps $\bm{r}$ and $\bm{r'}$ that have the same Casimir invariant, then we can also form an eigenstate of the strong-coupling Hamiltonian that has a superposition of $\bm{r}$ and $\bm{r'}$ on a link. In this case, with a slight abuse of notation, we let $\bm{r}_1$ and/or $\bm{r}_2$ be reducible representations, e.g. $\bm{r}_1 = \bm{r}\oplus \bm{r'}$.} The energy density in such a state is
\begin{equation}\label{eq:leading_epsilon}
    \varepsilon(\bm{r}_1,\bm{r}_2) = \frac{g^2}{4}\Big(C_2(\bm{r}_1) + C_2(\bm{r}_2)\Big)\,,
\end{equation}
where $C_2(\bm{r})$ is the quadratic Casimir invariant of $\bm{r}$. Note that because of the degrees of freedom in \eqref{eq:phys_states} other than the irreps on links (namely, the multiplicity of Clebsch-Gordan symbols and/or the multiple copies of $\bm{R}(\bm{\lambda})$ on sites), there may be infinitely many states with this pattern of link representations; this infinite degeneracy is broken at subleading orders in strong-coupling perturbation theory. 

The simple description of energy eigenstates at strong coupling provides us with a natural way of labeling the lattice vacua: we can adiabatically follow them to the strong-coupling limit and see what pattern $\{\bm{r}_1,\bm{r}_2\}$ of link representations they have in this limit. This is how we labeled the lattice vacua in Figure~\ref{fig:energydifferences}. For instance, in the $p = 0$ sector of the $\SU(2)+\psi_{\bm 5}$ theory, the lower-energy state has the irrep pattern $\{\bm{3},\bm{2}\}$ at strong coupling, and the higher-energy state has the pattern $\{\bm{1},\bm{4}\}$. And indeed, using \eqref{eq:leading_epsilon} we see that the difference in their energy densities is $\frac{1}{4}$ at leading order in strong coupling, which is consistent with Figure~\ref{fig:energydifferences_su25}. Note a possible confusion: while we described above that there are infinitely many states in the strong-coupling limit with a particular representation pattern, e.g. $\{\bm{1},\bm{4}\}$, when we refer to the $\{\bm{1},\bm{4}\}$ state we mean a \emph{single} one-parameter family of states distinguished by having this pattern at strong-coupling and being a local minimum of energy density at any finite $ga$.

\begin{figure}
    \begin{subfigure}[b]{0.5\textwidth}
        \centering
        \includegraphics[width=\textwidth]{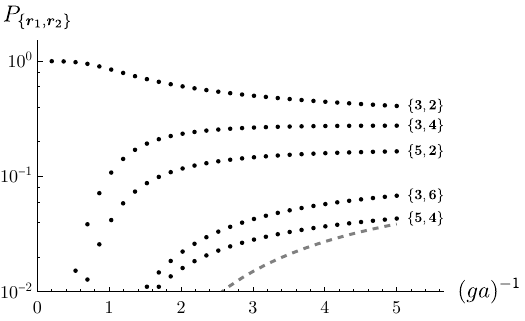}
        \caption{}
    \end{subfigure}
    \hfill
    \begin{subfigure}[b]{0.5\textwidth}
        \centering
        \includegraphics[width=\textwidth]{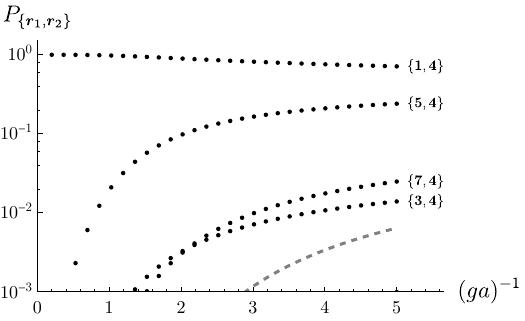}
        \caption{}
    \end{subfigure}
    \caption{The probability of finding representations $\{\bm r_1,\bm r_2\}$ on consecutive links in the two $p=0$ lattice vacua of the $\SU(2)+\psi_{\bm 5}$ theory. The gray dashed line represents all representations not explicitly plotted.}
    \label{fig:repsinsu25}
\end{figure}

For the four theories we consider in this paper, we have found that this adiabatic continuation procedure always proceeds smoothly. Indeed, in Figure~\ref{fig:repsinsu25}, we show how the wavefunctions of the states in Figure~\ref{fig:energydifferences_su25} evolve from the strong-coupling limit to the continuum limit by plotting the probability of finding the representations $\{\bm{r}_1,\bm{r}_2\}$ on a given consecutive pair of links. This adiabatic continuity allows us to label all the degenerate vacua listed in Table~\ref{tab:vacua} by their link representations $\{\bm{r}_1,\bm{r}_2\}$ at strong coupling. We list our results in Table~\ref{tab:lattice_vacua}, using a coloring scheme determined by the action of invertible symmetries on the vacua, described in Appendix~\ref{app:symmetries} (see Figure~\ref{fig:symmetries_on_vacua}). Note that Table~\ref{tab:lattice_vacua} lists all local minima found by VUMPS; we have not observed any local minima that fail to become degenerate global minima in the continuum limit.

The strong-coupling states in Table~\ref{tab:lattice_vacua} are \emph{not} limited to the lowest-energy states in each flux tube sector in the strong-coupling limit. For instance, in the $\SU(2)+\psi_{\bm{5}}$ theory, only $\{\bm{3},\bm{2}\}$ and $\{\bm{2},\bm{3}\}$ are ground states in this limit; as we saw in Figure~\ref{fig:energydifferences_su25}, the $\{\bm{1},\bm{4}\}$ and $\{\bm{4},\bm{1}\}$ states have higher energy density. In fact, in each theory we have studied, we find some lattice vacua that are not minimal-energy states with respect to \eqref{eq:strong_ham} in their flux tube sectors:
\begin{flushleft}
\begin{itemize}
    \item $\SU(2)+\psi_{\bm{5}}$: $\{\bm{1},\bm{4}\}$ and $\{\bm{4},\bm{1}\}$
    \item $\SU(2)\times\SU(2)+\psi_{(\bm{3},\bm{3})}$: $\{(\bm{1},\bm{1}),(\bm{2},\bm{4})\oplus(\bm{4},\bm{2})\}$ and $\{(\bm{2},\bm{4})\oplus(\bm{4},\bm{2}),(\bm{1},\bm{1})\}$
    \item $\SU(2)\times\SU(2)+\psi_{(\bm{2},\bm{4})}$: $\{(\bm{2},\bm{2}),(\bm{2},\bm{2})\}$, $\{(\bm{1},\bm{2}),(\bm{2},\bm{3})\oplus(\bm{3},\bm{2}))\}$, and $\{(\bm{2},\bm{3})\oplus(\bm{3},\bm{2}),(\bm{1},\bm{2})\}$
    \item $\SU(4)+\psi_{\bm{15}}$: $\{\bm{1},\bm{64}\}$ and $\{\bm{64},\bm{1}\}$
\end{itemize}
\end{flushleft}
In Section~\ref{sec:lattice_decay}, we will explain the criterion that identifies all the lattice vacua in Table~\ref{tab:lattice_vacua} by studying their strong-coupling labels. 

\begin{table}[t]
\centering
    \begin{tabular}{ccc}
        \toprule
        Theory & Flux tube sector & Lattice vacua\\
        \midrule
        $\SU(2)+\psi_{\bm{5}}$ & $p=0$ & {\color{green!50!black} $\{\bm{1},\bm{4}\}$} and {\color{blue} $\{\bm{3},\bm{2}\}$} \\
        & $p=1$ & {\color{green!50!black} $\{\bm{4},\bm{1}\}$} and {\color{blue} $\{\bm{2},\bm{3}\}$} \\
        & & \\
        $\SU(2)\times\SU(2) + \psi_{(\bm{3},\bm{3})}$ & $p=(0,0)$ & {\color{blue} $\{(\bm{1},\bm{1}), (\bm{2},\bm{4})\oplus (\bm{4},\bm{2})\}$} and {\color{green!50!black} $\{(\bm{1},\bm{3})\oplus (\bm{3},\bm{1}), (\bm{2},\bm{2})\}$} \\
        & $p=(0,1)$ & {\color{red!80!black} $\{(\bm{1},\bm{2}), (\bm{2},\bm{3})\}$} and {\color{red!80!black} $\{(\bm{3},\bm{2}), (\bm{2},\bm{1})\}$} \\
        & $p=(1,0)$ & {\color{red!80!black} $\{(\bm{2},\bm{1}), (\bm{3},\bm{2})\}$} and {\color{red!80!black} $\{(\bm{2},\bm{3}), (\bm{1},\bm{2})\}$} \\
        & $p=(1,1)$ & {\color{blue} $\{(\bm{2},\bm{4})\oplus (\bm{4},\bm{2}), (\bm{1},\bm{1})\}$} and {\color{green!50!black} $\{(\bm{2},\bm{2}), (\bm{1},\bm{3})\oplus (\bm{3},\bm{1})\}$} \\
        & & \\
        $\SU(2)\times\SU(2)+\psi_{(\bm{2},\bm{4})}$ & $p=0$ &  {\color{red!80!black} $\{(\bm{1},\bm{1}), (\bm{3},\bm{1})\}$}, {\color{red!80!black} $\{(\bm{3},\bm{1}), (\bm{1},\bm{1})\}$}, and {\color{red!50!blue} $\{(\bm{2},\bm{2}), (\bm{2},\bm{2})\}$} \\
        & $p=1$ & {\color{green!50!black}$\{(\bm{1},\bm{2}), (\bm{2},\bm{3})\oplus (\bm{3},\bm{2})\}$}, {\color{green!50!black}$\{(\bm{2},\bm{3})\oplus (\bm{3},\bm{2}), (\bm{1},\bm{2})\}$},\\
        && and {\color{blue}$\{(\bm{2},\bm{1}), (\bm{2},\bm{1})\}$} \\
        & & \\
        $\SU(4)+\psi_{\bm{15}}$ & $p=0$ & {\color{green!50!black}$\{\bm{1},\bm{64}\}$} and {\color{blue} $\{\bm{15},\bm{6}\}$} \\
        & $p=1$ & {\color{red!80!black} $\{\bm{\bar{20}},\bm{4}\}$} and {\color{red!80!black} $\{\bm{\bar{4}},\bm{20}\}$} \\
        & $p=2$ & {\color{green!50!black} $\{\bm{64},\bm{1}\}$} and {\color{blue}$\{\bm{6},\bm{15}\}$} \\
        & $p=3$ & {\color{red!80!black} $\{\bm{20},\bm{\bar{4}}\}$} and {\color{red!80!black} $\{\bm{4}, \bm{\bar{20}}\}$} \\
        \bottomrule
    \end{tabular}
    \caption{The strong-coupling labels of lattice vacua for the continuum gauge theories considered in this paper, ordered by their flux tube sectors. The notation $\bm r \oplus \bm r'$ indicates that the strong-coupling vacuum is a superposition of $\bm r$ and $\bm r'$. States of the same color are related by exact lattice symmetries, as described in Appendix~\ref{app:symmetries} (see Figure~\ref{fig:symmetries_on_vacua}).}
    \label{tab:lattice_vacua}
\end{table}

Once we know the strong-coupling limit $\ket{\psi_j(ga\to\infty)}$ of the $j$th lattice vacuum, we can be more numerically efficient by initializing VUMPS with a uMPS that has a large overlap with $\ket{\psi_j(ga\to\infty)}$. This is effective because, as we can see in Figure~\ref{fig:repsinsu25}, even as we approach the continuum limit the lattice vacuum $\ket{\psi_j(ga)}$ will have a relatively large overlap with $\ket{\psi_j(ga\to\infty)}$. Thus, when we use an initial state that is close to $\ket{\psi_j(ga\to\infty)}$, VUMPS will reliably converge to $\ket{\psi_j(ga)}$.

\subsection{Lattice decay rule}\label{sec:lattice_decay}

In Section~\ref{sec:adiabatic}, we observed that the local minima found by VUMPS at finite lattice spacing are adiabatically connected to the degenerate vacua in the continuum as $ga\to 0$, and also to simple states in the lattice strong-coupling limit $ga\to\infty$, characterized by pairs of link representations $\{\bm{r}_1,\bm{r}_2\}$. Because the strong-coupling states that are adiabatically connected to the continuum vacua include states beyond the ground states of the strong-coupling Hamiltonian, it is not straightforward to identify which strong-coupling states become vacua of the theory in the continuum. In the following, we will develop a rule that allows us to identify these strong-coupling states, and in particular all states in Table~\ref{tab:lattice_vacua}, without any numerical calculations.

Our approach will be based on a notion of decay on the lattice. Consider the following process, in which we start from a translation-invariant strong-coupling state $\{\bm{r}_1,\bm{r}_2\}$ and then nucleate a bubble of a different state $\{\bm{r'}_1,\bm{r'}_2\}$ one link at a time:
\begin{equation}
    \begin{tikzpicture}[xscale=1.4,baseline=-2cm]
        \foreach \y in {0,-1,-2,-3,-4} {
            \draw[ultra thick] (-2.5,\y) -- (6.5,\y);
        };
        \foreach \y in {0,-1,-2,-3} {
            \draw[thick,-latex] (-3,{\y-0.1}) -- (-3,{\y-0.9});
        };

        \draw[ultra thick,blue!50] (1,-1) -- (2,-1);
        \draw[ultra thick,blue!50] (1,-2) -- (3,-2);
        \draw[ultra thick,blue!50] (0,-3) -- (3,-3);
        \draw[ultra thick,blue!50] (0,-4) -- (4,-4);

        \foreach \y in {0,-1,-2,-3,-4} {
            \foreach \x in {-2,-1,...,6} {
                \node[fill,circle,minimum size=5,inner sep=0] at (\x,\y) {};
            };
        };

        \foreach \x/\y in {-2/0,0/0,2/0,4/0,-2/-1,0/-1,2/-1,4/-1,-2/-2,0/-2,4/-2,-2/-3,4/-3,-2/-4,4/-4} {
            \node[above] at ({\x + .5},\y) {$\bm{r}_1$};
        };
        \foreach \x/\y in {-1/0,1/0,3/0,5/0,-1/-1,3/-1,5/-1,-1/-2,3/-2,5/-2,-1/-3,3/-3,5/-3,-1/-4,5/-4} {
            \node[above] at ({\x + .5},\y) {$\bm{r}_2$};
        };
        \foreach \x/\y in {2/-2,0/-3,2/-3,0/-4,2/-4} {
            \node[above,blue!50] at ({\x + .5},\y) {$\bm{r'}_1$};
        };
        \foreach \x/\y in {1/-1,1/-2,1/-3,1/-4,3/-4} {
            \node[above,blue!50] at ({\x + .5},\y) {$\bm{r'}_2$};
        };
    \end{tikzpicture}
\end{equation}
In order for such a process to be kinematically possible, $\bm{r'}_2$ must appear in $\bm{r}_1 \otimes \bm{R}(\bm{\lambda})$ and $\bm{r'}_1$ must appear in $\bm{r}_2\otimes \bm{R}(\bm{\lambda})$. We will additionally impose a dynamical constraint: the energy must be non-increasing at each individual step, so $C_2(\bm{r'}_1) \leq C_2(\bm{r}_1)$ and $C_2(\bm{r'}_2) \leq C_2(\bm{r}_2)$.

We can consider all strong-coupling states, and ask which ones can decay into which others. We claim that the endpoints of decay chains are the strong-coupling limits of the degenerate vacua in the continuum. More precisely:
\begin{itemize}
    \item If a state $\{\bm{r}_1,\bm{r}_2\}$ does not decay to any other state, then it is the strong-coupling limit of one of the continuum vacua.
    \item If there are states $\{\bm{r}_1,\bm{r}_2\}$ and $\{\bm{r'}_1,\bm{r}_2\}$ that can decay into each other, but not into any lower-energy states, then in the notation of Table~\ref{tab:lattice_vacua}, $\{\bm{r}_1\oplus \bm{r'}_1,\bm{r}_2\}$ is a strong-coupling limit of one of the continuum vacua. A similar statement holds for larger groups of states that can decay into each other but not into any lower-energy states.
\end{itemize}

Let us explain how the lattice deacy rule works in the simplest example, the $\SU(2)$ theory with matter in the $\bm{5}$ representation. We can check that the four strong-coupling states we listed in Table~\ref{tab:lattice_vacua} cannot decay into any other states. For instance, consider the {\color{green!50!black} $\{\bm{1},\bm{4}\}$} state. Since we require the energy to be non-increasing every time we change a link representation, we must set $\bm{r'}_1 = \bm{1}$. And then, since we must have $\bm{r'}_2$ appearing in $\bm{r'}_1 \otimes \bm{R}(\bm{\lambda})$, and $\bm{R}(\bm{\lambda}) = \bm{4}$ in this case, we must set $\bm{r'}_2 = \bm{4}$. So, indeed, {\color{green!50!black} $\{\bm{1},\bm{4}\}$} cannot decay to any other state. Similar arguments follow for the other three states.

For any other state, however, we can find a decay. For example, starting from $\{\bm{3},\bm{4}\}$, we could decay to {\color{green!50!black} $\{\bm{1},\bm{4}\}$}, or starting from $\{\bm{2},\bm{5}\}$, we could decay to {\color{blue} $\{\bm{2},\bm{3}\}$}. To visualize these decays, it is helpful to define an elementary decay as one that cannot proceed through an intermediate state. That is, $\{\bm{r}_1,\bm{r}_2\} \to \{\bm{r'}_1,\bm{r'}_2\}$ is elementary if there does not exist any decay pathway $\{\bm{r}_1,\bm{r}_2\} \to \{\bm{r''}_1,\bm{r''}_2\} \to \{\bm{r'}_1,\bm{r'}_2\}$. Note that this implies either $\bm{r'}_1 = \bm{r}_1$ or $\bm{r'}_2 = \bm{r}_2$.

In Figure~\ref{fig:su2_5_decay}, we show the elementary decays that are possible for the lowest-lying translation-invariant strong-coupling states of the $\SU(2)+\psi_{\bm{5}}$ theory (the two flux tube sectors are marked with different colors in this figure). We clearly see that there are exactly four states (marked with double boxes in Figure~\ref{fig:su2_5_decay}) that cannot decay into any other states, and that these are the states listed in Table~\ref{tab:lattice_vacua}.

\begin{figure}
    \centering
    \begin{tikzpicture}
        \draw[thick,fill=gray!10] (5.7,2.2) rectangle (7.3,3.3);
        \node[scale=.7,fill=99_1,circle,minimum size=7,inner sep=0,label=0:{\small $p = 0$}] at (6,3) {};
        \node[scale=.7,fill=99_2,circle,minimum size=7,inner sep=0,label=0:{\small $p = 1$}] at (6,2.5) {};

        \draw[ultra thick,->] (-7.5,2) -- node[above,rotate=90] {Energy density} (-7.5,10);
        
        \clip (-8,1) rectangle (8,10);
        \begin{scope}[scale=1.2]
        \node[draw,very thick,align=center,fill=white,fill opacity=0.8,double] at (-2.65165,2.65165) (1) {$\{\bm{1},\bm{4}\}$ \tikz[baseline=-0.5ex,inner sep=0pt]{\node[fill=99_1,circle,minimum size=8] (a) {}; \path (-.75ex,0) -- (.75ex,0);} };
\node[draw,very thick,align=center,fill=white,fill opacity=0.8,double] at (-0.883883,1.94454) (2) {$\{\bm{2},\bm{3}\}$ \tikz[baseline=-0.5ex,inner sep=0pt]{\node[fill=99_2,circle,minimum size=8] (a) {}; \path (-.75ex,0) -- (.75ex,0);} };
\node[draw,very thick,align=center,fill=white,fill opacity=0.8] at (-3.71231,4.77297) (3) {$\{\bm{2},\bm{5}\}$ \tikz[baseline=-0.5ex,inner sep=0pt]{\node[fill=99_2,circle,minimum size=8] (a) {}; \path (-.75ex,0) -- (.75ex,0);} };
\node[draw,very thick,align=center,fill=white,fill opacity=0.8,double] at (0.883883,1.94454) (4) {$\{\bm{3},\bm{2}\}$ \tikz[baseline=-0.5ex,inner sep=0pt]{\node[fill=99_1,circle,minimum size=8] (a) {}; \path (-.75ex,0) -- (.75ex,0);} };
\node[draw,very thick,align=center,fill=white,fill opacity=0.8] at (-1.23744,4.06586) (5) {$\{\bm{3},\bm{4}\}$ \tikz[baseline=-0.5ex,inner sep=0pt]{\node[fill=99_1,circle,minimum size=8] (a) {}; \path (-.75ex,0) -- (.75ex,0);} };
\node[draw,very thick,align=center,fill=white,fill opacity=0.8] at (-4.77297,7.6014) (6) {$\{\bm{3},\bm{6}\}$ \tikz[baseline=-0.5ex,inner sep=0pt]{\node[fill=99_1,circle,minimum size=8] (a) {}; \path (-.75ex,0) -- (.75ex,0);} };
\node[draw,very thick,align=center,fill=white,fill opacity=0.8,double] at (2.65165,2.65165) (7) {$\{\bm{4},\bm{1}\}$ \tikz[baseline=-0.5ex,inner sep=0pt]{\node[fill=99_2,circle,minimum size=8] (a) {}; \path (-.75ex,0) -- (.75ex,0);} };
\node[draw,very thick,align=center,fill=white,fill opacity=0.8] at (1.23744,4.06586) (8) {$\{\bm{4},\bm{3}\}$ \tikz[baseline=-0.5ex,inner sep=0pt]{\node[fill=99_2,circle,minimum size=8] (a) {}; \path (-.75ex,0) -- (.75ex,0);} };
\node[draw,very thick,align=center,fill=white,fill opacity=0.8] at (-1.59099,6.89429) (9) {$\{\bm{4},\bm{5}\}$ \tikz[baseline=-0.5ex,inner sep=0pt]{\node[fill=99_2,circle,minimum size=8] (a) {}; \path (-.75ex,0) -- (.75ex,0);} };
\node[draw,very thick,align=center,fill=white,fill opacity=0.8] at (-5.83363,11.1369) (10) {$\{\bm{4},\bm{7}\}$ \tikz[baseline=-0.5ex,inner sep=0pt]{\node[fill=99_2,circle,minimum size=8] (a) {}; \path (-.75ex,0) -- (.75ex,0);} };
\node[draw,very thick,align=center,fill=white,fill opacity=0.8] at (3.71231,4.77297) (11) {$\{\bm{5},\bm{2}\}$ \tikz[baseline=-0.5ex,inner sep=0pt]{\node[fill=99_1,circle,minimum size=8] (a) {}; \path (-.75ex,0) -- (.75ex,0);} };
\node[draw,very thick,align=center,fill=white,fill opacity=0.8] at (1.59099,6.89429) (12) {$\{\bm{5},\bm{4}\}$ \tikz[baseline=-0.5ex,inner sep=0pt]{\node[fill=99_1,circle,minimum size=8] (a) {}; \path (-.75ex,0) -- (.75ex,0);} };
\node[draw,very thick,align=center,fill=white,fill opacity=0.8] at (-1.94454,10.4298) (13) {$\{\bm{5},\bm{6}\}$ \tikz[baseline=-0.5ex,inner sep=0pt]{\node[fill=99_1,circle,minimum size=8] (a) {}; \path (-.75ex,0) -- (.75ex,0);} };
\node[draw,very thick,align=center,fill=white,fill opacity=0.8] at (4.77297,7.6014) (14) {$\{\bm{6},\bm{3}\}$ \tikz[baseline=-0.5ex,inner sep=0pt]{\node[fill=99_2,circle,minimum size=8] (a) {}; \path (-.75ex,0) -- (.75ex,0);} };
\node[draw,very thick,align=center,fill=white,fill opacity=0.8] at (1.94454,10.4298) (15) {$\{\bm{6},\bm{5}\}$ \tikz[baseline=-0.5ex,inner sep=0pt]{\node[fill=99_2,circle,minimum size=8] (a) {}; \path (-.75ex,0) -- (.75ex,0);} };
\node[draw,very thick,align=center,fill=white,fill opacity=0.8] at (-2.2981,14.6725) (16) {$\{\bm{6},\bm{7}\}$ \tikz[baseline=-0.5ex,inner sep=0pt]{\node[fill=99_2,circle,minimum size=8] (a) {}; \path (-.75ex,0) -- (.75ex,0);} };
\node[draw,very thick,align=center,fill=white,fill opacity=0.8] at (5.83363,11.1369) (17) {$\{\bm{7},\bm{4}\}$ \tikz[baseline=-0.5ex,inner sep=0pt]{\node[fill=99_1,circle,minimum size=8] (a) {}; \path (-.75ex,0) -- (.75ex,0);} };
\node[draw,very thick,align=center,fill=white,fill opacity=0.8] at (2.2981,14.6725) (18) {$\{\bm{7},\bm{6}\}$ \tikz[baseline=-0.5ex,inner sep=0pt]{\node[fill=99_1,circle,minimum size=8] (a) {}; \path (-.75ex,0) -- (.75ex,0);} };
\draw[-latex,thick,99_2,transform canvas={xshift=0.cm}] (3) -- (2);
\draw[-latex,thick,99_1,transform canvas={xshift=0.cm}] (5) -- (1);
\draw[-latex,thick,99_1,transform canvas={xshift=0.cm}] (5) -- (4);
\draw[-latex,thick,99_1,transform canvas={xshift=0.cm}] (6) -- (5);
\draw[-latex,thick,99_2,transform canvas={xshift=0.cm}] (8) -- (2);
\draw[-latex,thick,99_2,transform canvas={xshift=0.cm}] (8) -- (7);
\draw[-latex,thick,99_2,transform canvas={xshift=0.cm}] (9) -- (3);
\draw[-latex,thick,99_2,transform canvas={xshift=0.cm}] (9) -- (8);
\draw[-latex,thick,99_2,transform canvas={xshift=0.cm}] (10) -- (9);
\draw[-latex,thick,99_1,transform canvas={xshift=0.cm}] (11) -- (4);
\draw[-latex,thick,99_1,transform canvas={xshift=0.cm}] (12) -- (5);
\draw[-latex,thick,99_1,transform canvas={xshift=0.cm}] (12) -- (11);
\draw[-latex,thick,99_1,transform canvas={xshift=0.cm}] (13) -- (6);
\draw[-latex,thick,99_1,transform canvas={xshift=0.cm}] (13) -- (12);
\draw[-latex,thick,99_2,transform canvas={xshift=0.cm}] (14) -- (8);
\draw[-latex,thick,99_2,transform canvas={xshift=0.cm}] (15) -- (9);
\draw[-latex,thick,99_2,transform canvas={xshift=0.cm}] (15) -- (14);
\draw[-latex,thick,99_2,transform canvas={xshift=0.cm}] (16) -- (10);
\draw[-latex,thick,99_2,transform canvas={xshift=0.cm}] (16) -- (15);
\draw[-latex,thick,99_1,transform canvas={xshift=0.cm}] (17) -- (12);
\draw[-latex,thick,99_1,transform canvas={xshift=0.cm}] (18) -- (13);
\draw[-latex,thick,99_1,transform canvas={xshift=0.cm}] (18) -- (17);
\node[draw,very thick,align=center,fill=white,fill opacity=0.8,double] at (-2.65165,2.65165) (1) {\textcolor{green!50!black}{$\{\bm{1},\bm{4}\}$} \tikz[baseline=-0.5ex,inner sep=0pt]{\node[fill=99_1,circle,minimum size=8] (a) {}; \path (-.75ex,0) -- (.75ex,0);} };
\node[draw,very thick,align=center,fill=white,fill opacity=0.8,double] at (-0.883883,1.94454) (2) {\textcolor{blue}{$\{\bm{2},\bm{3}\}$} \tikz[baseline=-0.5ex,inner sep=0pt]{\node[fill=99_2,circle,minimum size=8] (a) {}; \path (-.75ex,0) -- (.75ex,0);} };
\node[draw,very thick,align=center,fill=white,fill opacity=0.8] at (-3.71231,4.77297) (3) {$\{\bm{2},\bm{5}\}$ \tikz[baseline=-0.5ex,inner sep=0pt]{\node[fill=99_2,circle,minimum size=8] (a) {}; \path (-.75ex,0) -- (.75ex,0);} };
\node[draw,very thick,align=center,fill=white,fill opacity=0.8,double] at (0.883883,1.94454) (4) {\textcolor{blue}{$\{\bm{3},\bm{2}\}$} \tikz[baseline=-0.5ex,inner sep=0pt]{\node[fill=99_1,circle,minimum size=8] (a) {}; \path (-.75ex,0) -- (.75ex,0);} };
\node[draw,very thick,align=center,fill=white,fill opacity=0.8] at (-1.23744,4.06586) (5) {$\{\bm{3},\bm{4}\}$ \tikz[baseline=-0.5ex,inner sep=0pt]{\node[fill=99_1,circle,minimum size=8] (a) {}; \path (-.75ex,0) -- (.75ex,0);} };
\node[draw,very thick,align=center,fill=white,fill opacity=0.8] at (-4.77297,7.6014) (6) {$\{\bm{3},\bm{6}\}$ \tikz[baseline=-0.5ex,inner sep=0pt]{\node[fill=99_1,circle,minimum size=8] (a) {}; \path (-.75ex,0) -- (.75ex,0);} };
\node[draw,very thick,align=center,fill=white,fill opacity=0.8,double] at (2.65165,2.65165) (7) {\textcolor{green!50!black}{$\{\bm{4},\bm{1}\}$} \tikz[baseline=-0.5ex,inner sep=0pt]{\node[fill=99_2,circle,minimum size=8] (a) {}; \path (-.75ex,0) -- (.75ex,0);} };
\node[draw,very thick,align=center,fill=white,fill opacity=0.8] at (1.23744,4.06586) (8) {$\{\bm{4},\bm{3}\}$ \tikz[baseline=-0.5ex,inner sep=0pt]{\node[fill=99_2,circle,minimum size=8] (a) {}; \path (-.75ex,0) -- (.75ex,0);} };
\node[draw,very thick,align=center,fill=white,fill opacity=0.8] at (-1.59099,6.89429) (9) {$\{\bm{4},\bm{5}\}$ \tikz[baseline=-0.5ex,inner sep=0pt]{\node[fill=99_2,circle,minimum size=8] (a) {}; \path (-.75ex,0) -- (.75ex,0);} };
\node[draw,very thick,align=center,fill=white,fill opacity=0.8] at (-5.83363,11.1369) (10) {$\{\bm{4},\bm{7}\}$ \tikz[baseline=-0.5ex,inner sep=0pt]{\node[fill=99_2,circle,minimum size=8] (a) {}; \path (-.75ex,0) -- (.75ex,0);} };
\node[draw,very thick,align=center,fill=white,fill opacity=0.8] at (3.71231,4.77297) (11) {$\{\bm{5},\bm{2}\}$ \tikz[baseline=-0.5ex,inner sep=0pt]{\node[fill=99_1,circle,minimum size=8] (a) {}; \path (-.75ex,0) -- (.75ex,0);} };
\node[draw,very thick,align=center,fill=white,fill opacity=0.8] at (1.59099,6.89429) (12) {$\{\bm{5},\bm{4}\}$ \tikz[baseline=-0.5ex,inner sep=0pt]{\node[fill=99_1,circle,minimum size=8] (a) {}; \path (-.75ex,0) -- (.75ex,0);} };
\node[draw,very thick,align=center,fill=white,fill opacity=0.8] at (-1.94454,10.4298) (13) {$\{\bm{5},\bm{6}\}$ \tikz[baseline=-0.5ex,inner sep=0pt]{\node[fill=99_1,circle,minimum size=8] (a) {}; \path (-.75ex,0) -- (.75ex,0);} };
\node[draw,very thick,align=center,fill=white,fill opacity=0.8] at (4.77297,7.6014) (14) {$\{\bm{6},\bm{3}\}$ \tikz[baseline=-0.5ex,inner sep=0pt]{\node[fill=99_2,circle,minimum size=8] (a) {}; \path (-.75ex,0) -- (.75ex,0);} };
\node[draw,very thick,align=center,fill=white,fill opacity=0.8] at (1.94454,10.4298) (15) {$\{\bm{6},\bm{5}\}$ \tikz[baseline=-0.5ex,inner sep=0pt]{\node[fill=99_2,circle,minimum size=8] (a) {}; \path (-.75ex,0) -- (.75ex,0);} };
\node[draw,very thick,align=center,fill=white,fill opacity=0.8] at (-2.2981,14.6725) (16) {$\{\bm{6},\bm{7}\}$ \tikz[baseline=-0.5ex,inner sep=0pt]{\node[fill=99_2,circle,minimum size=8] (a) {}; \path (-.75ex,0) -- (.75ex,0);} };
\node[draw,very thick,align=center,fill=white,fill opacity=0.8] at (5.83363,11.1369) (17) {$\{\bm{7},\bm{4}\}$ \tikz[baseline=-0.5ex,inner sep=0pt]{\node[fill=99_1,circle,minimum size=8] (a) {}; \path (-.75ex,0) -- (.75ex,0);} };
\node[draw,very thick,align=center,fill=white,fill opacity=0.8] at (2.2981,14.6725) (18) {$\{\bm{7},\bm{6}\}$ \tikz[baseline=-0.5ex,inner sep=0pt]{\node[fill=99_1,circle,minimum size=8] (a) {}; \path (-.75ex,0) -- (.75ex,0);} };
        \end{scope}
    \end{tikzpicture}
    \caption{The elementary lattice decays in the $\SU(2)+\psi_{\bm{5}}$ theory. States are labeled by $\{\bm{r}_1,\bm{r}_2\}$, with colored circles indicating their flux tube sectors. The states identified by the lattice decay rule are marked with double boxes and colored according to the scheme in Table~\ref{tab:lattice_vacua}.}
    \label{fig:su2_5_decay}
\end{figure}

We can repeat this logic for any of the theories we are considering. In Figures~\ref{fig:su2su2_33_decay}, \ref{fig:su2su2_24_decay}, and~\ref{fig:su4_15_decay}, we show similar decay diagrams for the other theories in Table~\ref{tab:theories}. Note that if two states have the same values of $C_2(\bm{r}_1)$ and $C_2(\bm{r}_2)$, then we put them in the same box; in each case for which two states are in the same box and in the same flux tube sector (indicated by the colored dots next to the state labels), they can decay into each other. However, if two states are in different flux tube sectors, then they can never decay into each other, and so appearing in the same box is inconsequential.

\begin{figure}
    \centering
    \begin{tikzpicture}

    \begin{scope}[scale=.75,shift={(1.5,5.5)}]
    \draw[thick,fill=gray!10] (5.7,1.7) rectangle (8.3,3.8);
    \node[scale=.7,fill=99_1,circle,minimum size=7,inner sep=0,label=0:{\footnotesize 
    $p = (0,0)$}] at (6,3.5) {};
    \node[scale=.7,fill=99_2,circle,minimum size=7,inner sep=0,label=0:{\footnotesize $p = (0,1)$}] at (6,3) {};
    \node[scale=.7,fill=99_3,circle,minimum size=7,inner sep=0,label=0:{\footnotesize $p = (1,0)$}] at (6,2.5) {};
    \node[scale=.7,fill=99_4,circle,minimum size=7,inner sep=0,label=0:{\footnotesize $p = (1,1)$}] at (6,2) {};
    \end{scope}

    \draw[ultra thick,->] (-7.5,6) -- node[above,rotate=90] {Energy density} (-7.5,10);
    
    \begin{scope}[every node/.style={scale=.38},xscale=1.97,yscale=2.5]
        \clip (-4.2,2) rectangle (4.2,5);
        \input{figures/su2su2_33.tikz}
    \end{scope}
    \end{tikzpicture}
    \caption{The elementary lattice decay pathways in the $\SU(2)\times\SU(2)+\psi_{(\bm{3},\bm{3})}$ theory. States are labeled by $\{\bm{r}_1,\bm{r}_2\}$, with colored circles indicating their flux tube sectors. The states identified by the lattice decay rule are marked with double boxes and colored according to the scheme in Table~\ref{tab:lattice_vacua}.}
    \label{fig:su2su2_33_decay}
\end{figure}

\begin{figure}
    \centering
    \begin{tikzpicture}

    \begin{scope}[scale=.75,shift={(1.5,1.5)}]
    \draw[thick,fill=gray!10] (5.7,2.2) rectangle (7.6,3.3);
    \node[scale=.7,fill=99_1,circle,minimum size=7,inner sep=0,label=0:{\footnotesize $p = 0$}] at (6,3) {};
    \node[scale=.7,fill=99_2,circle,minimum size=7,inner sep=0,label=0:{\footnotesize $p = 1$}] at (6,2.5) {};
    \end{scope}

    \draw[ultra thick,->] (-7.5,3) -- node[above,rotate=90] {Energy density} (-7.5,10);
    
    \begin{scope}[every node/.style={scale=.4},xscale=1.95,yscale=2.5]
        \clip (-4.2,.9) rectangle (4.2,4.8);
        \input{figures/su2su2_24.tikz}
    \end{scope}
    \end{tikzpicture}
    \caption{The elementary lattice decay pathways in the $\SU(2)\times\SU(2)+\psi_{(\bm{2},\bm{4})}$ theory. States are labeled by $\{\bm{r}_1,\bm{r}_2\}$, with colored circles indicating their flux tube sectors. The states identified by the lattice decay rule are marked with double boxes and colored according to the scheme in Table~\ref{tab:lattice_vacua}.}
    \label{fig:su2su2_24_decay}
\end{figure}

\begin{figure}
    \centering
    \begin{tikzpicture}
    \begin{scope}[scale=.75,shift={(.5,10.5)}]
    \draw[thick,fill=gray!10] (5.7,1.7) rectangle (7.6,3.8);
    \node[scale=.7,fill=99_1,circle,minimum size=7,inner sep=0,label=0:{\footnotesize $p = 0$}] at (6,3.5) {};
    \node[scale=.7,fill=99_2,circle,minimum size=7,inner sep=0,label=0:{\footnotesize $p = 1$}] at (6,3) {};
    \node[scale=.7,fill=99_3,circle,minimum size=7,inner sep=0,label=0:{\footnotesize $p = 2$}] at (6,2.5) {};
    \node[scale=.7,fill=99_4,circle,minimum size=7,inner sep=0,label=0:{\footnotesize $p = 3$}] at (6,2) {};
    \end{scope}

    \draw[ultra thick,->] (-8,9.5) -- node[above,rotate=90] {Energy density} (-8,15);
    
    \begin{scope}[every node/.style={scale=.4},xscale=1.3,yscale=2]
        \clip (-5.7,4.3) rectangle (5.7,8);
        \input{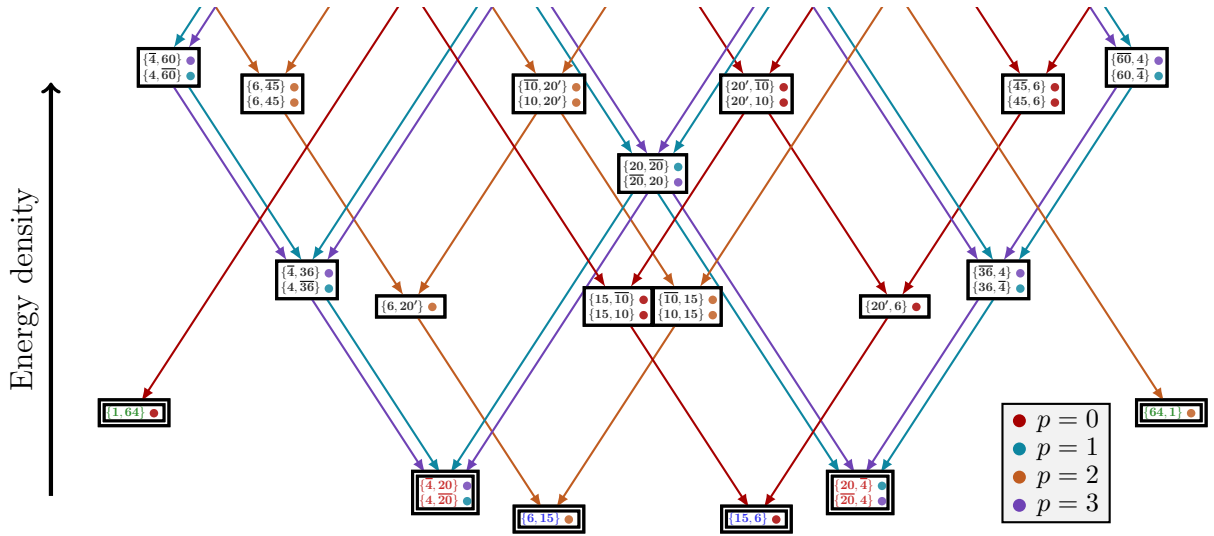}
    \end{scope}
    \end{tikzpicture}
    \caption{The elementary lattice decay pathways in the $\SU(4)+\psi_{\bm{15}}$ theory. States are labeled by $\{\bm{r}_1,\bm{r}_2\}$, with colored circles indicating their flux tube sectors. The states identified by the lattice decay rule are marked with double boxes and colored according to the scheme in Table~\ref{tab:lattice_vacua}.}
    \label{fig:su4_15_decay}
\end{figure}

In each case, the states we find from the decay rule are the same as those we found in Table~\ref{tab:lattice_vacua} from the strong-coupling limits of the lattice vacua we obtained numerically. We conjecture that this rule can be used in any gapped $(1+1)$D gauge theory to find the lattice strong-coupling limits of the degenerate vacua.

In Appendix~\ref{app:count_checks}, we repeat this exercise in many other examples (namely in the $\text{G}_2 + \psi_{\bf 14}$, $\Spin(9) + \psi_{\bf 16}$, $\text{F}_4 + \psi_{\bf 26}$, and $\USp(8) + \psi_{\bf 42}$ theories), and also independently calculate the number of vacua using the branching of coset characters as described in Section~\ref{subsec:continuum_action}. In every case we find that the counts match. For instance, in the $\USp(8)+\psi_{\bm{42}}$ theory, our decay rule finds 24 strong-coupling states, and from the coset characters we indeed find 24 degenerate vacua. It is surprising that we can compute the vacuum degeneracy in this manner, even though the degeneracy is lifted in the lattice strong-coupling limit.

\subsection{Lattice strong-coupling expansion}\label{sec:strong_coupling}

The lattice decay rule we explained in Section~\ref{sec:lattice_decay} identifies the lattice vacua in the strict $ga\to\infty$ limit. To understand how these states change as we make $ga$ finite and reduce it toward 0, we need to work harder. We will describe here how to calculate the expectation value of a local operator in some state in an expansion in $1/(ga)$, and we will use this method to calculate $\langle \bar\psi_a \psi_a \rangle/g$ up to order $1/(ga)^3$. This will serve as a useful check of our numerical results in Section~\ref{sec:results}.

In the strict strong-coupling limit $ga\gg 1$, the lattice Hamiltonian \eqref{eq:lattice_ham} reduces to the gauge kinetic term \eqref{eq:strong_ham}. To systematically study corrections in $1/(ga)$, we split the full massless lattice Hamiltonian \eqref{eq:lattice_ham} as $H=H_0+V$, with 
\begin{equation}
    H_0 = \frac{ag^2}{2}\sum_n L^A_nL^A_n\,,\qquad V=-\frac{i}{2a}\sum_n\lambda_n^aU^{ab}_n\lambda^b_{n+1}\,,
\end{equation}
and treat $V$ as a perturbation to $H_0$. We can then use standard Rayleigh-Schr\"odinger perturbation theory to calculate the expectation value of a local operator as a series in $1/(ga)$:
\begin{equation}\label{eq:pert}
    \braket{\lambda_m|\mathcal O|\lambda_m} = \braket{\lambda^{(0)}_m|\mathcal O|\lambda^{(0)}_m} + 2\Re\left[\sum_{n\neq m}\frac{\braket{\lambda^{(0)}_m|\mathcal O|\lambda^{(0)}_n}\braket{\lambda^{(0)}_n|V|\lambda^{(0)}_m}}{E^{(0)}_n-E^{(0)}_m}\right]+\cdots\,,
\end{equation}
where $\ket{\lambda^{(0)}_m}$ is the zeroth-order state and $H_0 \ket{\lambda^{(0)}_m} = E^{(0)}_m \ket{\lambda^{(0)}_m}$.

We will use this expression to compute the chiral condensate, defined on the lattice by the expectation value of\footnote{Note that for a translation-invariant state on an infinite lattice, it suffices to set $N = 2$.}
\begin{equation}\label{eq:chiral_cond}
    \mathcal{O} = \bar\psi_a \psi_a \equiv \frac{1}{Na} \frac{\partial H}{
    \partial m} = -\frac{i}{2Na} \sum_{n=1}^N (-1)^n \chi^a_n U^{ab}_n\chi^b_{n+1} \,.
\end{equation}
The strong-coupling expansion of this quantity was carried out in \cite{Dempsey:2023fvm,Dempsey:2024alw} for the case of adjoint QCD$_2$. There as well as here, the most difficult aspect is that each energy level in the strong-coupling limit is highly degenerate, on account of the quantum numbers in \eqref{eq:phys_states} that do not appear in $H_0$ (i.e., the coupling scheme $e_n$ and multiplicity $i$). Thus, following the usual procedure for degenerate perturbation theory, we first need to find the ground state of $V$ projected onto a degenerate eigenspace of $H_0$, and then use that state as $\ket{\lambda_m^{(0)}}$ in \eqref{eq:pert}. (Note that for all the theories we consider, the states in Table~\ref{tab:lattice_vacua} are colored so that states of the same color are related by exact symmetries on the lattice, and so we will only compute the strong-coupling expansion for one representative of each color. See Figure~\ref{fig:symmetries_on_vacua}.)

We can start with the $\SU(2)+\psi_{\bm{5}}$ theory. In this case, $n_0(\bm{5}) = 1$ and so, for an $N$-site lattice, the multiplicity label for basis states \eqref{eq:phys_states} runs from 1 to $2^{N/2}$. This extra degree of freedom can be encoded in a Majorana fermion $\lambda_n$ on every site;  see Appendix~\ref{app:su2_5} for further details of this construction.  When we project $V$ onto the degenerate subspaces with link representations \textcolor{green!50!black}{$\{\bm{1},\bm{4}\}$} or \textcolor{blue}{$\{\bm{3},\bm{2}\}$}, we find
\begin{equation}\label{eq:su25_eff}
    \underline{\SU(2)+\psi_{\bm 5}}:\qquad
        V_{\color{green!50!black} \{\bm{1},\bm{4}\}} = -\frac{5i}{4a}\sum_n \lambda_{2n}\lambda_{2n+1}\,,\qquad  V_{\color{blue} \{\bm{3},\bm{2}\}} = -\frac{5i}{8a}\sum_n \lambda_{2n}\lambda_{2n+1}\,,
\end{equation}
Since $V_{(\bm{1},\bm{4})}$ and $V_{(\bm{3},\bm{2})}$ are both sums of commuting terms, it is straightforward to calculate their ground states. Using these ground states along with \eqref{eq:pert}, we can compute the strong-coupling expansion of the chiral condensate in these states as
\begin{equation}\label{eq:su25_vev}
    \underline{\SU(2)+\psi_{\bm 5}}:\qquad\begin{aligned}
        \braket{\bar\psi_a\psi_a}_{\color{green!50!black} \{\bm{1},\bm{4}\}}/g &= -\frac{5}{8}(ga)^{-1} + \frac{5}{48}(ga)^{-3}+\mathcal O((ga)^{-5})\,,\\
        \braket{\bar\psi_a\psi_a}_{\color{blue} \{\bm{3},\bm{2}\}}/g &= \frac{5}{16}(ga)^{-1} - \frac{85}{384}(ga)^{-3}+\mathcal O((ga)^{-5})\,.
        \end{aligned}
\end{equation}
Using a one-site translation on the lattice, under which \eqref{eq:chiral_cond} changes sign, one can show that $\braket{\bar\psi_a\psi_a}_{\color{green!50!black} \{\bm{4},\bm{1}\}} = -\braket{\bar\psi_a\psi_a}_{\color{green!50!black} \{\bm{1},\bm{4}\}}$ and $\braket{\bar\psi_a\psi_a}_{\color{blue} \{\bm{2},\bm{3}\}} = -\braket{\bar\psi_a\psi_a}_{\color{blue} \{\bm{3},\bm{2}\}}$.

In the $\SU(2)\times \SU(2)+\psi_{(\bm 3,\bm 3)}$ theory, we again have a Majorana fermion on each site, as described in Appendix~\ref{app:lattice_detail}. For the {\color{red!80!black} $\{(\bm{1},\bm{2}), (\bm{2},\bm{3})\}$} state, these are the only degrees of freedom leading to strong-coupling degeneracy; the projection of $V$ onto this degenerate subspace is
\begin{equation}
    \underline{\SU(2)\times \SU(2)+\psi_{(\bm 3,\bm 3)}}:\qquad V_{\color{red!80!black} \{(\bm{1},\bm{2}), (\bm{2},\bm{3})\}} = -\frac{25i}{24a}\sum_n \lambda_{2n}\lambda_{2n+1}\,.
\end{equation}
For the {\color{blue} $\{(\bm{1},\bm{1}), (\bm{2},\bm{4})\oplus (\bm{4},\bm{2})\}$} state, we have these Majorana fermions as well as a two-dimensional Hilbert space on every link, spanned by states in which the $\SU(2) \times \SU(2)$ irreps $(\bm{2},\bm{4})$ and $(\bm{4},\bm{2})$, respectively, appear on that link. The projection of $V$ onto this subspace involves both of these degrees of freedom; we find
\begin{equation}\label{eq:su2su233_eff1}
    \underline{\SU(2)\times \SU(2)+\psi_{(\bm 3,\bm 3)}}:\qquad V_{\color{blue} \{(\bm{1},\bm{1}), (\bm{2},\bm{4})\oplus (\bm{4},\bm{2})\}} = -\frac{i}{a}\sum_n \lambda_{2n}\lambda_{2n+1} (1+{\textstyle \frac{5}{4}X_{2n}})\,,
\end{equation}
where $X_{2n}$ is the usual Pauli matrix acting on the two-dimensional space on link $2n$. In a similar manner, for {\color{green!50!black} $\{(\bm{1},\bm{3})\oplus (\bm{3},\bm{1}), (\bm{2},\bm{2})\}$} (for which there is a two-dimensional Hilbert space on every odd link) we find
\begin{equation}\label{eq:su2su233_eff2}
\begin{split}
    \underline{\SU(2)\times \SU(2)+\psi_{(\bm 3,\bm 3)}}:\qquad V_{\color{green!50!black} \{(\bm{1},\bm{3})\oplus (\bm{3},\bm{1}), (\bm{2},\bm{2})\}} ={}& -\frac{2i}{3a}\sum_n \lambda_{2n-1}\lambda_{2n} X_{2n-1}\\
    &-\frac{25i}{36a}\sum_n \lambda_{2n}\lambda_{2n+1}Z_{2n-1}Z_{2n+1}\,.
\end{split}
\end{equation}
After finding the ground states of all of these operators, we can use \eqref{eq:pert} to find the strong-coupling expansions
\begin{equation}\label{eq:sc_su2su233}
    \underline{\SU(2)\times \SU(2)+\psi_{(\bm 3,\bm 3)}}:\quad\begin{aligned}
        \braket{\bar\psi_a\psi_a}_{\color{blue} \{(\bm{1},\bm{1}), (\bm{2},\bm{4})\oplus (\bm{4},\bm{2})\}}/g &= -\frac{9}{8}(ga)^{-1} + \frac{9}{32}(ga)^{-3}+\mathcal O((ga)^{-5})\,,\\
        \braket{\bar\psi_a\psi_a}_{\color{green!50!black} \{(\bm{1},\bm{3})\oplus (\bm{3},\bm{1}), (\bm{2},\bm{2})\}}/g &= -\frac{1}{72}(ga)^{-1} + \frac{737}{2592}(ga)^{-3}+\mathcal O((ga)^{-5})\,,\\
        \braket{\bar\psi_a\psi_a}_{\color{red!80!black} \{(\bm{1},\bm{2}), (\bm{2},\bm{3})\}}/g &= -\frac{25}{48}(ga)^{-1} - \frac{8953}{17280}(ga)^{-3}+\mathcal O((ga)^{-5})\,.\\
        \end{aligned}
\end{equation}
The remaining five vacua appearing in Table \ref{tab:lattice_vacua} are related to the above three vacua by one-site translations and by permuting the two $\SU(2)$ factors in the gauge group.

For $\SU(2)\times \SU(2)+\psi_{(\bm 2,\bm 4)}$, the states {\color{red!50!blue} $\{(\bm{2},\bm{2}), (\bm{2},\bm{2})\}$} and {\color{blue} $\{(\bm{2},\bm{1}), (\bm{2},\bm{1})\}$} are invariant under one-site translations, and so the chiral condensate vanishes in these states. For {\color{red!80!black} $\{(\bm{1},\bm{1}), (\bm{3},\bm{1})\}$}, there is no strong-coupling degeneracy, because for this theory we do not have Majorana fermions on sites (see Appendix~\ref{app:lattice_detail}). Thus, we can apply ordinary perturbation theory. Only in the case of {\color{green!50!black}$\{(\bm{1},\bm{2}), (\bm{2},\bm{3})\oplus (\bm{3},\bm{2})\}$} do we need to use degenerate perturbation theory to account for the two-dimensional Hilbert space on every even link (spanned by states in which the irrep on that link is $(\bm{2},\bm{3})$ or $(\bm{3},\bm{2})$, respectively). The projection of $V$ onto the {\color{green!50!black}$\{(\bm{1},\bm{2}), (\bm{2},\bm{3})\oplus (\bm{3},\bm{2})\}$} subspace is given by
\begin{equation}
    \underline{\SU(2)\times \SU(2)+\psi_{(\bm 2,\bm 4)}}:\qquad 
V_{\color{green!50!black} \{(\bm{1},\bm{2}), (\bm{2},\bm{3})\oplus (\bm{3},\bm{2})\}} = \frac{3}{4a}\sum_n Y_{2n}\,,
\end{equation}
where $Y_{2n}$ is the usual Pauli matrix acting on the two-dimensional space on link $2n$. Using the ground state of this operator along with \eqref{eq:pert}, we find the strong-coupling expansions
\begin{equation}\label{eq:sc_su2su224}
    \underline{\SU(2)\times \SU(2)+\psi_{(\bm 2,\bm 4)}}:\qquad\begin{aligned}
        \braket{\bar\psi_a\psi_a}_{\color{red!80!black} \{(\bm{1},\bm{1}), (\bm{3},\bm{1})\}}/g &=-\frac{44}{45}(ga)^{-3}+\mathcal O((ga)^{-5})\,,\\
        \braket{\bar\psi_a\psi_a}_{\color{green!50!black} \{(\bm{1},\bm{2}), (\bm{2},\bm{3})\oplus (\bm{3},\bm{2})\}}/g &=-\frac{3}{8}(ga)^{-1}-\frac{503}{384}(ga)^{-3}+\mathcal O((ga)^{-5})\,.
        \end{aligned}
\end{equation}

The strong-coupling results for $\SU(4)+\psi_{\bm{15}}$ were derived in \cite{Dempsey:2024alw}.\footnote{Note that the result given there for the strong-coupling ground state of the $p = 0$ universe, which is denoted in this paper by $\{\bm{15},\bm{6}\}$, has an overall sign error.} Due to the high rank of $\SU(4)$ the details are cumbersome and we will only reproduce the result for the chiral condensate
\begin{equation}\label{eq:sc_su4}
    \underline{\SU(4)+\psi_{\bm{15}}}:\qquad\begin{aligned}
        \braket{\bar\psi_a\psi_a}_{\color{blue} \{\bm{15},\bm{6}\}}/g &= 0.502073(ga)^{-1} - \frac{1159631}{3456000}(ga)^{-3} + \mathcal O((ga)^{-5})\,,\\
        \braket{\bar\psi_a\psi_a}_{\color{red!80!black} \{\bm{20},\bm{\bar{4}}\}}/g &= 0.684369(ga)^{-1} + \frac{9459}{256000}(ga)^{-3} + \mathcal O((ga)^{-5})\,,\\
        \braket{\bar\psi_a\psi_a}_{\color{green!50!black} \{\bm{1},\bm{{64}}\}}/g &= -\frac{15}{8}(ga)^{-1} + \frac{15}{32}(ga)^{-3} + \mathcal O((ga)^{-5})\,.
        \end{aligned}
\end{equation}
Note that the last of these expansions did not appear in \cite{Dempsey:2024alw}; it follows a pattern
\begin{equation}\label{eq:condensate_1R}
    \frac{\langle \bar\psi_a \psi_a\rangle_{\{\bm{1},\bm{R}(\adj)\}}}{g\sqrt{N_c}} = \frac{N_c^2-1}{8} \left(-\frac{1}{ ga\sqrt{N_c}} + \frac{1}{(ga\sqrt{N_c})^3} + \mathcal O((ga)^{-5})\right)\,,
\end{equation}
which also holds for the $\SU(2)$ and $\SU(3)$ adjoint theories as shown in \cite{Dempsey:2023fvm,Dempsey:2024alw}.

\section{Lattice results}\label{sec:results}
As shown in Figure~\ref{fig:energydifferences}, the VUMPS algorithm can find the all vacua in Table~\ref{tab:lattice_vacua}, and these vacua become degenerate in the continuum limit.  In this section, we explore various other properties of these vacua.  
In particular, in Section~\ref{sec:condensate}, we characterize the vacua by numerically computing the chiral condensate, and in Section~\ref{sec:particles} we investigate the spectrum of massive excitations. 
Finally, we report further results for $\SU(2)+\psi_{\bm 5}$ at non-zero mass in Section~\ref{sec:su25_extra}.

\subsection{Chiral condensate}\label{sec:condensate}
\begin{figure}[t]
	\centering
	\includegraphics[width=\linewidth]{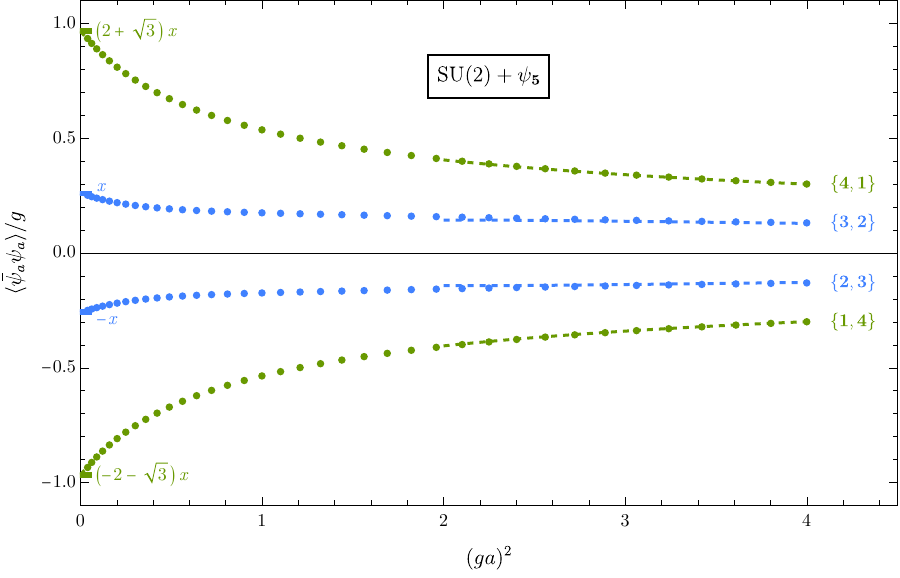}
	\caption{The expectation value of $\bar\psi_a \psi_a$ in the four states that become the degenerate vacua of the $\SU(2)+\psi_{\bm{5}}$ theory as $ga\to 0$. After extrapolating to the continuum limit, the ratios of these expectation values agree well with the values \eqref{eq:su2_ratios} predicted in \cite{Cordova:2024nux}.}
	\label{fig:su2_vevs}
\end{figure}

\begin{figure}[t]
	\centering
	\includegraphics[width=\linewidth]{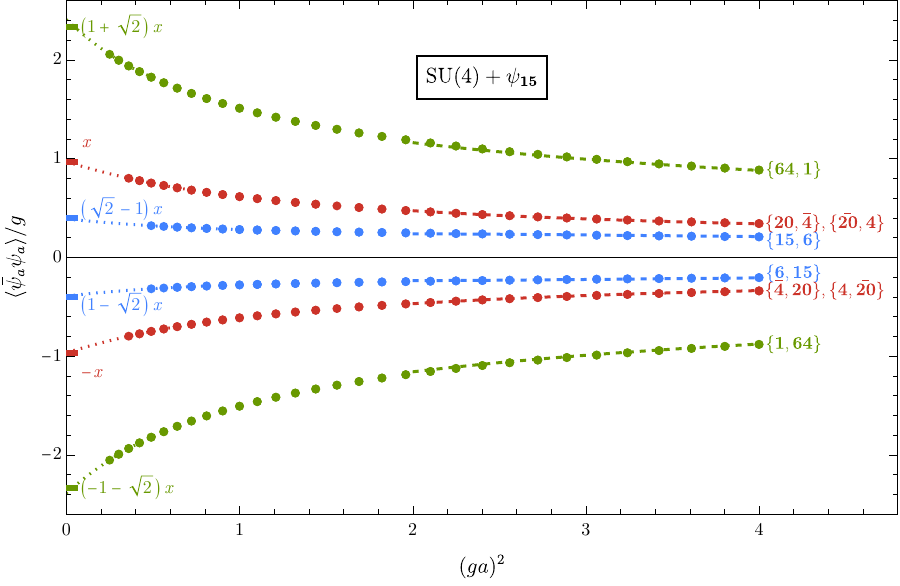}
	\caption{The expectation value of $\bar\psi_a \psi_a$ in the eight states that become the degenerate vacua of the $\SU(4)+\psi_{\bm{15}}$ theory as $ga\to 0$. The dashed lines are the strong-coupling expansions of these expectation values, given in \eqref{eq:sc_su4}. The dotted lines are extrapolations of the lattice data to $ga\to 0$.  The tick marks on the vertical axis correspond to the $ga\to 0$ limit of those extrapolations.}
	\label{fig:su4_vevs}
\end{figure}

\begin{figure}
    \centering
    \begin{subfigure}[t]{\linewidth}
        \includegraphics[width=\linewidth]{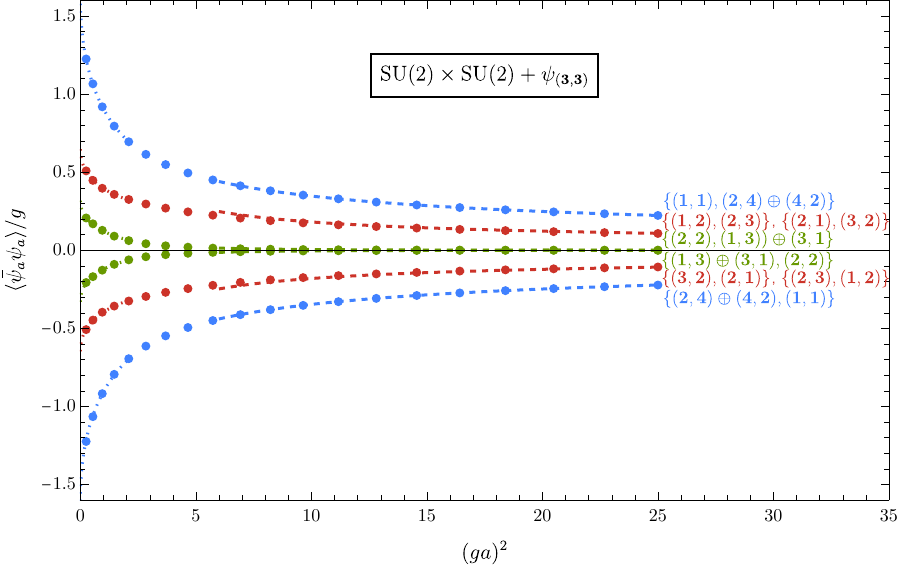}
    \end{subfigure}
    \begin{subfigure}[t]{\linewidth}
        \centering
        \includegraphics[width=\linewidth]{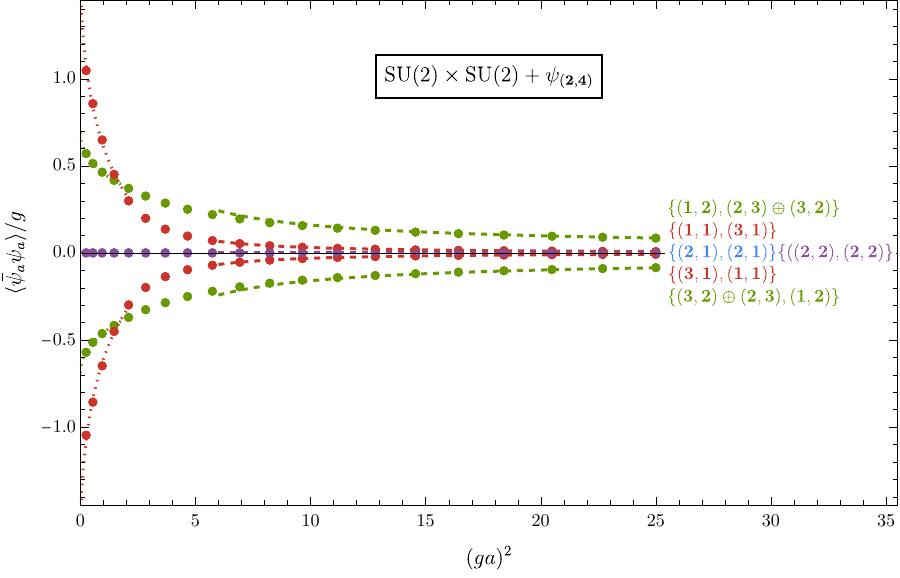}
    \end{subfigure}
    \caption{The expectation value of $\bar\psi_a \psi_a$ in the states that become the degenerate vacua of the two $\SU(2)\times \SU(2)$ theories we studyas $ga\to 0$. The dashed lines represent the strong-coupling expansions of these expectation values, given in \eqref{eq:sc_su2su233} and \eqref{eq:sc_su2su224}. The dotted lines are extrapolations of the lattice data to $ga\to 0$.}
    \label{fig:su2su2_vevs}
\end{figure}

One of the simplest observables that can be used to distinguish the degenerate vacua is the chiral condensate $\braket{\bar\psi_a\psi_a}$, which we defined on the lattice as the expectation value of the operator on the right-hand side of \eqref{eq:chiral_cond}. For the $\grSU(2) + \psi_{\bf 5}$, $\SU(4)+\psi_{\bm{15}}$, and two $\grSU(2) \times \grSU(2)$ theories (all listed in Table~\ref{tab:theories}), we plot $\braket{\bar\psi_a\psi_a}$ in terms of $(ga)^2$ in Figures~\ref{fig:su2_vevs}, \ref{fig:su4_vevs}, \ref{fig:su2su2_vevs}, respectively. In these figures, we also plot the strong-coupling expansions that we derived in Section~\ref{sec:strong_coupling}.  For large $ga$, we find good agreement between the strong-coupling expansions and the numerical results, which provides additional evidence that the strong-coupling vacua that we identified in Section~\ref{sec:strong_coupling} are indeed the adiabatic continuations of the continuum vacua.

We can also extrapolate our numerical results to the continuum limit ($ga\to 0$), where we can compare our results with quantitative predictions made with non-invertible symmetries. In \cite{Cordova:2024nux} and \cite{Komargodski:2020mxz}, the non-invertible symmetry lines of $\SU(2)+\psi_{\bm 5}$ and $\SU(4)+\psi_{\bm{15}}$, respectively,  were studied in detail. Using the algebra of these symmetry lines, the ratios of vacuum expectation values of local operators in those theories were calculated, including those of the operator $\bar\psi_a\psi_a$. In the case of $\SU(2)+\psi_{\bm 5}$, it was found in \cite{Cordova:2024nux} that the chiral condensates in the four vacua should be proportional to 
\begin{equation}\label{eq:su2_ratios}
\begin{split}
	\underline{p = 0}&: \quad 2 + \sqrt{3}\,, \quad -1\,, \\ \underline{p = 1}&: \quad -2 -\sqrt{3}\,, \quad 1\,.
\end{split}
\end{equation}
Similarly, for $\SU(4)+\psi_{\bm{15}}$, it was shown in \cite{Komargodski:2020mxz} that the  chiral condensates should be proportional to
\begin{equation}\label{eq:su4_ratios}
\begin{split}
	\underline{p = 0}&: \quad -1 \pm \sqrt{2}\,, \\ \underline{p = 1}&:\quad \pm 1\,, \\ \underline{p = 2}&:\quad 1 \pm \sqrt{2}\,, \\ \underline{p = 3}&:\quad \pm 1\,.
\end{split}
\end{equation}
In Figures~\ref{fig:su2_vevs} and \ref{fig:su4_vevs}, we use polynomial fits to extrapolate our lattice data to the continuum limit, and we find excellent agreement with \eqref{eq:su2_ratios} and \eqref{eq:su4_ratios} respectively. 
Numerically, our extrapolated values are
\begin{equation}\label{eq:su2_ratios_numerical}
\begin{split}
	\langle \bar\psi_a\psi_a\rangle_{\color{green!50!black} \{\bm{4},\bm{1}\}} = -\langle \bar\psi_a\psi_a\rangle_{\color{green!50!black} \{\bm{1},\bm{4}\}} &\approx 0.97g\,, \\
	\langle \bar\psi_a\psi_a\rangle_{\color{blue} \{\bm{3},\bm{2}\}} = -\langle \bar\psi_a\psi_a\rangle_{\color{blue} \{\bm{2},\bm{3}\}} &\approx 0.26g\,.
\end{split}
\end{equation}
for $\SU(2)+\psi_{\bm 5}$ and
\begin{equation}\label{eq:su4_ratios_numerical}
    \begin{split}
        \langle \bar\psi_a\psi_a\rangle_{\color{blue} \{\bm{15},\bm6\}} = - \langle \bar\psi_a\psi_a\rangle_{\color{blue} \{\bm6,\bm{15}\}} &\approx 0.39g\,,\\
        \langle \bar\psi_a\psi_a\rangle_{\color{red!80!black} \{\bm{20},\bm{\bar{4}}\}} = \langle \bar\psi_a\psi_a\rangle_{\color{red!80!black} \{\bm{\bar{20}},\bm{{4}}\}}= -\langle \bar\psi_a\psi_a\rangle_{\color{red!80!black} \{\bm{\bar{4}},\bm{20}\}} = - \langle \bar\psi_a\psi_a\rangle_{\color{red!80!black} \{\bm{{4}},\bm{\bar{20}}\}} &\approx 0.97g\,,\\
        \langle \bar\psi_a\psi_a\rangle_{\color{green!50!black} \{\bm{64},\bm1\}} = -\langle \bar\psi_a\psi_a\rangle_{\color{green!50!black} \{\bm1,\bm{64}\}}  &\approx 2.39g\,.
    \end{split}
\end{equation}
for $\SU(4)+\psi_{\bm{15}}$. Note especially that the first line follows the pattern
\begin{equation}
    \langle \bar\psi_a\psi_a\rangle_{(\bm{1},\bm{R}(\adj))}/g \approx -0.08(N_c^2 - 1)\sqrt{N_c}
\end{equation}
observed for $\SU(N_c)$ adjoint QCD$_2$ with $N_c = 2,3$ in \cite{Dempsey:2025wia} (and also seen in \cite{Bergner:2024ttq}).

For the two $\SU(2)\times \SU(2)$ theories we consider, the chiral condensate ratios have not yet been calculated in the literature. Our extrapolated values for the $\SU(2)\times\SU(2)+\psi_{(\bm{3},\bm{3})}$ are 
\begin{equation}\label{eq:su22_33_ratios_numerical}
\begin{split}
	\langle \bar\psi_a\psi_a\rangle_{\color{blue} \{(\bm{2},\bm{4})\oplus (\bm{4},\bm{2}), (\bm{1},\bm{1})\}} = -\langle \bar\psi_a\psi_a\rangle_{\color{blue} \{(\bm{1},\bm{1}), (\bm{2},\bm{4})\oplus (\bm{4},\bm{2})\}} &\approx 1.61g\,, \\
     \langle \bar\psi_a\psi_a\rangle_{\color{red!80!black} \{(\bm{1},\bm{2}),(\bm{2},\bm{3})\}} = \langle \bar\psi_a\psi_a\rangle_{\color{red!80!black} \{(\bm{2},\bm{1}),(\bm{3},\bm{2})\}} = -\langle \bar\psi_a\psi_a\rangle_{\color{red!80!black} \{(\bm{3},\bm{2}),(\bm{2},\bm{1})\}} = -\langle \bar\psi_a\psi_a\rangle_{\color{red!80!black} \{(\bm{2},\bm{3}),(\bm{1},\bm{2})\}} &\approx 0.65g\,, \\ 	
    \langle \bar\psi_a\psi_a\rangle_{\color{green!50!black} \{(\bm{1},\bm{3})\oplus (\bm{3},\bm{1}),(\bm{2},\bm{2})\}} = -\langle \bar\psi_a\psi_a\rangle_{\color{green!50!black} \{(\bm{2},\bm{2}),(\bm{1},\bm{3})\oplus(\bm{3},\bm{1})\}} &\approx 0.32g\,,
    \end{split}
\end{equation}
which have a ratio of roughly $5:2:1$. For $\SU(2)\times\SU(2)+\psi_{(\bm{2},\bm{4})}$, the nonzero values are
\begin{equation}\label{eq:su22_24_ratios_numerical}
    \begin{split}
        \langle \bar\psi_a\psi_a\rangle_{\color{red!80!black} \{(\bm{1},\bm{1}),(\bm{3},\bm{1})\}} = -\langle \bar\psi_a\psi_a\rangle_{\color{red!80!black} \{(\bm{3},\bm{1}),(\bm{1},\bm{1})\}}  &\approx 1.41g\,, \\ \langle \bar\psi_a\psi_a\rangle_{\color{green!50!black} \{(\bm{1},\bm{2}),(\bm{2},\bm{3})\oplus(\bm{3},\bm{2})\}} = -\langle \bar\psi_a\psi_a\rangle_{\color{green!50!black} \{(\bm{2},\bm{3})\oplus(\bm{3},\bm{2}),(\bm{1},\bm{2})\}} &\approx 0.71g\,,\\
    \end{split}
\end{equation}
which have a ratio of roughly $2:1$.

\subsection{Excitations and particle degeneracy}\label{sec:particles}
\begin{figure}[t]
	\centering
	\includegraphics[width=.8\linewidth]{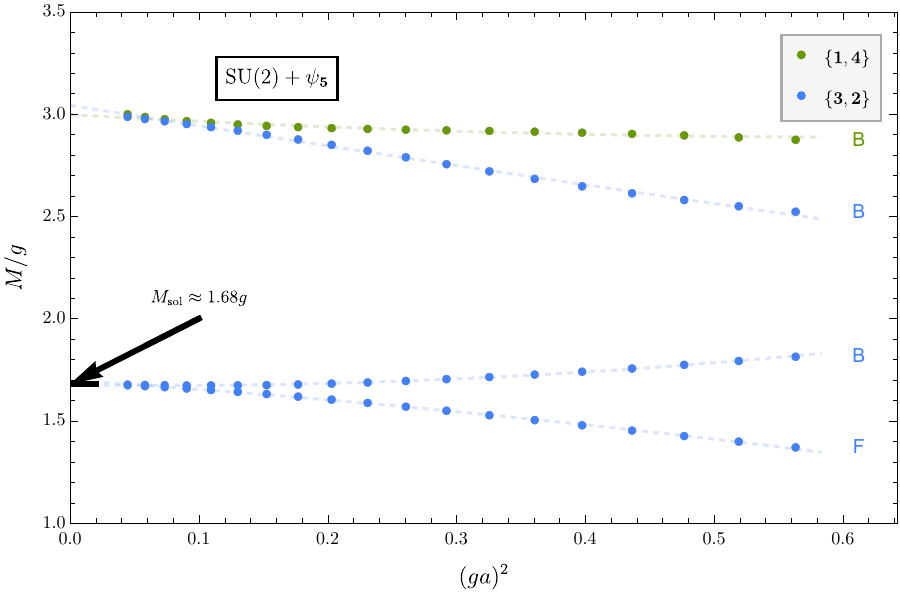}
	\caption{The energies of quasiparticle excitations above the two degenerate vacua of the $\SU(2)+\psi_{\bm{5}}$ theory in the $p = 0$ universe at $m=0$. These have degeneracies in the continuum limit that fall into the patterns predicted in \cite{Cordova:2024nux}. We also mark the soliton mass inferred from the data in \cite{Narayanan:2023jmi} (see \eqref{eq:soliton_mass}). This soliton appears to be degenerate with the lightest fermion and the lightest boson above the {\color{green!50!black} $\{\bm1, \bm 4\}$} vacuum.}
	\label{fig:su2_particles}
\end{figure}

\begin{figure}[t]
    \centering
    \includegraphics[width=.8\linewidth]{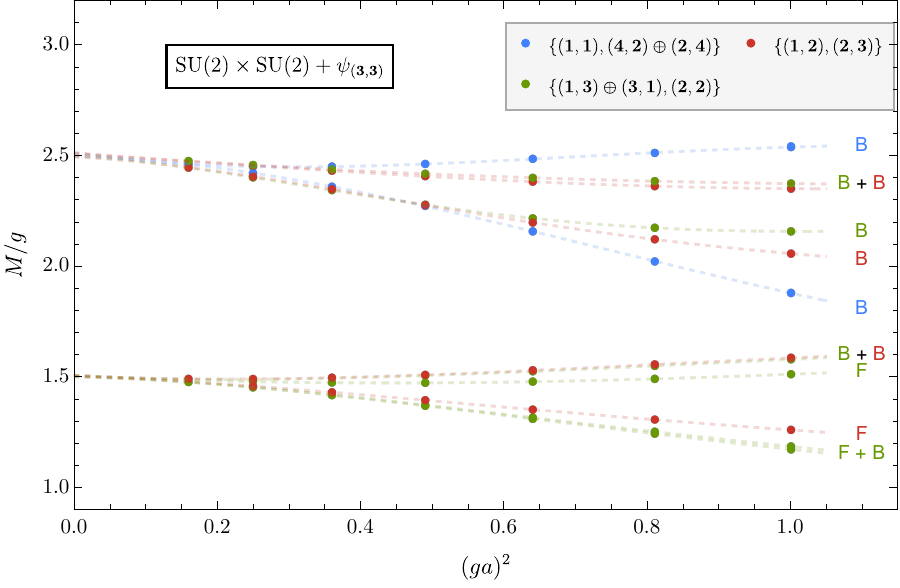}
    \caption{The energies of quasiparticle excitations above three degenerate vacua in the $p=(0,0)$ and $p=(0,1)$ flux tube sectors of the $\SU(2)\times\SU(2)+\psi_{(\bm{3},\bm{3})}$ theory at $m=0$. The dashed lines correspond to polynomial fits extrapolating the data to the continuum limit, intended to guide the eye.}
    \label{fig:su2su233_particles}
\end{figure}

\begin{figure}[t]
    \centering
    \includegraphics[width=.8\linewidth]{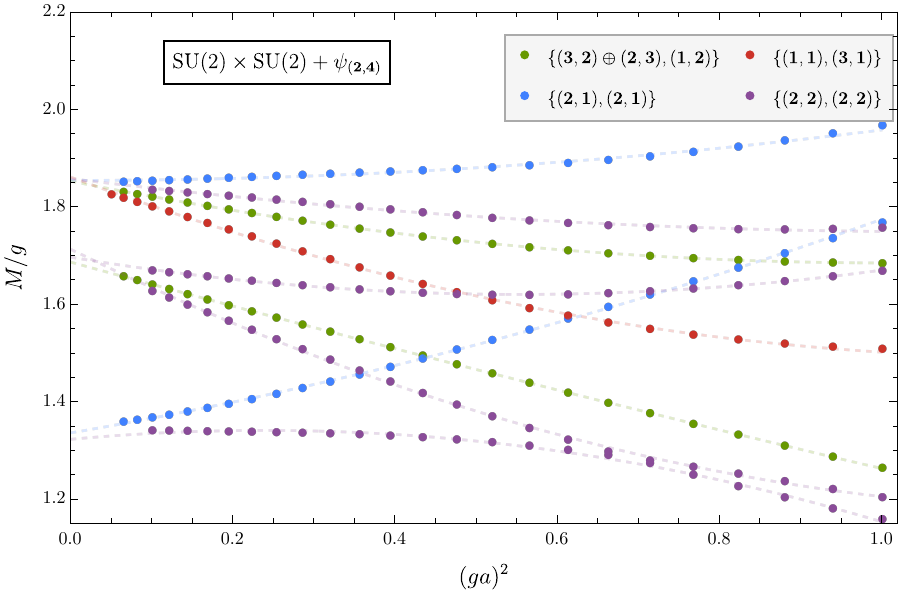}
    \caption{The energies of quasiparticle excitations above four degenerate vacua in the $p=0$ and $p=1$ flux tube sectors of the $\SU(2)\times\SU(2)+\psi_{(\bm{2},\bm{4})}$ theory at $m=0$. The dashed lines correspond to polynomial fits extrapolating the data to the continuum limit, intended to guide the eye. Note that because fermion parity is gauged, there are only bosonic excitations.}
    \label{fig:su2su224_particles}
\end{figure}

In \cite{Cordova:2024vsq,Cordova:2024nux}, it is shown that the non-invertible symmetries present in gapped $(1+1)$D gauge theories lead to nontrivial degeneracy patterns between particle and soliton excitations above the various vacua. We can investigate this on the lattice using the quasi-particle ansatz of~\cite{Haegeman:2013xcv}, adapted to work with LEMPOs. In Figures \ref{fig:su2_particles}, \ref{fig:su2su233_particles}, and \ref{fig:su2su224_particles}, we plot particle spectra for the theories in Table~\ref{tab:theories} (except for $\SU(4)+\psi_{\bm{15}}$, for which it is too numerically taxing).

In each case, we see that as $ga\to 0$, various groups of particle excitations, usually above different vacua, become degenerate.\footnote{Note that there are also degeneracies as a consequence of invertible symmetries, but these are exact on the lattice rather than emerging only in the continuum limit. We have only plotted spectra above vacua not related by invertible symmetries (i.e., one representative from each color in Table~\ref{tab:lattice_vacua}), and so all the emerging degeneracies shown in Figures \ref{fig:su2_particles}, \ref{fig:su2su233_particles}, and \ref{fig:su2su224_particles} are enforced by the non-invertible symmetries.} For instance, Figure~\ref{fig:su2_particles}, shows these degeneracies for the $p = 0$ flux tube sector of the $\SU(2)+\psi_{\bm{5}}$ theory. We see evidence for a degenerate boson-fermion\footnote{Note that here and in the other excitation figures, the vacuum is taken to be bosonic and the fermion parity of states is defined relative to the vacuum.} pair above the {\color{blue} $\{\bm{3},\bm{2}\}$} vacuum at $M\approx 1.68g$ and degenerate bosons (one above the {\color{green!50!black} $\{\bm{1},\bm{4}\}$} vacuum and one above the {\color{blue} $\{\bm{3},\bm{2}\}$} vacuum) at $M\approx 3.02g$. Since the {\color{green!50!black} $\{\bm{4},\bm{1}\}$} vacuum has the same excitation spectrum as the {\color{green!50!black} $\{\bm{1},\bm{4}\}$} vacuum, and likewise for {\color{blue} $\{\bm{2},\bm{3}\}$} and {\color{blue} $\{\bm{3},\bm{2}\}$}, the $p = 0$ and $p = 1$ spectra are identical. Thus, if we include both sectors, each of these multiplets consist of a total of four degenerate particles.

For the $\SU(2)+\psi_{\bm{5}}$ theory, the possible patterns of degeneracies in the $p = 0$ flux tube sector were determined in \cite{Cordova:2024nux} by constructing representations of a strip algebra. Their results imply two nontrivial degeneracy patterns: two degenerate particles above the two distinct vacua, or two degenerate particles above one of the vacua along with a soliton between the two vacua. The first of these patterns is clearly realized by the particles at $M \approx 3.02g$. Our lattice results also give strong evidence for the particles of the second pattern at $M \approx 1.68g$.\footnote{Note also that \cite{Cordova:2024nux} argues that this second pattern must be realized at least once in the spectrum, so it is an important consistency check that we find it.} 

We are presently not able to reliably estimate the mass of solitonic states using our numerical lattice setup. However, the $\SU(2)+\psi_{\bm{5}}$ theory has also been studied using discretized lightcone quantization (DLCQ) in \cite{Narayanan:2023jmi}. Their Figure~2 shows a spectrum of excited states above some Fock vacuum (with their masses plotted as $\pi (M/g)^2$). The lightest state they find is a boson of mass $M/g\approx \sqrt{28.6/\pi} \approx 3.02$; comparing this value with that in our Figure~\ref{fig:su2_particles}, we infer that their Fock vacuum is the {\color{green!50!black} $\{\bm{1},\bm{4}\}$} vacuum. The next feature in their spectrum consists of several trajectories that all appear to tend toward $\pi(M/g)^2 \approx 35.8$. This value does not correspond to a two-particle threshold of their lightest state, so the authors of \cite{Narayanan:2023jmi} interpreted these states as several closely-spaced particles. However, in view of our results and the predictions of \cite{Cordova:2024nux}, there is another possible explanation for this feature in their spectrum: it could be a two-soliton continuum corresponding to the soliton degenerate with the two $M\approx 1.68g$ particles above the {\color{blue} $\{\bm{3},\bm{2}\}$} vacuum. And indeed, if we assume the dense set of trajectories in \cite{Narayanan:2023jmi} corresponds to a two-soliton continuum, we would infer a soliton mass of
\begin{equation}\label{eq:soliton_mass}
    M_\text{sol}/g \approx \frac{1}{2}\sqrt{35.8/\pi} \approx 1.68\,.
\end{equation}
This completes the degeneracy pattern predicted in \cite{Cordova:2024nux}. In the following subsection, we will study the spectrum of the $\SU(2)+\psi_{\bm{5}}$ theory with a nonzero mass for the fermion, which sheds some light on why DLCQ computes the spectrum above the {\color{green!50!black} $\{\bm{1},\bm{4}\}$} vacuum.

We also see intricate patterns of particle degeneracy for the two $\SU(2)\times \SU(2)$ theories we consider. In these cases, we have no way of estimating soliton masses, but we can tabulate the particle degeneracy patterns we observe. For $\SU(2)\times \SU(2)+\psi_{(\bm{3},\bm{3})}$ (see Figure~ \ref{fig:su2su233_particles}), we identify the two lowest-energy multiplets:
\begin{itemize}
    \item At $M \approx 1.5g$, above {\color{red!80!black} $\{(\bm{1},\bm{2}),(\bm{2},\bm{3})\}$} we have two degenerate particles (a boson and a fermion), and above the {\color{green!50!black} $\{(\bm{1},\bm{3})\oplus(\bm{3},\bm{1}),(\bm{2},\bm{2})\}$} we have four degenerate particles (two bosons and two fermions). After accounting for the other vacua related by invertible symmetries according to the color coding in Table~\ref{tab:lattice_vacua}, this gives a total of\footnote{Here and below, the number of vacua related by exact lattice symmetries is colored according to the coloring scheme for the vacua in Table~\ref{tab:lattice_vacua}.} ${\color{red!80!black} 4}\times 2 + {\color{green!50!black} 2}\times 4 = 16$ degenerate particles.
    \item At $M\approx 2.5g$, we have two bosons each above the {\color{red!80!black} $\{(\bm{1},\bm{2}),(\bm{2},\bm{3})\}$}, {\color{blue} $\{(\bm{1},\bm{1}),(\bm{2},\bm{4})\oplus (\bm{4},\bm{2})\}$}, and {\color{green!50!black} $\{(\bm{1},\bm{3})\oplus(\bm{3},\bm{1}),(\bm{2},\bm{2})\}$} vacua. If we account for the other vacua, this becomes ${\color{red!80!black} 4}\times 2 + {\color{blue} 2}\times 2 + {\color{green!50!black} 2}\times 2 = 16$ degenerate particles.
\end{itemize}
For $\SU(2)\times \SU(2)+\psi_{(\bm{2},\bm{4})}$ (see Figure~\ref{fig:su2su224_particles}), we identify the three lowest-energy multiplets (all bosons, since fermion parity is gauged in this theory):
\begin{itemize}
    \item At $M\approx 1.3g$, we have two degenerate particles: one above {\color{blue} $
    \{(\bm{2},\bm{1}),(\bm{2},\bm{1})\}$} and one above {\color{red!50!blue} $\{(\bm{2},\bm{2}),(\bm{2},\bm{2})\}$}.
    \item At $M\approx 1.7g$, above {\color{red!50!blue} $\{(\bm{2},\bm{2}),(\bm{2},\bm{2})\}$} we have two particles, and above {\color{green!50!black} $\{(\bm{3},\bm{2})\oplus (\bm{2},\bm{3}),(\bm{1},\bm{2})\}$} we have one particle. Accounting for the other vacua, this becomes ${\color{red!50!blue} 1}\times 2+{\color{green!50!black} 2}\times 1 = 4$ degenerate particles.
    \item At $M\approx 1.9g$, we have six degenerate particles, one above each of the six vacua in Table~\ref{tab:lattice_vacua}.
\end{itemize}
The multiplets we observe in each theory are summarized in Table~\ref{tab:multiplets}.

For the two $\SU(2)\times\SU(2)$ theories, the strip algebra representations and the corresponding multiplets have not been computed. It would be interesting to do so and compare with our numerical findings. In addition, we do not currently have a reliable way of estimating soliton masses in these theories; it would be interesting to find the solitons numerically, and again to compare these findings with the multiplets predicted using non-invertible symmetries.

\begin{table}[t]
    \centering
    \begin{tabular}{ccrr}
        \toprule
        Theory & Multiplet ($M/g$) & Vacuum & Particles \\
        \midrule
        $\SU(2)+\psi_{\bm{5}}$ & $\approx 1.68$ & {\color{blue} $\{\bm{3},\bm{2}\}$} ($\times$2) & $\text{B}+\text{F}$ \\[0.5em]
        & $\approx 3.0\phantom{8}$ & {\color{blue} $\{\bm{3},\bm{2}\}$} ($\times$2) & B \\
        & & {\color{green!50!black} $\{\bm{1},\bm{4}\}$} ($\times$2) & B \\
        \midrule
        $\SU(2)\times\SU(2) + \psi_{(\bm{3},\bm{3})}$ & $\approx 1.5\phantom{8}$ & {\color{red!80!black} $\{(\bm{1},\bm{2}),(\bm{2},\bm{3})\}$} ($\times$4) & $\text{B}+\text{F}$ \\
        & & {\color{green!50!black} $\{(\bm{1},\bm{3})\oplus(\bm{3},\bm{1}),(\bm{2},\bm{2})\}$} ($\times$2) & $2\text{B}+2\text{F}$ \\[0.5em]
        & $\approx 2.5\phantom{8}$ & {\color{red!80!black} $\{(\bm{1},\bm{2}),(\bm{2},\bm{3})\}$} ($\times$4) & 2B \\
        & & {\color{green!50!black} $\{(\bm{1},\bm{3})\oplus(\bm{3},\bm{1}),(\bm{2},\bm{2})\}$} ($\times$2) & 2B \\
        & & {\color{blue} $\{(\bm{1},\bm{1}), (\bm{2},\bm{4})\oplus(\bm{4},\bm{2})\}$} ($\times$2) & 2B \\
        \midrule
        $\SU(2)\times\SU(2) + \psi_{(\bm{2},\bm{4})}$ & $\approx 1.3\phantom{8}$ & {\color{blue} $\{(\bm{2},\bm{1}),(\bm{2},\bm{1})\}$} & B \\
        & & {\color{red!50!blue} $\{(\bm{2},\bm{2}),(\bm{2},\bm{2})\}$} & B \\[0.5em]
        & $\approx 1.7\phantom{8}$ & {\color{red!50!blue} $\{(\bm{2},\bm{2}),(\bm{2},\bm{2})\}$} & 2B \\
        & & {\color{green!50!black} $\{(\bm{3},\bm{2})\oplus(\bm{2},\bm{3}),(\bm{1},\bm{2})\}$} ($\times$2) & B \\[0.5em]
        & $\approx 1.9\phantom{8}$ & {\color{red!50!blue} $\{(\bm{2},\bm{2}),(\bm{2},\bm{2})\}$} & B \\
        & & {\color{blue} $\{(\bm{2},\bm{1}),(\bm{2},\bm{1})\}$} & B \\
        & & {\color{green!50!black} $\{(\bm{3},\bm{2})\oplus(\bm{2},\bm{3}),(\bm{1},\bm{2})\}$} ($\times$2) & B \\
        & & {\color{red!80!black} $\{(\bm{1},\bm{1}),(\bm{3},\bm{1})\}$} ($\times$2) & B \\
        \bottomrule
    \end{tabular}
    \caption{The particle degeneracy observed in the multiplets shown in Figure~\ref{fig:su2_particles}, \ref{fig:su2su233_particles}, and \ref{fig:su2su224_particles}. Each vacuum plotted in these figures is listed along with the number of vacua in Table~\ref{tab:lattice_vacua} that are related to it by the invertible symmetries described in Appendix~\ref{app:symmetries}; each such vacuum has an identical spectrum.}
    \label{tab:multiplets}
\end{table}

\subsection{Additional results for $\SU(2)+\psi_{\textbf{5}}$ with $m\neq0$}\label{sec:su25_extra}

In this paper we have primarily focused on theories with massless matter, since they have non-invertible symmetries and many degenerate vacua. When we turn on a mass, these symmetries are broken and there is a unique vacuum. To determine which of the degenerate vacua at $m = 0$ becomes the true vacuum at $m>0$, we can use the chiral condensates we computed in Section~\ref{sec:condensate}. The energy densities are modified as $\epsilon_{(\bm{r}_1,\bm{r}_2)} + m\langle \bar\psi_a \psi_a\rangle + \mathcal{O}(m^2)$, so whichever vacuum has the most negative value of $\langle \bar\psi_a \psi_a\rangle$ at $m = 0$ will become the true vacuum at $m>0$.

\begin{figure}
    \centering    
    \includegraphics[width=.6\linewidth]{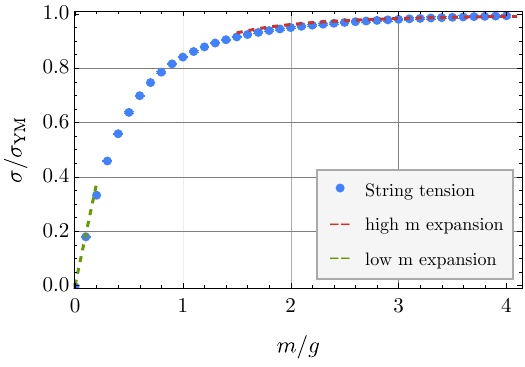}
    \caption{The string tension in the $\SU(2)+\psi_{\bm{5}}$ theory, normalized by its value in pure $\SU(2)$ Yang-Mills theory, as a function of the fermion mass $m$.}
    \label{fig:su2-5-stringtension}
\end{figure}

We can apply this to the $\SU(2)+\psi_{\bm{5}}$ theory using Figure~\ref{fig:su2_vevs}. We see that {\color{green!50!black} $\{\bm{1},\bm{4}\}$}, which is in the $p = 0$ flux tube sector, becomes the true vacuum, and {\color{blue} $\{\bm{2},\bm{3}\}$} becomes the lowest state in the $p = 1$ flux tube sector.  The string tension, $\sigma \equiv \varepsilon_{p=1} - \varepsilon_{p = 0}$, can be estimated at small mass using the chiral condensates:
\begin{equation}\label{eq:sigma_low_mass}
    \sigma = m\left(\langle \bar\psi_a \psi_a\rangle_{\color{blue} \{\bm{2},\bm{3}\}} - \langle \bar\psi_a \psi_a\rangle_{\color{green!50!black} \{\bm{1},\bm{4}\}}\right) \approx 0.71 mg\,,
\end{equation}
where we have used the numerical estimates in \eqref{eq:su25_vev}. At large mass, the leading-order string tension is $\sigma_\text{YM} = \frac{3}{8}g^2$ (from the quadratic Casimir of the fundamental representation), and the one-loop correction to the gauge coupling gives
\begin{equation}\label{eq:sigma_high_mass}
    \sigma = \sigma_\text{YM}\left(1 - \frac{1}{2\pi}\frac{g^2}{m^2} + \ldots\right)
\end{equation}
at large $m/g$ (see \cite{Dempsey:2023gib} for a very similar calculation). In Figure~\ref{fig:su2-5-stringtension}, we plot the string tension estimates from the lattice (after extrapolating $ga\to 0$) as a function of $m/g$, showing good agreement with both \eqref{eq:sigma_low_mass} and \eqref{eq:sigma_high_mass}.

\begin{figure}
    \centering
    \begin{subfigure}[t]{\textwidth}
    \centering
    \includegraphics[width=.9\textwidth]{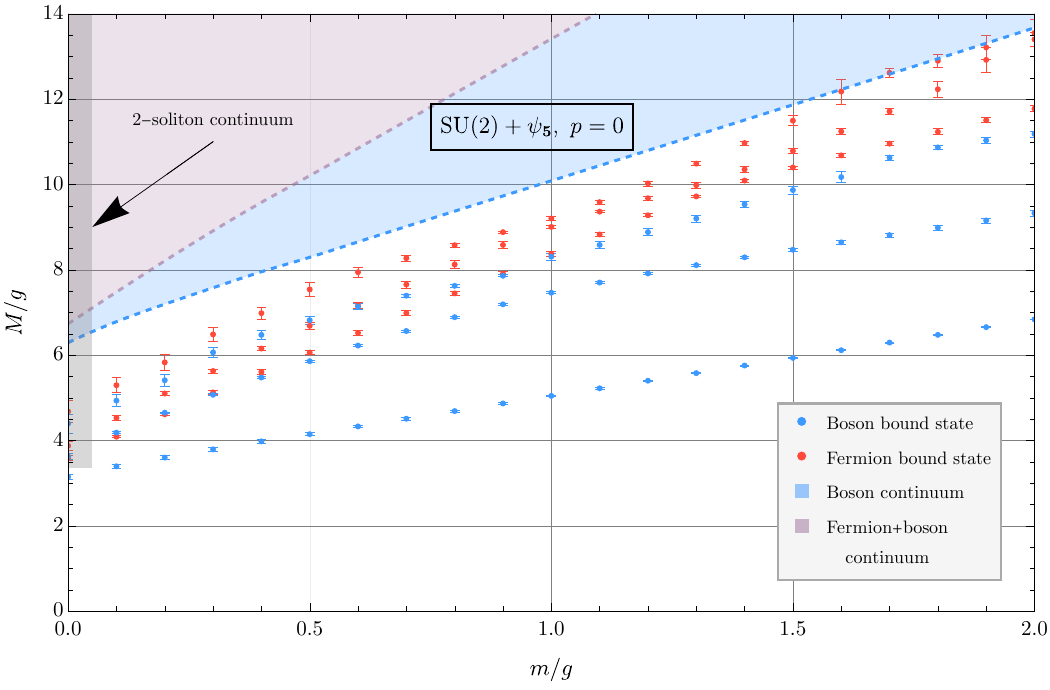}
    \caption{}
    \label{fig:su2-5-p=0-spectrum}
    \end{subfigure}\\[1em]
    \begin{subfigure}[t]{\textwidth}
    \centering
    \includegraphics[width=.9\textwidth]{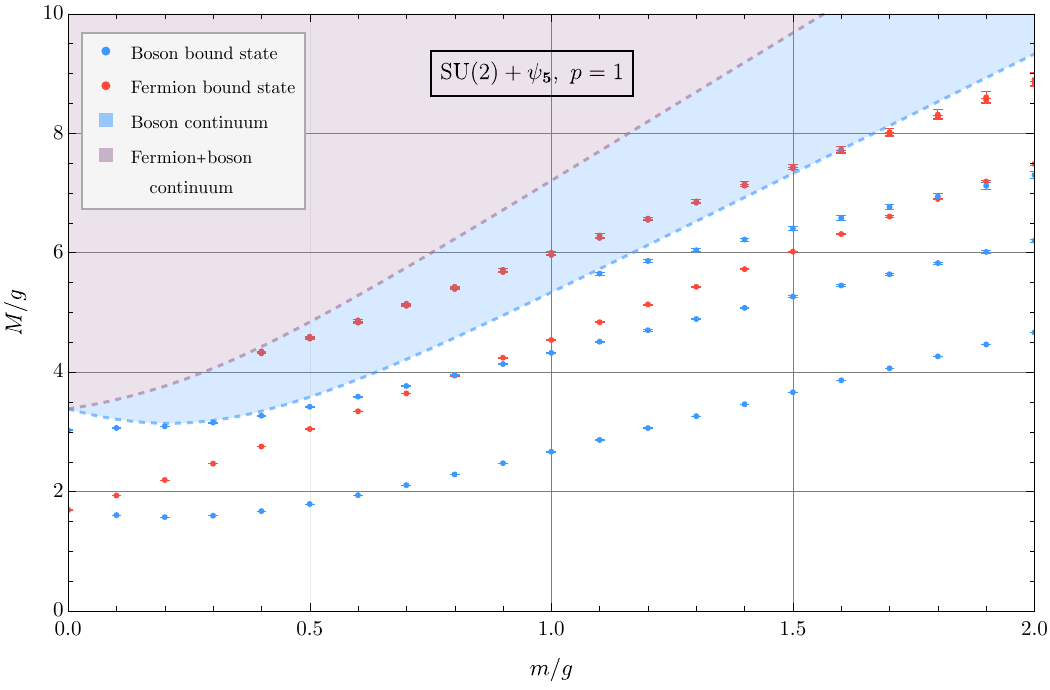}
    \caption{}
    \label{fig:su2-5-p=1-spectrum}
    \end{subfigure}
    \caption{The lowest three bosonic and fermionic excitations in the (a) $p=0$ and (b) $p=1$ flux tube sectors of the $\SU(2)+\psi_{\bm{5}}$ theory for $m\geq0$ as a function of the dimensionless mass.}
\end{figure}

We can also use the quasi-particle ansatz of \cite{Haegeman:2013xcv} to calculate bound state spectra at nonzero mass and see how the spectrum in Figure~\ref{fig:su2_particles} changes at $m>0$. In Figures~\ref{fig:su2-5-p=0-spectrum} and \ref{fig:su2-5-p=1-spectrum}, we plot the bound state spectra in the $p = 0$ and $p = 1$ flux tube sectors, respectively, at $m > 0$.

In the $p = 0$ sector, the lightest boson is well-separated from the rest of the spectrum, but many of the quasi-particle masses above it cannot be extrapolated very cleanly to $ga\to 0$, leading to relatively large error bars in Figure~\ref{fig:su2-5-p=1-spectrum}. At $m = 0$, this happens because (as we discussed in the previous subsection) a two-soliton continuum begins at $M = 2M_\text{sol} \approx 3.34g$; above this mass, the quasi-particle ansatz is no longer an adequate description. For $m > 0$, there is a unique vacuum and hence no longer a finite-energy soliton state, but nevertheless the spectrum must become increasingly dense as $m$ approaches 0, which presents a challenge for the numerical methods. We see signs of the spectrum becoming cleaner and more sparse as we increase $m$.

In the $p = 1$ sector, we see the degenerate boson-fermion pair at $M \approx 1.68g$ splitting as we turn on the mass. The second- and third-lightest fermions, which are stable for $m \gtrsim 0.4g$, appear to be nearly (or possibly even exactly) degenerate; this is surprising and would be interesting to study further.

\section{Discussion}\label{sec:discussion}

In this paper, we have demonstrated an example of the ``unreasonable effectiveness'' of Hamiltonian lattice gauge theory for studying gapped $(1+1)$D gauge theories. Despite the fact that the lattice model we used breaks the non-invertible symmetries of the continuum theory, which are crucial for protecting the vacuum degeneracy in the continuum, we were able to identify states on the lattice that are adiabatically connected to all of the continuum vacua. Moreover, we found that this adiabatic continuity extends into the lattice strong-coupling limit, where we could give a simple characterization of these states. This feature allowed us to determine the vacuum degeneracy of the continuum theory by working in the lattice strong-coupling limit, for which the symmetry protecting this degeneracy is most strongly violated.

Our ability to distinguish the lattice states that are adiabatically connected to the degenerate vacua in the continuum relied crucially upon the use of the uMPS ansatz, which guarantees exact translation-invariance and cluster decomposition. To implement the Hamiltonian of the lattice gauge theory as a tensor network, we used the LEMPO construction recently developed in \cite{Dempsey:2025wia}. For any finite lattice spacing, we used the VUMPS algorithm \cite{Zauner-Stauber:2017eqw} to find a uMPS that approximates the lattice state of interest very well, and then we extrapolated observables computed from these states to the continuum limit. This procedure allowed us to numerically test many qualitative and quantitative predictions made in previous works using non-invertible symmetries \cite{Komargodski:2020mxz,Cordova:2024nux}, and we found excellent agreement with all of these predictions. 

There are several important questions stemming from this work that we leave to future studies. On the analytic side, it is of great interest to better understand the correspondence between the local minima of energy density in the lattice strong-coupling limit, labeled by pairs of Lie algebra representations and described by a lattice decay rule in Section~\ref{sec:lattice_decay}, and the continuum vacua, which are in one-to-one correspondence with affine algebra representations (see Appendix~\ref{app:count_checks}). In particular, we would like to understand exactly what pieces of data of the continuum theory are calculable from the lattice strong-coupling expansion. As the lattice decay rule correctly predicts the number of continuum vacua, it is natural to ask whether other structures related to the non-invertible symmetries, such as the fusion rules of the symmetry lines, can be derived from considering the strong-coupling limit of the lattice model. This question naturally extends beyond the gapped $(1+1)$D gauge theories considered in this paper: the lattice strong-coupling limit may shed light on $(1+1)$D gauge theories with gapless excitations (such as two-flavor adjoint QCD$_2$ \cite{Gopakumar:2012gd,Isachenkov:2014zua,Damia:2024kyt}) or on higher-dimensional theories (see, for instance, \cite{Banks:1976ia}).

It would also be interesting to have a better understanding of how the non-invertible symmetries are restored in the continuum limit of our lattice model, despite prior suggestions that lattice regularizations of adjoint QCD$_2$ could exhibit four-fermion deformations that spoil this property \cite{Cherman:2019hbq}. Our numerical results contain clear evidence that this is not the case for us; in particular, the energy differences in Figure~\ref{fig:energydifferences} serve as order parameters for non-invertible symmetry breaking, and they all vanish in the continuum limit. It is not clear whether the restoration of the non-invertible symmetry should be attributed to details of the renormalization group flow of these four-fermion operators (as studied in \cite{Cherman:2024onj} in the case of adjoint QCD$_2$), or whether it should be expected more generally. A more precise understanding of the challenges, and their resolution, is in order.

On the numerical side, it would be interesting to have a better handle on the particle-soliton degeneracy predicted in \cite{Cordova:2024nux}. Although we observe clear evidence of this degeneracy in the $\SU(2)+\psi_{\bm{5}}$ theory, so far we can only obtain the mass of the lightest soliton indirectly by looking for the two-soliton continuum, either on our lattice or from results of discretized light-cone quantization \cite{Narayanan:2023jmi}. In principle, we should be able to identify these states directly on the lattice using a solitonic version of the quasiparticle ansatz \cite{Haegeman:2013xcv}. In addition, our study of the spectrum of the $\SU(2)+\psi_{\bm{5}}$ theory reveals an intriguing near-exact degeneracy of two fermions, even at finite mass (see Figure~\ref{fig:su2-5-p=1-spectrum}). This is a curious puzzle worthy of further study.

Relatedly, the continuum in the DLCQ spectrum of \cite{Narayanan:2023jmi} and its explanation as a two-soliton continuum is reminiscent of similar puzzles in other DLCQ studies. In particular, \cite{Gross:1997mx,Dempsey:2021xpf,Dempsey:2022uie} identified continua in DLCQ spectra for adjoint QCD$_2$ that appear at the two-particle mass threshold for the lightest fermion, and yet are present in symmetry sectors that do not correspond to two-particle states of this fermion. It would be interesting to investigate whether these continua can be explained in terms of two-soliton states comprised of solitons that are degenerate with the lightest fermion.

\section*{Acknowledgments} 
We thank Aleksey Cherman, Clay Cordova, Liam Fitzpatrick, Diego Garc\'ia-Sep\'ulveda, Nicholas Holfester, Ami Katz, Sahand Seifnashri, and Shu-Heng Shao for useful discussions, and Igor Klebanov for useful discussions and collaboration on related work. All numerical calculations have been carried out using the Julia packages \texttt{MPSKit.jl} \cite{MPSKit2025} and \texttt{TensorKit.jl} \cite{Devos:2025yoj}. The work of BTS, AMG, and SSP is supported in part by the U.S.~Department~of~Energy under Award No.~DE-SC0007968. The work of RD is supported by a Pappalardo Fellowship in Physics at MIT.

\appendix

\section{Internal invertible symmetries}\label{app:symmetries}

In this appendix, we give the internal invertible symmetries of the continuum theory \eqref{eq:contiuum_action} with $m = 0$, and the mixed 't Hooft anomalies of these symmetries, for each of the cases in Table~\ref{tab:theories}.

The symmetries are
\begin{equation}\label{eq:symmetries}
\begin{aligned}
    \SU(2)+\psi_{\bm 5}:&\qquad\mathbb Z^{[1]}_2\times(\mathbb Z_2)_F\times(\mathbb Z_2)_\chi\,,\\
    \SU(2)\times\SU(2)+\psi_{(\bm 3,\bm 3)}:&\qquad [(\mathbb Z_2\times \mathbb Z_2)^{[1]} \rtimes(\mathbb Z_2)_P] \times(\mathbb Z_2)_F\times(\mathbb Z_2)_\chi\,,\\
    \SU(2)\times\SU(2)+\psi_{(\bm 2,\bm 4)}:&\qquad \mathbb Z^{[1]}_2\times(\mathbb Z_2)_\chi\,,\\
    \SU(4)+\psi_{\bm{15}}:&\qquad [ \mathbb Z_4^{[1]} \rtimes(\mathbb Z_2)_C] \times(\mathbb Z_2)_F\times(\mathbb Z_2)_\chi\,.
\end{aligned}
\end{equation}
Here, $(\mathbb Z_2)_F$ and $(\mathbb Z_2)_\chi$ are fermion parity and chiral symmmetry, generated by $\psi\to -\psi$ and $\psi\to \gamma^5 \psi$, respectively. Note that $\SU(2)\times\SU(2)+\psi_{(\bm 2,\bm 4)}$ does not have a fermion parity symmetry, because the element $(I,-I)$ of the center of $\SU(2) \times \SU(2)$ acts as $\psi \to -\psi$, and so fermion parity is gauged. The one-form symmetries for these theories were given also in Table~\ref{tab:vacua}; for the $\SU(4)+\psi_{\bm{15}}$ theory, the one-form symmetry is acted upon by the $(\Z_2)_C$ charge conjugation symmetry, and in the $\SU(2)\times \SU(2) + \psi_{(\bm{3},\bm{3})}$ theory, it is acted upon by the $(\mathbb{Z}_2)_P$ permutation symmetry that exchanges the two $\SU(2)$ factors. In all cases, if a mass term is added, the chiral symmetry is explicitly broken and the factor $(\mathbb Z_2)_\chi$ should be dropped from the symmetry group.

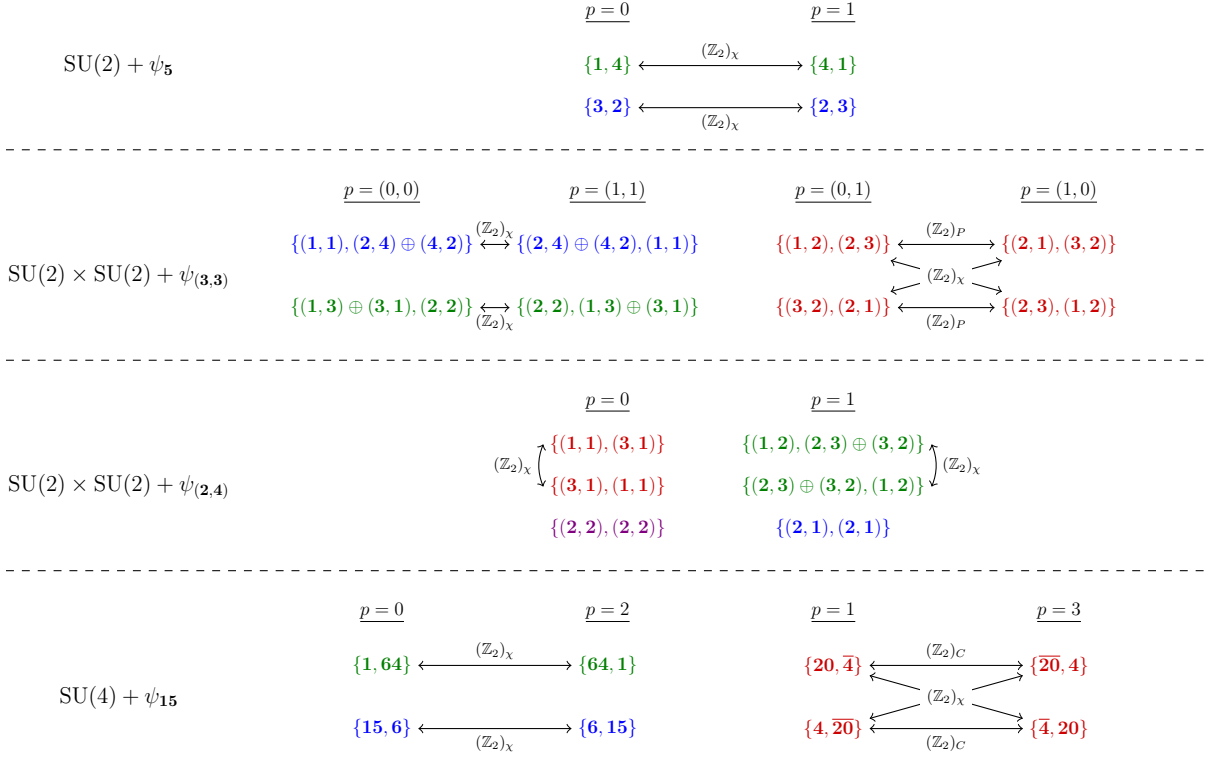
\begin{figure}[t]
    \centering
    \begin{tikzpicture}[every node/.style={scale=.6},yscale=1.4]
        \node at (-2.5,3) {\large $\SU(2)+\psi_{\bm{5}}$};
        \node at (-2.5,1) {\large $\SU(2)\times \SU(2)+\psi_{(\bm{3},\bm{3})}$};
        \node at (-2.5,-1) {\large $\SU(2)\times \SU(2)+\psi_{(\bm{2},\bm{4})}$};
        \node at (-2.5,-3) {\large $\SU(4)+\psi_{\bm{15}}$};

        \begin{scope}[yshift=.3cm]
        \node at (4,3.2) {$\underline{p = 0}$};
        \node at (7,3.2) {$\underline{p = 1}$};

        \node[green!50!black] at (4,2.7) (a1) {$\{\bm{1},\bm{4}\}$};
        \node[blue] at (4,2.3) (a2) {$\{\bm{3},\bm{2}\}$};
        \node[blue] at (7,2.3) (a3) {$\{\bm{2},\bm{3}\}$};
        \node[green!50!black] at (7,2.7) (a4) {$\{\bm{4},\bm{1}\}$};

        \draw[<->] (a1) -- node[above] {\footnotesize $(\mathbb{Z}_2)_\chi$} (a4);
        \draw[<->] (a2) -- node[below] {\footnotesize $(\mathbb{Z}_2)_\chi$} (a3);

        \end{scope}
        \draw[dashed] (-4,2.2) -- (12,2.2);

        \node at (1,1.8) {$\underline{p = (0,0)}$};
        \node at (7,1.8) {$\underline{p = (0,1)}$};
        \node at (10,1.8) {$\underline{p = (1,0)}$};
        \node at (4,1.8) {$\underline{p = (1,1)}$};

        \begin{scope}[yshift=-0cm]
        \node[blue] at (1,1.3) (b1) {$\{(\bm{1},\bm{1}),(\bm{2},\bm{4})\oplus(\bm{4},\bm{2})\}$};
        \node[green!50!black] at (1,.7) (b2) {$\{(\bm{1},\bm{3})\oplus(\bm{3},\bm{1}),(\bm{2},\bm{2})\}$};
        \node[red!80!black] at (7,1.3) (b3) {$\{(\bm{1},\bm{2}),(\bm{2},\bm{3})\}$};
        \node[red!80!black] at (7,.7) (b4) {$\{(\bm{3},\bm{2}),(\bm{2},\bm{1})\}$};
        \node[red!80!black] at (10,1.3) (b5) {$\{(\bm{2},\bm{1}),(\bm{3},\bm{2})\}$};
        \node[red!80!black] at (10,.7) (b6) {$\{(\bm{2},\bm{3}),(\bm{1},\bm{2})\}$};
        \node[blue] at (4,1.3) (b7) {$\{(\bm{2},\bm{4})\oplus(\bm{4},\bm{2}),(\bm{1},\bm{1})\}$};
        \node[green!50!black] at (4,.7) (b8) {$\{(\bm{2},\bm{2}),(\bm{1},\bm{3})\oplus(\bm{3},\bm{1})\}$};

        \draw[<->] (b1) to node[above] {\footnotesize $(\mathbb{Z}_2)_\chi$} (b7);
        \draw[<->] (b2) to node[below] {\footnotesize $(\mathbb{Z}_2)_\chi$} (b8);
        \draw[<->] (b3) -- node[above] {\footnotesize $(\mathbb{Z}_2)_P$} (b5);
        \draw[<->] (b4) -- node[below] {\footnotesize $(\mathbb{Z}_2)_P$} (b6);
        \draw[<->] (b3) -- (b6);
        \draw[<->] (b4) -- (b5);
        \node[fill=white,fill opacity=1] at (8.5,1) {\footnotesize $(\mathbb{Z}_2)_\chi$};
        
        \draw[dashed] (-4,.2) -- (12,.2);
        
        \node at (4,-.2) {$\underline{p = 0}$};
        \node at (7,-.2) {$\underline{p = 1}$};

        \node[red!80!black] at (4,-.6) (c1) {$\{(\bm{1},\bm{1}),(\bm{3},\bm{1})\}$};
        \node[red!80!black] at (4,-1) (c2) {$\{(\bm{3},\bm{1}),(\bm{1},\bm{1})\}$};
        \node[red!50!blue] at (4,-1.4) (c3) {$\{(\bm{2},\bm{2}),(\bm{2},\bm{2})\}$};
        \node[green!50!black] at (7,-.6) (c4) {$\{(\bm{1},\bm{2}),(\bm{2},\bm{3})\oplus(\bm{3},\bm{2})\}$};
        \node[green!50!black] at (7,-1) (c5) {$\{(\bm{2},\bm{3})\oplus(\bm{3},\bm{2}),(\bm{1},\bm{2})\}$};
        \node[blue] at (7,-1.4) (c6) {$\{(\bm{2},\bm{1}),(\bm{2},\bm{1})\}$};

        \draw[<->] (c1.west) to[bend right=30] node[left] {\footnotesize $(\mathbb{Z}_2)_\chi$} (c2.west);
        \draw[<->] (c4.east) to[bend left=30] node[right] {\footnotesize $(\mathbb{Z}_2)_\chi$} (c5.east);

        \draw[dashed] (-4,-1.8) -- (12,-1.8);

        \node at (1,-2.2) {$\underline{p = 0}$};
        \node at (4,-2.2) {$\underline{p = 2}$};
        \node at (7,-2.2) {$\underline{p = 1}$};
        \node at (10,-2.2) {$\underline{p = 3}$};

        \node[green!50!black] at (1,-2.7) (d1) {$\{\bm{1},\bm{64}\}$};
        \node[blue] at (1,-3.3) (d2) {$\{\bm{15},\bm{6}\}$};
        \node[red!80!black] at (7,-2.7) (d3) {$\{\bm{20},\bm{\overline{4}}\}$};
        \node[red!80!black] at (7,-3.3) (d4) {$\{\bm{4},\bm{\overline{20}}\}$};
        \node[green!50!black] at (4,-2.7) (d5) {$\{\bm{64},\bm{1}\}$};
        \node[blue] at (4,-3.3) (d6) {$\{\bm{6},\bm{15}\}$};
        \node[red!80!black] at (10,-2.7) (d7) {$\{\bm{\overline{20}},\bm{4}\}$};
        \node[red!80!black] at (10,-3.3) (d8) {$\{\bm{\overline{4}},\bm{20}\}$};

        \draw[<->] (d1) -- node[above] {\footnotesize $(\mathbb{Z}_2)_\chi$} (d5);
        \draw[<->] (d2) -- node[below] {\footnotesize $(\mathbb{Z}_2)_\chi$} (d6);
        \draw[<->] (d3) -- node[above] {\footnotesize $(\mathbb{Z}_2)_C$} (d7);
        \draw[<->] (d4) -- node[below] {\footnotesize $(\mathbb{Z}_2)_C$} (d8);
        \draw[<->] (d3) -- (d8);
        \draw[<->] (d4) -- (d7);
        \node[fill=white] at (8.5,-3) {\footnotesize $(\mathbb{Z}_2)_\chi$};
        \end{scope}
    \end{tikzpicture}
    \caption{The action of symmetry generators on the lattice vacua in Table~\ref{tab:lattice_vacua}, showing how the degeneracies indicated by the color coding in that table are derived. }
    \label{fig:symmetries_on_vacua}
\end{figure}

The internal symmetries may have mixed anomalies. In particular, it was explained in \cite{Cherman:2019hbq} (in the case of adjoint QCD$_2$) that the mixed anomaly between the $(\mathbb{Z}_2)_\chi$ and the one-form center symmetry can be described a mod 2 index, which specifies how the fermion measure changes sign under a chiral transformation in the presence of a background two-form gauge field for the one-form symmetry. 

To compute this mod 2 index, it is easiest to work on the lattice, where a nontrivial index manifests as the lattice chiral symmetry generator anticommuting (rather than commuting) with a lattice one-form symmetry generator. In \cite{Dempsey:2024alw}, it was shown (again in the context of adjoint QCD$_2$) that this can occur whenever $\bm{R}(\bm{\lambda})$ acts non-trivially on $Z_{\bm{\lambda}}(G)$. It is straightforward to check this criterion using the data in Table~\ref{tab:lattice}. Denoting the chiral symmetry generator by $\mathcal V$ and the one-form symmetry generator by\footnote{For $\SU(2)\times\SU(2)+\psi_{(\bm 3,\bm 3)}$, we denote the two generators of $\Z_2 \times \Z_2$ by $\mathcal{U}_1$ and $\mathcal{U}_2$} $\mathcal U(x)$, we find
\begin{equation}\label{eq:anomalies}
\begin{aligned}
    \SU(2)+\psi_{\bm 5}:&\qquad \mathcal U(x)\mathcal V=-\mathcal V\mathcal U(x) \,,\\
    \SU(2)\times\SU(2)+\psi_{(\bm 3,\bm 3)}:&\qquad \mathcal U_i(x)\mathcal V=-\mathcal V\mathcal U_i(x)\,,\\
    \SU(2)\times\SU(2)+\psi_{(\bm 2,\bm 4)}:&\qquad \mathcal U(x)\mathcal V=+\mathcal V\mathcal U(x)\,,\\
    \SU(4)+\psi_{\bm{15}}:&\qquad \mathcal U(x)\mathcal V=-\mathcal V\mathcal U(x)\,.
\end{aligned}
\end{equation}
That is, each theory we consider except for $\SU(2)\times \SU(2)+\psi_{(\bm{2},\bm{4})}$ has a mixed chiral-center anomaly.

The symmetries of \eqref{eq:symmetries} imply many exact degeneracies among the lattice vacua we study in this paper, as we indicated with the color coding in Table~\ref{tab:lattice_vacua}. In Figure~\ref{fig:symmetries_on_vacua}, we show explicitly which symmetries imply which degeneracies. Note that $(\mathbb{Z}_2)_P$ and $(\mathbb{Z}_2)_C$ can relate states in different flux tube sectors because of the semidirect product structure in \eqref{eq:symmetries}, while for $(\mathbb{Z}_2)_\chi$ this occurs because of the anomalies in \eqref{eq:anomalies}. Furthermore, as the chiral condensate $\braket{\bar\psi_a\psi_a}$ is odd under $(\mathbb{Z}_2)_\chi$ (which is realized on the lattice as one-site translation), the chiral condensates of vacua with the same color coding can also be inferred from one another.

\section{Irrep decomposition of the fermionic Hilbert space}\label{app:matter_rep}

For a lattice with $N$ sites and fermions transforming in the representation $\bm{\lambda}$ of the gauge group, the Hilbert space of fermions transforms in the spinor representation of $\so(N\cdot \dim\bm{\lambda})$. In this section, we will show how to determine the decomposition of this representation under the embedding of the $N$ copies of the gauge algebra into $\so(N\cdot \dim\bm{\lambda})$.

In Appendix A of \cite{Dempsey:2023fvm}, it is shown that under the embedding $\so(\dim\bm{\lambda})^N \hookrightarrow \so(N\cdot \dim\bm{\lambda})$, the spinor representation branches into the $N$ copies of the spinor of $\so(\dim \bm{\lambda})$, times a factor of $2^{N/2}$ if $\dim \bm{\lambda}$ is odd. Using this fact, it suffices to determine how the spinor of $\so(\dim \bm{\lambda})$ branches under the embedding of the gauge algebra into $\so(\dim \bm{\lambda})$.

The embedding is fixed by the requirement that the vector of $\so(\dim \bm{\lambda})$ branches into $\bm{\lambda}$. In a basis where the weights of the vector representation are
\begin{equation}
	\big(\underbrace{0,0,\ldots,\pm 1,\ldots,0,0}_{\lfloor\dim\bm{\lambda}/2\rfloor}\big)\,,
\end{equation}
and additionally a zero weight if $\dim\bm{\lambda}$ is odd, this means we can take the projection matrix for the embedding to be
\begin{equation}\label{eq:rep_projection}
	P = \begin{pmatrix} \vec{\omega}_1 & \vec{\omega}_2 & \cdots & \vec{\omega}_{n_p} & 0_{(\dim \bm{\lambda})\times \lfloor n_0(\bm{\lambda})/2\rfloor} \end{pmatrix}\,.
\end{equation}
Here $\vec{\omega}_1,\ldots,\vec{\omega}_{n_p}$ are half of the nonzero weights of $\bm{\lambda}$, chosen so that $\vec{\omega}_i \neq -\vec{\omega}_j$. The number of such weights is $n_p = (\dim\bm{\lambda} - n_0(\bm{\lambda}))/2$, where $n_0(\bm{\lambda})$ is the number of zero weights contained in $\bm{\lambda}$.

\begingroup
\renewcommand{\arraystretch}{1.3}
\begin{table}[t]
	\centering
	\begin{tabular}{crcr}
	\toprule
	Gauge algebra & Real $\bm{\lambda}$ & $n_0(\bm{\lambda})$ & $\bm{R}(\bm{\lambda})$ \\
	\midrule
	$\mathfrak{g}$ & adjoint & $\rk \mathfrak{g}$ & $[11\cdots 1]$ \\
	\arrayrulecolor{gray}\hline
	$\su(2)$ & $\bm{5}$ & 1 & $\bm{4}$ \\
	$\su(2)$ & $\bm{7}$ & 1 & $\bm{1}\oplus\bm{7}$ \\
	$\su(4)$ & $\bm{6} = [010]$ & 0 & $\bm{4}\oplus\bm{\overline{4}} = [100]\oplus[001]$ \\
	$\su(4)$ & $\bm{20'} = [020]$ & 2 & $\bm{256}\oplus\bm{\overline{256}} = [113]\oplus[311]$ \\
	\hline
	$\so(2n+1)$ & $\text{vector} = [10\cdots 0]$ & 1 & $\text{spinor} = [0\cdots 01]$ \\
	$\so(2n+1)$ & $[20\cdots 0]$ & $n$ & $[1\cdots 13]$ \\
	$\so(9)$ & $\bm{16} = [0001]$ & 0 & $\bm{44}\oplus\bm{84}\oplus\bm{128} = [2000]\oplus [0010]\oplus[1001]$ \\
	\hline
	$\so(2n)$ & $\text{vector} = [10\cdots 0]$ & 0 & $\text{spinor} = [0\cdots 001]\oplus [0\cdots 010]$ \\	
	$\so(2n)$ & $[20\cdots 0]$ & $n-1$ & $[1\cdots 113]\oplus [1\cdots 131]$ \\	
	\hline
	$\sp(2n)$ & $[010\cdots 0]$ & $n-1$ & $[1\cdots 10]$ \\
    $\sp(8)$ & $[0001]$ & $2$ & $[1130]\oplus [3111]\oplus [5110]$ \\
	\hline
	$\mathfrak{g}_2$ & $\bm{7} = [10]$ & 1 & $\bm{1}\oplus\bm{7} = [00]\oplus[10]$ \\
	\hline
	$\mathfrak{f}_4$ & $\bm{26} = [0001]$ & 2 & $\bm{4096}=[0011]$ \\
	\arrayrulecolor{black}\bottomrule
	\end{tabular}
	\caption{For a gauge algebra and a real representation $\bm{\lambda}$, we give the number of zero weights $n_0(\bm{\lambda})$ and the representation $\bm{R}(\bm{\lambda})$. The dimensions satisfy $2^{\dim \bm{\lambda}} = 2^{n_0(\bm{\lambda})} \left(\dim \bm{R}(\bm{\lambda})\right)^2$.}
	\label{tab:repR}
\end{table}
\endgroup

In the same basis, the weights of the spinor representation of $\so(\dim\bm{\lambda})$ are
\begin{equation}
	\Big(\underbrace{\pm\frac{1}{2},\pm\frac{1}{2},\ldots,\pm\frac{1}{2}}_{\lfloor\dim\bm{\lambda}/2\rfloor}\Big)\,.
\end{equation}
By acting with \eqref{eq:rep_projection} on these weights, we obtain $2^{\lfloor n_0(\bm{\lambda})/2\rfloor}$ times the character of a new representation of the gauge algebra, $\bm{R}(\bm{\lambda})$. 

We can write the relationship between $\bm{\lambda}$ and $\bm{R}(\bm{\lambda})$ compactly as
\begin{equation}\label{eq:Rlam_character}
	\prod_{w\in \bm{\lambda}} \left(e^{w/2} + e^{-w/2}\right) = 2^{n_0(\bm{\lambda})} \left(\chi\left\lbrack\bm{R}(\bm{\lambda})\right\rbrack\right) ^2\,,
\end{equation}
where $\chi\left\lbrack\bm{R}(\bm{\lambda})\right\rbrack = \sum_{w\in \bm{R}(\bm{\lambda})} e^w$. When $\bm{\lambda} = \adj$, $\bm{R}(\bm{\lambda}) = [11\cdots 1]$, as shown in \cite{Dempsey:2023fvm}; for other $\bm{\lambda}$, we have to work out $\bm{R}(\bm{\lambda})$ case-by-case. For the theories we study in this paper, the values of $\bm{R}(\bm{\lambda})$ are listed in Table~\ref{tab:lattice}. Some other examples are listed in Table~\ref{tab:repR}.

\section{Details of the lattice model}\label{app:lattice_detail}
In this appendix, we will explain how to represent the fermion operators $\chi_n^a$ on the fermionic Hilbert space described in Section~\ref{subsec:lattice_hamiltonian}. In particular, we will express these operators in terms of Clebsch-Gordan coefficients, which makes gauge covariance manifest and is required for numerical simulations using LEMPOs. 

For each theory in Table~\ref{tab:theories}, we will start from the most general expression for an operator $\chi^a_n$ that acts on the fermionic degrees of freedom on the lattice which is highly constrained by being a invariant tensor of the gauge group. The fermionic Hilbert space is organized into on-site copies of the representation $\bm{R}(\bm{\lambda})$, as well as multiplicity degrees of freedom in cases for which $n_0(\bm{\lambda})>0$. The details of how we write a general ansatz for $\chi^a_n$ depend on $\bm{R}(\bm{\lambda})$ (in particular, whether it is reducible) and on $n_0(\bm{\lambda})$, as we will explain below. In each case, we fix any undetermined parameters using the Clifford algebra
\begin{equation}\label{eq:clifford}
    \left\lbrace \chi^a_m, \chi^b_n\right\rbrace = \delta_{mn} h^{ab} \mathbbm{I}\,,
\end{equation}
where $h^{ab}$ is a metric\footnote{In the main text, we work in a basis for which $h^{ab} = \delta^{ab}$. Here, we will use the standard Cartan basis for $\SU(2)$ irreps, for which $h^{ab}$ will be a non-trivial metric in each example, as given below. One can then perform a change of basis to recover the $\chi^a_n$ used in the main text.} on the $\bm{\lambda}$-indices, and $\mathbbm{I}$ on the right-hand side acts on the on-site space.

For $\SU(2)$ representations, we will denote Clebsch-Gordan coefficients by
\begin{equation}\label{eq:CGC}
    (C^{\mu_1\mu_2\mu_3}_{m_1})_{m_2m_3} \equiv C^{\mu_1\mu_2\mu_3}_{m_1m_2 m_3}\,,
\end{equation}
where $C^{\mu_1\mu_2\mu_3}_{m_1m_2 m_3}$ are the standard $\SU(2)$ Clebsch-Gordan coefficients defined by
\begin{equation}
\ket{\mu_3,m_3}=\sum_{m_1,m_2}C^{\mu_1\mu_2\mu_3}_{m_1m_2 m_3} \ket{\mu_1,m_1}\ket{\mu_2,m_2}\,.
\end{equation}
The notation of \eqref{eq:CGC} is more suggestive of changing the basis of a wavefunction $\ket{\psi} = \sum \alpha_{m_3}\ket{\mu_3,m_3}$:
\begin{equation}
    \big(\bra{\mu_2,m_2}\bra{\mu_1,m_1}\big)\ket{\psi} = \left(C^{\mu_1 \mu_2 \mu_3}_{m_1}\right)_{m_2 m_3} \alpha_{m_3}\,.
\end{equation}

\subsection{\texorpdfstring{$\SU(2)+\psi_{\bm 5}$}{SU(2) + 5}}\label{app:su2_5}
For $G=\SU(2)$ and $\lambda =\bm 5$, the on-site representation and zero-weight counts are
\begin{equation}\label{eq:Rsu25}
    \bm{R}(\bm 5) = \bm 4 \,, \qquad n_0(\bm{5}) =1 \,.
\end{equation}
Following \cite{Dempsey:2024alw}, we can naturally encode the multiplicity using a single uncharged Majorana fermion $\lambda_n$ on each site; we will take these to satisfy $\{\lambda_m, \lambda_n\} = 2\delta_{mn}$. The on-site degrees of freedom are then a copy of $\bm{4}$ along with this Majorana fermion, and the most general expression for $\chi^a_n$ is
\begin{equation}
    \chi^a_n = C_{a;n}\lambda_n\,,
\end{equation}
where $C_{a;n}$ is a Clebsch-Gordan symbol for $\bm{5}\times\bm{4} \to \bm{4}$, acting on the copy of $\bm{4}$ at site $n$ (and we have only made the $\bm{5}$ index explicit). There is only one such Clebsch-Gordan symbol up to overall normalization, and so it is this normalization that we fix by imposing the Clifford algebra on $\chi^a_n$. We find (using the notation \eqref{eq:CGC})
\begin{equation}
    \chi^a_n = \sqrt{\frac{5}{2}} C^{\bm{544}}_{a} \lambda_n\,,
\end{equation}
where (here and throughout) the Clebsch-Gordan symbol acts on the Hilbert space at site $n$. Note that if we use the standard Cartan basis for $\SU(2)$, \eqref{eq:clifford} is satisfied with $h^{ab} = (-1)^a \delta^{a,-b}$.

Note also that, much like the theories discussed in \cite{Dempsey:2024alw}, the fermion parity operator can be expressed solely in terms of $\lambda_n$:
\begin{equation}\label{eq:fp}
    \mathcal{F} = \prod_n (-1)^{1/4} \lambda_n\,.
\end{equation}
The phase in this expression is fixed by requiring that $\mathcal{F}^2 = 1$.

\subsection{\texorpdfstring{$\SU(2)\times\SU(2)+\psi_{(\bm 3,\bm 3)}$}{SU(2) x SU(2) + (3,3)}}
For $G = \SU(2)\times\SU(2)$ and $\bm\lambda =(\bm 3,\bm 3)$ representation, the on-site representation and the zero weight count are given by
\begin{equation}
    R((\bm3,\bm3))= (\bm4,\bm2)\oplus (\bm2,\bm4)\,,\qquad n_0((\bm3,\bm3)) =1 \,.
\end{equation}
Since $n_0 = 1$, the multiplicity degrees of freedom can be encoded using a single Majorana fermion on each site, like in Appendix~\ref{app:su2_5}. Thus, in total, on each site we have a copy of $(\bm4,\bm2)\oplus (\bm2,\bm4)$ along with this Majorana fermion. The most general expression for $\chi^{a_1,a_2}_n$ (where $a_1$ and $a_2$ are both $\bm{3}$-indices) consists of a Clebsch-Gordan symbol for $(\bm{3},\bm{3}) \otimes ((\bm4,\bm2)\oplus (\bm2,\bm4)) \to (\bm4,\bm2)\oplus (\bm2,\bm4)$ multiplied by $\lambda_n$. There are four arbitrary coefficients in such a Clebsch-Gordan symbol, corresponding to the four different fusion channels; we fix these coefficients by imposing \eqref{eq:clifford}. The result is
\begin{equation}
    \chi^{a_1,a_2}_n = \begin{pmatrix}
        \sqrt{\frac 52}C^{\bm 3\bm 4\bm4}_{a_1}\otimes C^{\bm 3\bm 2\bm2}_{a_2} & \sqrt{2}C^{\bm 3\bm 4\bm2}_{a_1}\otimes C^{\bm 3\bm 2\bm4}_{a_2}\\
        \sqrt{2}C^{\bm 3\bm 2\bm4}_{a_1}\otimes C^{\bm 3\bm 4\bm2}_{a_2} & -\sqrt{\frac 52}C^{\bm 3\bm 2\bm2}_{a_1}\otimes C^{\bm 3\bm 4\bm4}_{a_2}
    \end{pmatrix}\lambda_n\,.
\end{equation}
Note that \eqref{eq:clifford} is satisfied with $h^{(a_1,a_2),(b_1,b_2)} = (-1)^{b_1+b_2} \delta^{a_1,-b_1} \delta^{a_2,-b_2}$, and the fermion parity operator is again given by \eqref{eq:fp}.

\subsection{\texorpdfstring{$\SU(2)\times\SU(2)+\psi_{(\bm 2,\bm 4)}$}{SU(2) x SU(2) + (2,4)}}

When $G = \SU(2)\times\SU(2)$ and $\bm\lambda = (\bm 2,\bm 4)$, the on-site representation and the zero-weight count are
\begin{equation}\label{eq:Rsu2su224}
    R((\bm2,\bm4))= (\bm2,\bm4)\oplus(\bm3,\bm1)\oplus (\bm 1,\bm5)\,, \qquad n_0((\bm2,\bm4)) = 0\,.
\end{equation}
Note that the total dimension is $8+3+5=16$, which corresponds to the Hilbert space generated by $2\times4 =8$ Majorana fermions. Since $n_0 = 0$, we do not have any multiplicity degrees of freedom to keep track of, and so on each site all we have is a copy of $(\bm2,\bm4)\oplus(\bm3,\bm1)\oplus (\bm 1,\bm5)$.

To build $\chi^{a_1,a_2}_n$ (where $a_1$ is a $\bm{2}$-index and $a_2$ is a $\bm{4}$-index), we first need a Clebsch-Gordan symbol for $(\bm{2},\bm{4}) \otimes ((\bm2,\bm4)\oplus(\bm3,\bm1)\oplus (\bm 1,\bm5)) \to (\bm2,\bm4)\oplus(\bm3,\bm1)\oplus (\bm 1,\bm5)$. There are four arbitrary coefficients in such a Clebsch-Gordan symbol, and we fix them as
\begin{equation}\label{eq:Csu2su224}
C^{a_1a_2} = \begin{pmatrix}
        0&2C^{\bm 2\bm 2\bm3}_{a_1}\otimes C^{\bm 4\bm 4\bm1}_{a_2} & -2C^{\bm 2\bm 2\bm1}_{a_1}\otimes C^{\bm 4\bm 4\bm5}_{a_2} \\
        \sqrt{\frac 32}C^{\bm 2\bm 3\bm2}_{a_1}\otimes C^{\bm 4\bm 1\bm4}_{a_2} & 0 &0\\
        \sqrt{\frac 52}C^{\bm 2\bm 1\bm2}_{a_1}\otimes C^{\bm 4\bm 5\bm4}_{a_2} &0 &0
    \end{pmatrix}\,,
\end{equation}
so that we have
\begin{equation}
    \{C^{a_1a_2},C^{b_1b_2}\} = (-1)^{1+b_1+b_2}\delta^{a_1,-b_1}\delta^{a_2,-b_2}\,.
\end{equation}

It is not enough to set $\chi^{a_1,a_2}_n$ equal to $C^{a_1,a_2}$ acting on the $n$th site, because then the operators on different sites would commute instead of anticommuting. To fix this, we use
\begin{equation}
    F = \begin{pmatrix}
        -1&0&0\\
        0&1&0\\
        0&0&1
    \end{pmatrix}\,,
\end{equation}
which anticommutes with $C^{a_1,a_2}$. The lattice fermions are then given by
\begin{equation}
\chi^{a_1a_2}_n = F_1F_2\cdots F_{n-1}C^{a_1a_2}_n\,.
\end{equation}
Note that the full fermion parity operator is $\mathcal{F} = F_1 F_2 \cdots F_N$, but the action of $\mathcal{F}$ coincides with a gauge transformation in this theory (see Appendix~\ref{app:symmetries}), so we will not be needing it.

\subsection{\texorpdfstring{$\SU(4)+\psi_{\bm 15}$}{SU(4) + 15}}
For gauge group $\SU(4)$ with adjoint matter $\lambda=\bm{15}$, one finds (in agreement with the general rule derived in \cite{Dempsey:2023fvm}) that the on-site representation and the zero-weight count are
\begin{equation}
    R(\bm {15}) = \bm{64}\, , \qquad n_0 = 3\,.
\end{equation}
In this case, each site has three Majorana fermions in addition to a copy of $\bm{64}$. The explicit gauge-covariant construction of the fermion operator is rather unwieldy; in particular, it depends on the three linearly independent Clebsch-Gordan symbols for the fusion $\bm{64}\subset\bm{15}\otimes\bm{64}$, which do not have a standard form. For this reason, we refer to \cite{Dempsey:2024alw} for details of the construction of the lattice model for adjoint QCD$_2$ with arbitrary gauge group.

\section{Further evidence for the lattice decay rule}
\label{app:count_checks}

In Section~\ref{subsec:continuum_action}, we explained that the infrared description of a gauge theory with simple gauge group $G$ and Majorana fermions in a representation $\bm \lambda$ of $G$ is the WZW coset model $\frac{\SO(\dim \bm \lambda)_1}{G_{I(\bm \lambda)}}$, where $I(\bm \lambda)$ is the Dynkin index of $\bm \lambda$. In this paper, we are focused on theories for which this infrared model is a TQFT, meaning that $c_\text{IR}$ (given in \eqref{eq:cir}) vanishes.

In this case, we can calculate the number of degenerate vacua as follows. The numerator WZW model has chiral primaries with characters $\chi_{\bm \Lambda}$, where $\bm \Lambda$ are integrable representations of $\so(\dim\bm \lambda)_1$. These representations correspond to the singlet and vector representations\footnote{In the bosonic model, for which $(-1)^F$ is gauged, we also have to consider the character corresponding to the spinor representation(s) of $\Spin(\dim\bm \lambda)$. Since we are considering the fermionic model in this paper, we will only decompose the singlet and vector characters.} of $\SO(\dim\bm \lambda)$, which we label by $\bm 1$ and $\bm v$. The sum of these characters is given by \cite{Delmastro:2021otj}
\begin{equation}\label{eq:numerator_characters}
\begin{split}
    \chi_{\bm 1}(q) + \chi_v(q) &= q^{-\dim \bm \lambda/48}\prod_{r=1}^\infty \left(1 + q^{r-1/2}\right)^{\dim \bm \lambda}\,.
\end{split}
\end{equation}
The terms with integer powers of $q$ multiplied by $q^{-\dim\bm \lambda/48}$ belong to $\chi_1(q)$, and the terms with half-integer powers multiplied by the same factor belong to $\chi_v(q)$.

For a general WZW coset model, the numerator characters can be decomposed into the denominator characters using branching functions that are themselves nontrivial functions of $q$. In the cases we are considering, for which the coset model is a TQFT, the branching functions are in fact just nonnegative integers $b_{\bm \Lambda|\bm \mu}$, giving the decompositions
\begin{equation}\label{eq:branching}
    \chi_{\bm \Lambda}(q) = \sum_{\bm \mu} b_{\bm \Lambda|\bm \mu} \chi_{\bm \mu}(q)\,,
\end{equation}
where $\bm \mu$ runs over the integrable representations of the affine algebra for the denominator theory. The number of degenerate vacua is equal to $\sum b_{\bm \Lambda|\bm \mu}^2$; in the cases we will consider, we always have $b_{\bm \Lambda|\bm \mu} \in \{0,1\}$, so we just need to count the values that are 1, i.e., the total number of terms on the right hand sides of these decompositions.

Thus, to calculate the number of degenerate vacua, we need to calculate the characters $\chi_{\bm \mu}(q)$ of the denominator theory. These characters take the form of a polynomial in $q$ multiplied by $q^{m_{\bm \mu}}$, where $m_{\bm \mu}$ (called the modular anomaly) is given by (see (14.158) of \cite{DiFrancesco:1997nk})
\begin{equation}\label{eq:modular_anomaly}
    m_{\bm \mu} \equiv \frac{|\bm \mu + \bm \rho|^2}{2(I(\bm \lambda) + h^\vee)} - \frac{|\bm \rho|^2}{2h^\vee}\,.
\end{equation}
Here $\bm \rho$ is the Weyl vector of $\mathfrak{g}$. By comparing this with \eqref{eq:numerator_characters}, we see that we only need to consider representations $\bm \mu$ for which $2m_{\bm \mu} + \frac{\dim \bm \lambda}{24}$ is an integer. For those representations, we will use SageMath \cite{sagemath} to calculate the denominator characters, and then compare with \eqref{eq:numerator_characters} to fix the branching coefficients in \eqref{eq:branching}.

Once we know which integrable representations appear on the right-hand side of \eqref{eq:branching}, we can also determine how the vacua decompose into the flux tube sectors of the theory. The center symmetry of the gauge group, which distinguishes the flux tube sectors, can be identified with the outer automorphism group of the affine algebra \cite{DiFrancesco:1997nk}. Thus, we just need to determine the action of the outer automorphism group on the set of integrable representations appearing.

We illustrate this procedure in some detail in Appendix~\ref{sec:branching_su2_5} for our simplest example theory, $\SU(2)+\psi_{\bm{5}}$. We then give the results for our other three theories, as well as a large set of other examples, in the following subsections.

\subsection{\texorpdfstring{$\SU(2)+\psi_{\mathbf{5}}$}{SU(2) + 5}}\label{sec:branching_su2_5}

The $\SU(2)$ theory with fermions in the $\mathbf{5}$ representation is described by the coset $\frac{\SO(5)_1}{\SU(2)_{10}}$. The affine algebra for the denominator has 11 integrable representations, corresponding to the spin-0, spin-$\frac{1}{2}$, \ldots, and spin-5 representations of $\SU(2)$. To be consistent with the notation used in the following examples, we will use the Dynkin labels $[0]$, $[1]$, \ldots, $[10]$ for these representations.

We can calculate the modular anomalies of these representations using \eqref{eq:modular_anomaly}:
\begin{center}
    \begin{tabular}{c|ccccccccccc}
         $\mu$ & $[0]$ & $[1]$ & $[2]$ & $[3]$ & $[4]$ & $[5]$ & $[6]$ & $[7]$ & $[8]$ & $[9]$ & $[10]$ \\
         \hline
         $m_\mu+\frac{5}{48}$ & $0$ & $\frac{1}{16}$ & $\frac{1}{6}$ & $\frac{5}{16}$ & $\frac{1}{2}$ & $\frac{15}{32}$ & $1$ & $\frac{19}{16}$ & $\frac{5}{3}$ & $\frac{33}{16}$ & $\frac{5}{2}$
    \end{tabular}
\end{center}
From this we see that only $\mu \in \{[0],[4],[6],[10]\}$ could possibly appear in the decomposition of \eqref{eq:numerator_characters}.

Using SageMath \cite{sagemath}, we can calculate these four denominator characters:
\begin{equation}
\begin{split}
    \chi_{[0]}(q) &= q^{-5/48}\left(1 + 3q + 9q^2 + 22q^3 + 51q^4 + \ldots\right)\,, \\
    \chi_{[4]}(q) &= q^{19/48}\left(5 + 15q + 45q^2 + 110q^3 + 255q^4 + \ldots\right)\,, \\
    \chi_{[6]}(q) &= q^{43/48}\left(7 + 21q + 63q^2 + 154q^3 + 357q^4 + \ldots\right)\,, \\
    \chi_{[10]}(q) &= q^{115/48}\left(11 + 20q + 60q^2 + 125q^3 + 275q^4 + \ldots\right)\,.
\end{split}
\end{equation}
We can then compare this with \eqref{eq:numerator_characters}, which in this case gives
\begin{equation}
\begin{split}
    \chi_{\bm 1}(q) &= q^{-5/48}\left(1 + 10q + 30q^2 + 85q^3 + 205q^4 + \ldots\right)\,,\\
    \chi_{\bm v}(q) &= q^{19/48}\left(5 + 15q + 56q^2 + 130q^3 + 315q^4 + \ldots\right)\,.
\end{split}
\end{equation}
We see that
\begin{equation}
    \chi_{\bm 1}(q) = \chi_{[0]}(q) + \chi_{[6]}(q), \qquad \chi_{\bm v}(q) = \chi_{[4]}(q) + \chi_{[10]}(q)\,.
\end{equation}
This is how we arrive at the conclusion, listed in Table~\ref{tab:vacua}, that the theory has four degenerate vacua. This matches the number we computed using the lattice decay rule, as illustrated in Figure~\ref{fig:su2_5_decay}. 

Furthermore, the $\mathbb{Z}_2$ outer automorphism group of $\su(2)_{10}$ exchanges $[k] \leftrightarrow [10-k]$. We see that the representation of this group on the characters appearing in the branching rules decomposes into two copies of $p = 0$ representation and two copies of the $p = 1$ representation, which is how we find the splitting listed in Table~\ref{tab:vacua}.

\subsection{\texorpdfstring{$\SU(2)\times\SU(2)+\psi_{(\mathbf{3}, \mathbf{3})}$}{SU(2) x SU(2) + (3,3)}}

The coset theory is $\frac{\SO(9)_1}{\SU(2)_6 \times \SU(2)_6}$. The scalar character of the numerator,
\begin{equation}
    \chi_{\bm 1}(q) = q^{-3/16}\left(1 + 36q + 207q^2 + 957q^3 + \ldots\right),
\end{equation}
branches into the sum of the following denominator characters:
\begin{equation}
    \begin{split}
        \chi_{([0],[0])}(q) &= q^{-3/16}\left(1 + 6q + 27q^2 + 98q^3 + \ldots\right)\,,\\
        \chi_{([2],[4])}(q) &= q^{-3/16}\left(15q + 90q^2 + 405q^3 + \ldots\right)\,,\\
        \chi_{([4],[2])}(q) &= q^{-3/16}\left(15q + 90q^2 + 405q^3 + \ldots\right)\,,\\
        \chi_{([6],[6])}(q) &= q^{-3/16}\left(49q^3 + \ldots\right)\,.
    \end{split}
\end{equation}

The vector character of the numerator,
\begin{equation}
    \chi_{\bm v}(q) = q^{5/16}\left(9 + 93q + 459q^2 + 1827q^3 + \ldots\right),
\end{equation}
branches into the sum of the following denominator characters:
\begin{equation}
    \begin{split}
        \chi_{([2],[2])}(q) &= q^{5/16}\left(9 + 54q + 243q^2 + 882q^3 + \ldots\right)\,,\\
        \chi_{([4],[4])}(q) &= q^{5/16}\left(25q + 150q^2 + 675q^3 + \ldots\right)\,,\\
        \chi_{([0],[6])}(q) &= q^{5/16}\left(7q + 33q^2 + 135q^3 + \ldots\right)\,,\\
        \chi_{([6],[0])}(q) &= q^{5/16}\left(7q + 33q^2 + 135q^3 + \ldots\right)\,.
    \end{split}
\end{equation}
This is how we arrive at the conclusion, listed in Table~\ref{tab:vacua}, that the theory has eight degenerate vacua. This matches the number we computed using the lattice decay rule, as illustrated in Figure~\ref{fig:su2su2_33_decay}.

Furthermore, the $\mathbb{Z}_2\times \mathbb{Z}_2$ outer automorphism group of $\su(2)_6\times \su(2)_6$ is generated by the exchanges $([k_1],[k_2]) \leftrightarrow ([6-k_1],[k_2])$ and $([k_1],[k_2]) \leftrightarrow ([k_1],[6-k_2])$. We see that the representation of this group on the characters appearing in the branching rules decomposes into two copies of each of the four irreps of $\mathbb{Z}_2\times\mathbb{Z}_2$, which is how we deduce the splitting listed in Table~\ref{tab:vacua}.

\subsection{\texorpdfstring{$\SU(2)\times\SU(2)+\psi_{(\mathbf{2}, \mathbf{4})}$}{SU(2) x SU(2) + (2,4)}}

The coset theory is $\frac{\SO(8)_1}{\SU(2)_2 \times \SU(2)_{10}}$. The scalar character of the numerator,
\begin{equation}
    \chi_{\bm 1}(q) = q^{-1/6}\left(1 + 28q + 134q^2 + 568q^3 + \ldots\right),
\end{equation}
branches into the sum of the following denominator characters:
\begin{equation}
    \begin{split}
        \chi_{([0],[0])}(q) &= q^{-1/6}\left(1+6 q+27 q^2+91 q^3+\ldots\right)\,,\\
        \chi_{([0],[6])}(q) &= q^{-1/6}\left(7 q+42 q^2+189 q^3+\ldots\right)\,,\\
        \chi_{([2],[4])}(q) &= q^{-1/6}\left(15 q+65 q^2+255 q^3+\ldots\right)\,,\\
        \chi_{([2],[10])}(q) &= q^{-1/6}\left(33q^3 + \ldots\right)\,.
    \end{split}
\end{equation}

The vector character of the numerator,
\begin{equation}
    \chi_{\bm v}(q) = q^{1/3}\left(8 + 64q + 288q^2 + 1024q^3 + \ldots\right)\,,
\end{equation}
branches into the sum of the following denominator characters:
\begin{equation}
    \begin{split}
        \chi_{([1],[3])}(q) &= q^{1/3}\left(8+48 q+192 q^2+640 q^3+\ldots\right)\,,\\
        \chi_{([1],[7])}(q) &= q^{1/3}\left(16 q+96 q^2+384 q^3 + \ldots\right)\,.
    \end{split}
\end{equation}
This is how we arrive at the conclusion, listed in Table~\ref{tab:vacua}, that the theory has six degenerate vacua. This matches the number we computed using the lattice decay rule, as illustrated in Figure~\ref{fig:su2su2_24_decay}.

Furthermore, the preserved $\mathbb{Z}_2$ subgroup of the center symmetry acts as $([k_1],[k_2]) \leftrightarrow ([2-k_1],[10-k_2])$ on integrable representations of $\su(2)_2\times \su(2)_{10}$. We see that the representation of this symmetry on the characters appearing in the branching rules decomposes into three copies of each irrep of $\mathbb{Z}_2$, which is how we deduce the splitting listed in Table~\ref{tab:vacua}.

\subsection{\texorpdfstring{$\SU(4)+\psi_{\mathbf{15}}$}{SU(4) + 15}}\label{sec:branching_su4_15}

The coset theory is $\frac{\SO(15)_1}{\SU(4)_4}$. The scalar character of the numerator,
\begin{equation}
\chi_{\bm 1}(q) = q^{-5/16} \left(1 + 105q + 1590q^{2} + \ldots\right)\,,
\end{equation}
branches into the sum of the following denominator characters:
\begin{equation}
\begin{split}
\chi_{[0,0,0]}(q) &= q^{-5/16} \left(1 + 15q + 135q^{2} + \ldots\right)\,, \\
\chi_{[0,1,2]}(q) &= q^{-5/16} \left(45q + 675q^{2} + \ldots\right)\,, \\
\chi_{[2,1,0]}(q) &= q^{-5/16} \left(45q + 675q^{2} + \ldots\right)\,, \\
\chi_{[0,4,0]}(q) &= q^{-5/16} \left(105q^{2} + \ldots\right)\,.
\end{split}
\end{equation}

The vector character of the numerator,
\begin{equation}
\chi_{\bm v}(q) = q^{11/16} \left(15 + 470q + \ldots\right)\,,
\end{equation}
branches into the sum of the following denominator characters:
\begin{equation}
\begin{split}
\chi_{[1,0,1]}(q) &= q^{11/16} \left(15 + 225q + \ldots\right)\,, \\
\chi_{[0,0,4]}(q) &= q^{11/16} \left(35q + \ldots\right)\,, \\
\chi_{[1,2,1]}(q) &= q^{11/16} \left(175q + \ldots\right)\,, \\
\chi_{[4,0,0]}(q) &= q^{11/16} \left(35q + \ldots\right)\,.
\end{split}
\end{equation}
This is how we arrive at the conclusion, listed in Table~\ref{tab:vacua}, that the theory has eight degenerate vacua. This matches the number we computed using the lattice decay rule, as illustrated in Figure~\ref{fig:su4_15_decay}.

Furthermore, the $\mathbb{Z}_4$ outer automorphism group of $\su(4)_4$ is generated by $[k_1,k_2,k_3] \mapsto [4-k_1-k_2-k_3,k_1,k_2]$. The representation of this group on the characters appearing in the branching rule decomposes into two copies of each of the irreps of $\mathbb{Z}_4$, which is how we deduce the splitting listed in Table~\ref{tab:vacua}.

\subsection{\texorpdfstring{$\mathrm{G}_2+\psi_{\mathbf{14}}$}{G2 + 14}}

Whenever the fermions are in the adjoint representation, there are $2^{\rk G}$ degenerate vacua. We saw an example of this fact in Appendix~\ref{sec:branching_su4_15}. Here we give another example, the gauge group $\mathrm{G}_2$ of rank $2$ with fermions in the 14-dimensional adjoint representation.

The coset theory is $\frac{\SO(14)_1}{(\mathrm{G}_2)_4}$. The scalar character of the numerator,
\begin{equation}
\chi_{\bm 1}(q) = q^{-7/24} \left(1 + 91q + 1197q^{2} + 8386q^{3} + \ldots\right)\,,
\end{equation}
branches into the sum of the following denominator characters:
\begin{equation}
\begin{split}
\chi_{[0,0]}(q) &= q^{-7/24} \left(1 + 14q + 119q^{2} + 770q^{3} + \ldots\right)\,, \\
\chi_{[3,0]}(q) &= q^{-7/24} \left(77q + 1078q^{2} + 7616q^{3} + \ldots\right)\,.
\end{split}
\end{equation}

The vector character of the numerator,
\begin{equation}
\chi_{\bm v}(q) = q^{5/24} \left(14 + 378q + 3290q^{2} + 20008q^{3} + \ldots\right)\,,
\end{equation}
branches into the sum of the following denominator characters:
\begin{equation}
\begin{split}
\chi_{[0,1]}(q) &= q^{5/24} \left(14 + 196q + 1666q^{2} + 10032q^{3} + \ldots\right)\,, \\
\chi_{[4,0]}(q) &= q^{5/24} \left(182q + 1624q^{2} + 9976q^{3} + \ldots\right)\,.
\end{split}
\end{equation}

In total, there are four degenerate vacua (and since $G_2$ has trivial center, they all belong to the same flux tube sector). To count the vacua using the lattice decay rule, we would use the fact that $\bm{R}(\adj) = [11\cdots 1]$ (see Table~\ref{tab:repR}). We then find that the strong-coupling vacua chosen by the lattice decay rule are those for which all Dynkin labels are either 0 or 1. This immediately gives four vacua in this case, with strong-coupling labels $\{\bm{1}, \bm{64}\}$, $\{\bm{7},\bm{14}\}$, $\{\bm{14},\bm{7}\}$, and $\{\bm{64}, \bm{1}\}$. (The representations $\bm{1}$, $\bm{7}$, $\bm{14}$, and $\bm{64}$ have Dynkin labels [00], [10], [01], and [11] respectively.) 

More generally, it is straightforward to convince oneself that the lattice decay rule will always give $2^{\rk G}$ vacua, namely those for which the Dynkin indices of link representations are all 0 or 1, for adjoint QCD$_2$ with any gauge group.

\subsection{\texorpdfstring{$\mathrm{Spin}(9) + \psi_{\mathbf{16}}$}{Spin(9) + 16}}

For a $\text{Spin}(9)$ theory with fermions in the $\bm{16}$ (spinor) representation, the coset theory is $\frac{\SO(16)_1}{\Spin(9)_2}$. The scalar character of the numerator,
\begin{equation}
\chi_{\bm 1}(q) = q^{-3/4} \left(1 + 120q + 2076q^{2} + 17344q^{3} + \ldots\right)\,,
\end{equation}
branches into the sum of the following denominator characters:
\begin{equation}
\begin{split}
\chi_{[0,0,0,0]}(q) &= q^{-3/4} \left(1 + 36q + 702q^{2} + 5764q^{3} + \ldots\right)\,, \\
\chi_{[0,0,1,0]}(q) &= q^{-3/4} \left(84q + 1374q^{2} + 11580q^{3} + \ldots\right)\,.
\end{split}
\end{equation}

The vector character of the numerator,
\begin{equation}
\chi_{\bm v}(q) = q^{-1/4} \left(16 + 576q + 6304q^{2} + 44416q^{3} + \ldots\right)\,,
\end{equation}
is equal to the following denominator character:
\begin{equation}
\begin{split}
\chi_{[0,0,0,1]}(q) &= q^{-1/4} \left(16 + 576q + 6304q^{2} + 44416q^{3} + \ldots\right)\,.
\end{split}
\end{equation}
Thus, in total, there are three degenerate vacua (and $Z_{\bm{16}}(\SO(9))$ is trivial, so they all belong to the same flux tube sector). Using the lattice decay rule with $\bm{R}(\bm{16}) = \bm{44}\oplus\bm{84}\oplus\bm{128}$ (see Table~\ref{tab:repR}), we also find three vacua, with strong-coupling labels $\{\bm{1},\bm{44}\oplus\bm{84}\oplus\bm{128}\}$, $\{\bm{44}\oplus\bm{84}\oplus\bm{128},\bm{1}\}$, and $\{\bm{9}, \bm{9}\}$. 

\subsection{\texorpdfstring{$\mathrm{F}_4 + \psi_{\mathbf{26}}$}{F4 + 26}}
For an $\text{F}_4$ theory with fermions in the $\bm{26}$ representation, the coset theory is $\frac{\SO(26)_1}{(F_4)_3}$. scalar character of the numerator,
\begin{equation}
\chi_{\bm 1}(q) = q^{-13/12} \left(1 + 325q + 15626q^{2} + 298831q^{3} + \ldots\right)\,,
\end{equation}
branches into the sum of the following denominator characters:
\begin{equation}
\begin{split}
\chi_{[0,0,0,0]}(q) &= q^{-13/12} \left(1 + 52q + 1430q^{2} + 27560q^{3} + \ldots\right)\,, \\
\chi_{[0,0,1,0]}(q) &= q^{-13/12} \left(273q + 14196q^{2} + 271271q^{3} + \ldots\right)\,.
\end{split}
\end{equation}

The vector character of the numerator,
\begin{equation}
\chi_{\bm v}(q) = q^{-7/12} \left(26 + 2626q + 74256q^{2} + 1063426q^{3} + \ldots\right)\,,
\end{equation}
branches into the sum of the following denominator characters:
\begin{equation}
\begin{split}
\chi_{[0,0,0,1]}(q) &= q^{-7/12} \left(26 + 1352q + 37180q^{2} + 531804q^{3} + \ldots\right)\,, \\
\chi_{[0,1,0,0]}(q) &= q^{-7/12} \left(1274q + 37076q^{2} + 531622q^{3} + \ldots\right)\,.
\end{split}
\end{equation}

In total, there are four degenerate vacua (and $F_4$ has trivial center, so they all belong to the same flux tube sector). Using the lattice decay rule, with $\bm{R}(\bm{26}) = \bm{4096}$ (see Table~\ref{tab:repR}), we indeed find four vacua with strong-coupling labels with $\{\bm{1},\bm{4096}\}$, $\{\bm{26},\bm{273}\}$, $\{\bm{273},\bm{26}\}$, and $\{\bm{4096},\bm{1}\}$.

\subsection{\texorpdfstring{$\USp(8) + \psi_{\mathbf{42}}$}{USp(8) + 42}}

The most complicated example we will consider is a $\USp(8)$ theory with fermions in the $\bm{42}$ representation. The coset theory is $\frac{\SO(42)_1}{\USp(8)_7}$, and the scalar character of the numerator,
\begin{equation}
\begin{split}
\chi_{\bm 1}(q) = q^{-3/4} \Big(1 &+ 861q + 113694q^{2} + 5730571q^{3} + 154985250q^{4} \\
&+ 2739120048q^{5} + 35804625717q^{6} + 373995678582q^{7} + \ldots\Big)\,,
\end{split}
\end{equation}
branches into the sum of the following denominator characters:
{\small
\begin{equation}
\begin{split}
\chi_{[0,0,0,0]}(q) &= q^{-3/4} \Big(1 + 36q + 702q^{2} + 9768q^{3} + 108225q^{4} + 1012284q^{5} \\
&\hspace{.8in}+ 8287038q^{6} + 60839064q^{7} + \ldots\Big)\,, \\
\chi_{[0,0,2,0]}(q) &= q^{-3/4} \Big(825q + 29700q^{2} + 579150q^{3} + 8058600q^{4} + 89285625q^{5}\\
&\hspace{.8in} + 835134300q^{6} + 6827245227q^{7} + \ldots\Big)\,, \\
\chi_{[0,3,0,1]}(q) &= q^{-3/4} \Big(41250q^{2} + 1485000q^{3} + 28957500q^{4}+ 402930000q^{5} \\
&\hspace{.8in} + 4442807070q^{6} + 41050263270q^{7} + \ldots\Big)\,, \\
\chi_{[1,0,3,0]}(q) &= q^{-3/4} \Big(42042q^{2} + 1513512q^{3} + 29513484q^{4}+ 410666256q^{5} \\
&\hspace{.8in} + 4534156550q^{6} + 41988243528q^{7} + \ldots\Big)\,, \\
\chi_{[1,4,1,0]}(q) &= q^{-3/4} \Big(594594q^{3} + 21405384q^{4} + 405050646q^{5}+ 5363237880q^{6} \\
&\hspace{.8in} + 55770858858q^{7} + \ldots\Big)\,, \\
\chi_{[2,2,0,2]}(q) &= q^{-3/4} \Big(833833q^{3} + 30017988q^{4} + 570534195q^{5}+ 7611484188q^{6} \\
&\hspace{.8in} + 79840343583q^{7} + \ldots\Big)\,, \\
\chi_{[0,2,2,1]}(q) &= q^{-3/4} \Big(714714q^{3} + 25729704q^{4} + 501729228q^{5} + 6865688102q^{6} \\
&\hspace{.8in}+ 73553002110q^{7} + \ldots\Big)\,, \\
\chi_{[1,4,1,1]}(q) &= q^{-3/4} \left(6441435q^{4} + 197475993q^{5} + 3282923358q^{6} + 39017734899q^{7} + \ldots\right)\,, \\
\chi_{[3,0,1,3]}(q) &= q^{-3/4} \left(2807805q^{4} + 88471383q^{5} + 1517133618q^{6} + 18574702146q^{7} + \ldots\right)\,, \\
\chi_{[0,3,0,3]}(q) &= q^{-3/4} \left(1945125q^{4} + 70024500q^{5} + 1284949575q^{6} + 16345888575q^{7} + \ldots\right)\,, \\
\chi_{[2,0,0,5]}(q) &= q^{-3/4} \left(1939938q^{5} + 58824038q^{6} + 965342196q^{7} + \ldots\right) \\
\chi_{[0,0,0,7]}(q) &= q^{-3/4} \left(1215126q^{7} + \ldots\right)\,.
\end{split}
\end{equation}
}

The vector character of the numerator,
\begin{equation}
\chi_{\bm v}(q) = q^{-1/4} \left(42 + 11522q + 886872q^{2} + 31751754q^{3} + 680921934q^{4} + 10225595982q^{5} + \ldots\right)\,,
\end{equation}
branches into the sum of the following denominator characters:
\begin{equation}
\begin{split}
\chi_{[0,0,0,1]}(q) &= q^{-1/4} \left(42 + 1512q + 29484q^{2} + 410256q^{3} + 4545450q^{4} + 42515928q^{5} + \ldots\right)\,, \\
\chi_{[0,1,2,0]}(q) &= q^{-1/4} \left(10010q + 360360q^{2} + 7027020q^{3} + 97777680q^{4} + 1083332250q^{5} + \ldots\right)\,, \\
\chi_{[0,5,0,0]}(q) &= q^{-1/4} \left(47124q^{2} + 1696464q^{3} + 33081048q^{4} + 454329624q^{5} + \ldots\right)\,, \\
\chi_{[1,2,1,1]}(q) &= q^{-1/4} \left(396396q^{2} + 14270256q^{3} + 278269992q^{4} + 3838462914q^{5} + \ldots\right)\,, \\
\chi_{[0,0,4,0]}(q) &= q^{-1/4} \left(53508q^{2} + 1926288q^{3} + 37562616q^{4} + 522666144q^{5} + \ldots\right)\,, \\
\chi_{[2,4,0,1]}(q) &= q^{-1/4} \left(1793220q^{3} + 56690920q^{4} + 975700440q^{5} + \ldots\right)\,, \\
\chi_{[4,0,0,3]}(q) &= q^{-1/4} \left(617100q^{3} + 19721394q^{4} + 343412784q^{5} + \ldots\right)\,, \\
\chi_{[0,4,2,0]}(q) &= q^{-1/4} \left(1222650q^{3} + 44015400q^{4} + 814571550q^{5} + \ldots\right)\,, \\
\chi_{[1,2,1,2]}(q) &= q^{-1/4} \left(2788500q^{3} + 100386000q^{4} + 1890460572q^{5} + \ldots\right)\,, \\
\chi_{[0,5,0,2]}(q) &= q^{-1/4} \left(4806802q^{4} + 134190420q^{5} + \ldots\right)\,, \\
\chi_{[2,1,0,4]}(q) &= q^{-1/4} \left(4064632q^{4} + 125558004q^{5} + \ldots\right)\,, \\
\chi_{[0,0,0,6]}(q) &= q^{-1/4} \left(395352q^{5} + \ldots\right)\,.
\end{split}
\end{equation}

In total, there are 24 degenerate vacua. The $\mathbb{Z}_2$ outer automorphism group of $\sp(8)_7$ acts as $[k_1,k_2,k_3,k_4] \leftrightarrow [k_3,k_2,k_1,7-k_1-k_2-k_3-k_4]$. We see that the representation of this group on the characters appearing in the branching rules decomposes into 12 copies of each irrep of $\mathbb{Z}_2$.

We can calculate the same number using the lattice decay rule. The lattice decay rule with $\bm{R}(\bm{42}) = [1130]\oplus [3111]\oplus [5110] = \bm{344064}\oplus\bm{524288}\oplus\bm{180224}$ also gives 24 vacua, 12 in each flux tube sector. Those in the trivial flux tube sector are
\begin{equation}
    \begin{aligned}
        &\{\bm{1}, \bm{344064}\oplus\bm{524288}\oplus\bm{180224}\}\,, & \{\bm{344064}\oplus\bm{524288}\oplus\bm{180224}, \bm{1}\}&\,, \\
        &\{\bm{27}, \bm{42042}\oplus\bm{63063}\oplus\bm{21021}\}\,, & \{\bm{42042}\oplus\bm{63063}\oplus\bm{21021}, \bm{27}\}&\,, \\
        &\{\bm{315}, \bm{10010}\oplus\bm{14300}\oplus\bm{4290}\}\,, & \{\bm{10010}\oplus\bm{14300}\oplus\bm{4290}, \bm{315}\}&\,,\\
        &\{\bm{4096}, \bm{825}\oplus\bm{1155}\oplus\bm{330}\}\,, & \{\bm{825}\oplus\bm{1155}\oplus\bm{330}, \bm{4096}\}&\,, \\
        &\{\bm{36}, \bm{36864}\}\,, & \{\bm{36864}, \bm{36}\}&\,, \\
        &\{\bm{594}, \bm{3696}\}\,, & \{\bm{3696}, \bm{594}\}&\,,
    \end{aligned}
\end{equation}
and those in the nontrivial flux tube sector are
\begin{equation}
    \begin{aligned}
        &\{\bm{8}, \bm{85800}\}\,, & \{\bm{85800}, \bm{8}\}&\,, \\
        &\{\bm{48}, \bm{47040}\}\,, & \{\bm{47040}, \bm{48}\}&\,, \\
        &\{\bm{120}, \bm{13728}\}\,, & \{\bm{13728}, \bm{120}\}&\,, \\
        &\{\bm{160}, \bm{9360}\}\,, & \{\bm{9360}, \bm{160}\}&\,, \\
        &\{\bm{792}, \bm{4752}\}\,, & \{\bm{4752}, \bm{792}\}&\,, \\
        &\{\bm{1232}, \bm{1728}\}\,, & \{\bm{1728}, \bm{1232}\}&\,.
    \end{aligned}
\end{equation}

\bibliographystyle{ssg}
\bibliography{refs}

\end{document}